\documentclass[twocolumn]{aastex63}
\usepackage{mathtools}
\usepackage[maxfloats=40]{morefloats} 
\usepackage{color}
\usepackage{amsmath}
\usepackage[figure,figure*]{hypcap}
\DeclarePairedDelimiter\abs{\lvert}{\rvert}%
\DeclarePairedDelimiter\norm{\lVert}{\rVert}%

\makeatletter
\let\oldabs\abs
\def\abs{\@ifstar{\oldabs}{\oldabs*}}
\let\oldnorm\norm
\def\norm{\@ifstar{\oldnorm}{\oldnorm*}}
\makeatother

\newcommand{\Rom}[1]{\uppercase\expandafter{\romannumeral #1\relax}}
\def \mjya {mJy~arcsec$^{-2}$}
\def \kms {km~s$^{-1}$}
\def\C18O{\textrm{C$^{18}$O}}
\def\13CO{\textrm{$^{13}$CO}}
\def\nh3{\textrm{NH$_{3}$}}
\def \IG {$\nabla$I}

\begin{document}


\title{Filamentary Network and Magnetic Field Structures Revealed with BISTRO in the High-Mass Star-Forming Region NGC2264 -- Global Properties and Local Magneto-Gravitational Configurations}

\author[0000-0002-6668-974X]{Jia-Wei Wang}
\affiliation{Academia Sinica Institute of Astronomy and Astrophysics, No.1, Sec. 4., Roosevelt Road, Taipei 10617, Taiwan}

\author[0000-0003-2777-5861]{Patrick M. Koch}
\affiliation{Academia Sinica Institute of Astronomy and Astrophysics, No.1, Sec. 4., Roosevelt Road, Taipei 10617, Taiwan}

\author[0000-0001-9751-4603]{Seamus D. Clarke}
\affiliation{Academia Sinica Institute of Astronomy and Astrophysics, No.1, Sec. 4., Roosevelt Road, Taipei 10617, Taiwan}

\author[0000-0001-8509-1818]{Gary Fuller}
\affiliation{Jodrell Bank Centre for Astrophysics, School of Physics and Astronomy, University of Manchester, Oxford Road, Manchester, M13 9PL, UK}
\affiliation{Physikalisches Institut, University of Cologne, Zülpicher Str. 77, D-50937 Köln, Germany}

\author[0000-0002-6893-602X]{Nicolas Peretto}
\affiliation{School of Physics and Astronomy, Cardiff University, The Parade, Cardiff, CF24 3AA, UK}

\author[0000-0002-0675-276X]{Ya-Wen Tang}
\affiliation{Academia Sinica Institute of Astronomy and Astrophysics, No.1, Sec. 4., Roosevelt Road, Taipei 10617, Taiwan}

\author[0000-0003-1412-893X]{Hsi-Wei Yen}
\affiliation{Academia Sinica Institute of Astronomy and Astrophysics, No.1, Sec. 4., Roosevelt Road, Taipei 10617, Taiwan}

\author[0000-0001-5522-486X]{Shih-Ping Lai}
\affiliation{Institute of Astronomy and Department of Physics, National Tsing Hua University, Hsinchu 30013, Taiwan}
\affiliation{Academia Sinica Institute of Astronomy and Astrophysics, No.1, Sec. 4., Roosevelt Road, Taipei 10617, Taiwan}

\author[0000-0003-0998-5064]{Nagayoshi Ohashi}
\affiliation{Academia Sinica Institute of Astronomy and Astrophysics, No.1, Sec. 4., Roosevelt Road, Taipei 10617, Taiwan}

\author[0000-0002-1959-7201]{Doris Arzoumanian}
\affiliation{Division of Science, National Astronomical Observatory of Japan, 2-21-1 Osawa, Mitaka, Tokyo 181-8588, Japan}

% ...


\begin{abstract}
We report 850 $\mu$m continuum polarization observations toward the filamentary high-mass star-forming
region NGC 2264, taken as part of the B-fields In STar forming
Regions Observations (BISTRO) large program on the James Clerk Maxwell Telescope (JCMT). These data reveal a
well-structured non-uniform magnetic field in the NGC 2264C and 2264D regions with a prevailing orientation
around 30\degr\ from north to east. Field strengths estimates and a virial analysis for the major clumps indicate that NGC 2264C is globally dominated by gravity while in 2264D magnetic, gravitational, and kinetic energies are roughly balanced. We present an analysis scheme that utilizes the locally resolved magnetic field structures, together with the locally measured gravitational vector field and the extracted filamentary network. From this, we infer statistical trends showing that this network consists of two main groups of filaments oriented approximately perpendicular to one another. Additionally, gravity shows one dominating converging direction that is
roughly perpendicular to one of the filament orientations, which is suggestive of mass accretion along this
direction. Beyond these statistical trends,
we identify two types of filaments. The type-I filament is perpendicular to the magnetic field with local gravity transitioning from parallel to perpendicular to the magnetic field from the outside to the filament ridge. The type-II
filament is parallel to the magnetic field and local gravity. We
interpret these two types of filaments as originating from the competition between radial
collapsing, driven by filament self-gravity, and the longitudinal collapsing, driven
by the region's global gravity.

\end{abstract}


\section{Introduction}
Observations over the last decade have shown that interstellar filaments are ubiquitous within molecular clouds, and that they can be major sites of star formation \citep{sc79,my09,an10,ar11,an14,chung19,ku20}. 
Understanding how these filaments form is therefore crucial to determining the initial stage of star formation. 
Hub-filament structures (HFSs), consisting of a massive hub connecting with several converging filaments, are special filamentary configurations that have recently drawn much attention, because they are now commonly identified in nearby massive star-forming regions \citep[e.g.,][]{my09,sc12,mo18,liuHL23}. \citet{ku20} show that all massive clumps within 2 kpc detected in the \textbf{\it Herschel} Hi-GAL survey are located in the hubs of HFSs. Recent ALMA observations suggest that self-similar hierarchical HFSs, where small-scale HFSs are embedded in the hub of large-scale HFSs, are possibly common in massive proto-clusters \citep{zh22}. All these observational results point to HFSs being an essential step during the formation of massive stars and clusters.

Theoretically, the formation mechanism of filaments and their assembly into a hub remains a topic of debate, with several proposed processes including layer fragmentation threaded by magnetic fields \citep[][]{my09,va14}, magnetic-field-channeled gravitational collapse \citep[e.g.,][]{na08}, shock/turbulence compression \citep[e.g.,][]{pa01a,fe16}, hierarchical collapse in molecular clouds \citep[e.g.,][]{go14,va17,go18}, and filament-filament collisions \citep[e.g.,][]{na14,do14,fr15,ku20}.The magnetic field can be one of the key factors determining the dominating mechanism in these processes. A strong large-scale magnetic field might be important in guiding large-scale accretion flows \citep{li14,pal13}, gravitational collapse driven by MHD turbulence \citep{na08}, or layer fragmentation \citep{my09,va14}.

Observationally, magnetic fields are commonly found to correlate with the orientation of interstellar filaments \citep{hey08,li08,bu13,pal13,pl16,hw22}, suggesting that they play an important role in the filament formation. The correlation between filaments and magnetic fields likely varies with column densities, along or across filaments \citep{pl16,ar20,pi20,kw22}, implying that the role of the magnetic field possibly evolves locally. These variations could be more significant in HFSs, because the local densities within filaments can dramatically change due to the strong gravitational field of a massive hub \citep{ar20,wa21,chung22}. However, studies addressing the local variations of the magnetic field on subparsec-scale are still rare, and hence the exact role of the magnetic field remains unclear.

Most studies on interstellar magnetic fields focus on global, averaged properties of magnetic fields, such as mass-to-magnetic flux ratio, virial parameters, and Alfv\'{e}nic Mach number. This is due to the difficulties and challenges in both probing the detailed magnetic field morphologies and extracting quantitative information. These analyses on averaged quantities are useful to describe relatively simple systems, but become insufficient for complex systems, such as hierarchical filamentary networks within HFSs.
Recently, 
new analytical methods have been proposed to extract local properties of magnetic fields.
These include the
polarization-intensity gradient method \citep{koch12a,koch12b,koch13}, the histogram of relative orientation analysis \citep{so13,pl16,soler19,kw22}, spatial distributions of magnetic field strengths \citep{wa20,hw21,hw22}, and relative orientations among gravity, magnetic fields, and density structures \citep{ko18,wa21,ko22,wa22,liu23}.
In this paper, we aim at 
building an analysis scheme highlighting the local physical conditions within HFSs, in order to reveal the variations in the role of the magnetic field.

NGC 2264 is an active cluster-forming region populated by hundreds of young stellar objects (YSOs) \citep{su09,ra14}, embedded in the Mon OB1 molecular cloud complex. It is located at a Gaia-DR2 determined distance of $715\substack{+81 \\ -42}$ pc \citep{zu20}. \citet{bu12} identified six parsec-scale filaments converging towards NGC 2264 from $^{12}$CO (3–2) and H$_2$ 1-0 S(1) wide-field images, revealing that NGC 2264 is the hub center of a 10-pc-sized HFS. \citet{ku20} further identified filamentary structures of young stars joining at NGC 2264. Since the radial velocities of these stars in different filaments fall into different velocity subgroups \citep{to15}, they proposed that NGC 2264 originated from a collision of stellar filaments. Within the hub center, 
high-resolution millimeter and sub-millimeter continuum observations reveal an HFS-like morphology, where filamentary structures with lengths around 0.5--2 pc converge towards dense cores 
\citep{pe06,bu15}. Hence, this likely points at self-similar hierarchical HFSs. Based on this, NGC 2264 is an ideal source to investigate the formation of hierarchical HFSs.

NGC 2264 primarily consists of two massive, supervirial cluster-forming clumps \citep{pe06}, embedded in the hub of the 10-pc sized HFS. The southern 1650 M$_{\sun}$ clump \citep{pe06} hosts the bright 9.5 M$_{\sun}$ B2 zero-age main-sequence (ZAMS) star IRAS 06384+0932 (IRS1), also known as Allen’s source \citep{al72}, which is associated with the molecular outflow 2264C \citep{ma88}. The northern 1310 M$_{\sun}$ clump \citep{pe06} contains a class I YSO, IRAS 06382+0939 (IRS2), which is associated with the molecular outflow NGC 2264D \citep{ma88}. These two clumps are commonly named after their two dominating sources (IRS1 and IRS2) or their associated outflows (NGC 2264C and NGC 2264D). For clarity, we use NGC 2264C and NGC 2264D (or 2264C and 2264D) when referring to the two massive clumps, and IRS1 and IRS2 when referring to the two sources.

In this paper, we report 850 $\mu$m continuum polarization observations toward NGC 2264, using the James Clerk Maxwell Telescope (JCMT) POL-2 polarimeter, as part of the B-fields In STar forming Regions Observations (BISTRO) large program \citep{wa17}. These observations, with a resolution of 14\arcsec allow us to probe the detailed magnetic field morphology within 2264C and 2264D. The goal of the paper is to study how a hierarchical HFS can be formed by investigating how gravity and magnetic fields interact and link to hubs and filaments. In \autoref{sec:obs}, we present the observations and data reduction. \autoref{sec:results} reports the observed intensity and magnetic field morphology. In \autoref{sec:ana}, we present how to extract spatial parameters from the observed data. Statistical analyses are performed to identify possible trends and correlations of these parameters, both using maps probing the local spatial properties and histograms highlighting the statistical properties. 
The roles of gravity and magnetic field in the formation of NGC 2264 are discussed in \autoref{sec:dis}. Our conclusions are summarized in \autoref{sec:con}.

\section{Observations}\label{sec:obs}
\subsection{JCMT POL-2 Observations}\label{sec:pol2}
We carried out polarization continuum observations toward NGC 2264 with SCUBA-2 and POL-2 mounted on the JCMT (project code M17BL011) between November 2017 and February 2019. The two-pointing mosaic observations targeted 2264C (south) and 2264D (north) with the reference positions (R.A., Dec.)=(6$^{h}$41$^{m}$11.97$^{s}$, 9\degr29\arcmin44\farcs5) and (6$^{h}$41$^{m}$01.5$^{s}$, 9\degr35\arcmin39\farcs5). We performed 21 sets of 40-minute observations for each pointing under a  weather condition with $\tau_{225 GHz}$ ranging from 0.02 to 0.06. The POL-2 DAISY scan mode \citep{fr16} was adopted, producing a fully sampled circular region with a diameter of 11\arcmin\ for each pointing and a resolution of 14.1\arcsec (corresponding to 0.05 pc at a source distance of 715 pc). 
Polarization was simultaneously observed in 
both the 450 $\mu$m and 850 $\mu$m continuum bands. This paper focuses on the 850 $\mu$m data.

The POL-2 850 $\mu$m polarization data were reduced with $pol2map$\footnote{http://starlink.eao.hawaii.edu/docs/sc22.pdf} in the \textsc{smurf} package\footnote{version 2021 Dec 2} \citep{be05,ch13}. The $skyloop$ mode was invoked, which is a script that runs map-making on the full set of observations in order to find a solution that minimizes residuals across the full set of maps. The \textit{MAPVARS} mode was activated to estimate the uncertainties from the standard deviation among the individual observations, accounting for the instrumental and map-making uncertainties. The details of the data reduction steps and procedure are described in previous BISTRO papers \citep[e.g.,][]{wa19,pa21}. The POL-2 data reduction was done with a 4\arcsec\ pixel size, because larger pixel sizes might increase the uncertainty during the map-making process.

The output Stokes I, Q, and U images were calibrated in units of Jy/arcsec$^{-2}$, using an aperture flux conversion factor (FCF) of 
2.79 Jy$~$pW$^{-1}$~arcsec$^{-2}$ (including a factor of 1.35 for POL-2, equivalent to 630 Jy$~$pW$^{-1}$~beam$^{-1}$) for extended sources \citep{ma21}, and binned to a pixel size of 12\arcsec\ to improve the sensitivity. The typical rms noise of the final Stokes I, Q, and U maps is $\sim$0.01 \mjya\ at the map center, and gradually increases to $\sim$ 0.04 \mjya\ near the edge of the mapped region. The Stokes I image has a higher rms noise of up to $\sim$0.07 \mjya\ for pixels with large intensities ($I>6$ \mjya) where the rms noise is dominated by the map-making uncertainties. The calculated polarization fraction $P$ was debiased using the asymptotic estimator \citep{wa74} as
\begin{equation}\label{eq:debias}
P=\frac{1}{I}\sqrt{(Q^2+U^2)-\frac{1}{2}(\sigma_{Q}^2+\sigma_{U}^2)}
\end{equation}
with a polarization uncertainty $\sigma_{P}$ calculated as
\begin{equation}\label{eq:dp}
\sigma_{P}=\sqrt{\frac{Q^2\sigma_{Q}^2+U^2\sigma_{U}^2}{(Q^2+U^2)I^2}+\frac{\sigma_{I}^2(Q^2+U^2)}{I^{4}}},
\end{equation}
where $\sigma_{I}$, $\sigma_{Q}$, and $\sigma_{U}$ are the uncertainties in the 
$I$, $Q$, and $U$ Stokes parameters.
The polarization position angle ($PA$) is
\begin{equation}\label{eq:PA}
PA=\frac{1}{2}\tan^{-1}(\frac{U}{Q}),
\end{equation}
and the corresponding uncertainty is estimated as
\begin{equation}\label{eq:ePA}
\delta PA=\frac{1}{2}\sqrt{\frac{(Q^2\delta U^2+U^2\delta Q^2)}{(Q^2+U^2)^2}} .
\end{equation}
The magnetic field orientations in this paper are inferred to be $PA +90\degr$.
 
\subsection{JCMT HARP Observations}\label{sec:harp}
$^{13}$CO (3-2) and C$^{18}$O (3-2) molecular line observations were carried out simultaneously using the Heterodyne Array Receiver Program (HARP) instrument \citep{bu09} on the JCMT toward 2264D in February 2022 (project code: M22AP045, PI: Jia-Wei Wang). The raster scan produced an 8.5\arcmin$\times$5.5\arcmin map, centered on (R.A., Dec.)=(6$^{h}$41$^{m}$02.0$^{s}$, 9\degr34\arcmin43\farcs2). 
The observations with a half-power
beam width (HPBW) of 14\farcs1 were carried out with a spectral resolution of 50 kHz, yielding 0.066 \kms, and a bandwidth of 625 MHz. 
The data reduction was performed using the Starlink package, using the default ORAC-DR pipeline \citep{ca08}. \textbf{To enhance the sensitivity, we binned 2$\times$2 pixel of the reduced data, originally Nyquist sampled with a pixel size of 7\arcsec, resulting in a pixel size of 14\arcsec. Additionally, we smoothed the spectral resolution with a Gaussian Kernel to 0.11 \kms.}

In order to obtain a complete $^{13}$CO (3-2) and C$^{18}$O (3-2) map over the entire NGC 2664 system, we additionally included archival JCMT HARP data toward 2264C (project code: M08BU18) for these two lines, with a spectral resolution of 0.11 \kms. These data cover an 8.5\arcmin$\times$5.5\arcmin\ area centered on (R.A., Dec.)=(6$^{h}$41$^{m}$10$^{s}$, 9\degr29\arcmin40\farcs2). The archival data were rebinned to the same pixel grid as our above 2264D data using the starlink package \textit{wcsalign}. The two data sets were then combined to obtain a full map over NGC 2264.
The final rms noise of these data is $\sim$0.5 K and 0.2 K for NGC 2264 C and D, respectively. 
This paper focuses on the velocity dispersion from these two lines, in order to estimate a magnetic field strength. A more detailed analysis of these velocity data sets will be reported in a later paper.

\section{Results}\label{sec:results}

\begin{figure*}
\includegraphics[width=\textwidth]{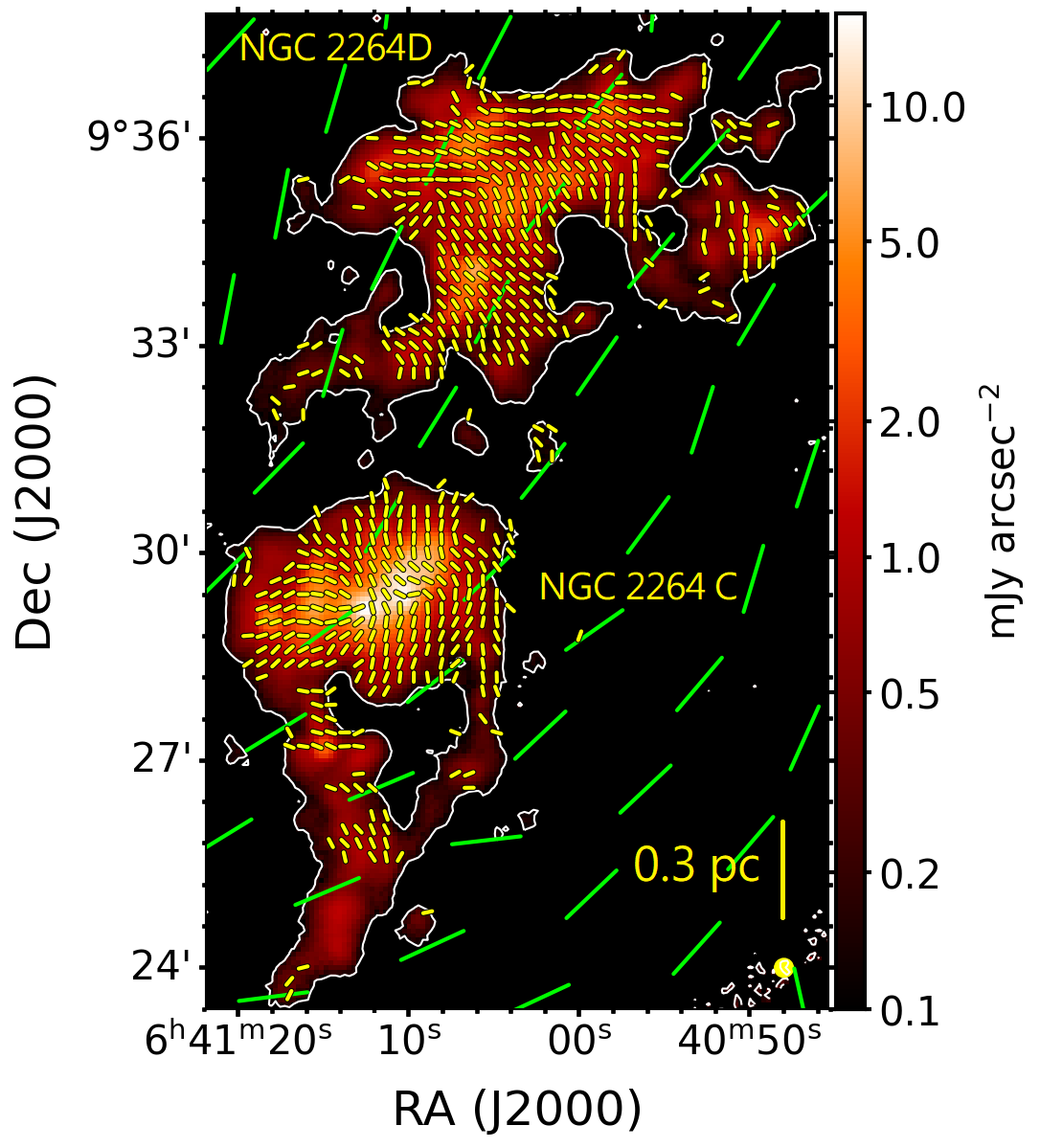}
\caption{
Magnetic field map (yellow segments) toward NGC 2264 based on POL-2 850$\mu$m continuum polarization observations, overlaid on Stokes I 850$\mu$m continuum in color. Segments are rotated by 90\degr\ with respect to the originally detected polarization orientations
(\autoref{fig:pfmap}).
White contours mark 10$\sigma$ (0.1 \mjya) in Stokes I.
Green segments are the larger-scale magnetic field from \textbf{\it Planck} at 353 GHz (850 $\mu$m).
A scale bar and the JCMT POL-2 beam are shown in the lower right corner.
}\label{fig:Bmap}
\end{figure*}

\begin{figure*}
\includegraphics[width=\textwidth]{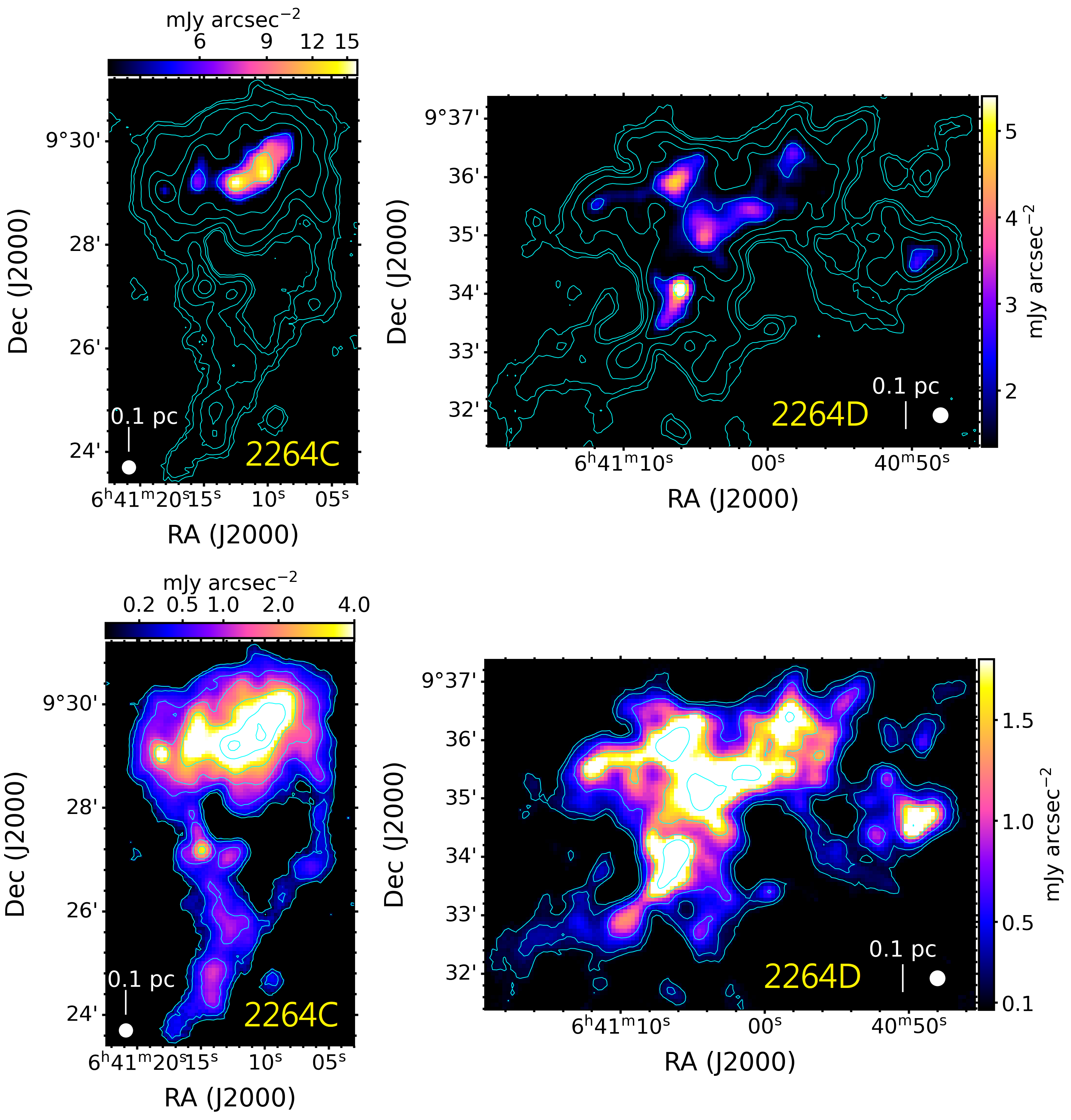}
\caption{NGC 2264 continuum maps on different intensity scales. The cyan contours mark 0.1, 0.2, 0.5, 1.0, 2.0, 5.0, and 10 \mjya.(Top panel) High-intensity compact clumps are distributed along a filamentary structure in 2264C. The bright clumps in 2264D are more scattered. (Bottom panel) A number of streamer-like structures are found in 2264C extending from the massive filament with the dense clumps. In 2264D, a more complex filamentary network is revealed connecting the bright clumps, with fainter streamers extending to the low-intensity outskirts.}\label{fig:Imap}
\end{figure*}

\subsection{POL-2 850 \texorpdfstring{$\mu$}m Continuum Map}
\autoref{fig:Bmap} shows the observed 850 $\mu$m continuum map in a logarithmic color scale, selected to display all reliable continuum detection ($>10\sigma_I$). The structures within NGC 2264 span over a wide dynamical range, from 0.1 \mjya (S/N = 10) in the faint outskirts to 16 \mjya (S/N = 1600) in the brightest clumps. To better visualize the 
filamentary and clumpy structures, \autoref{fig:Imap} shows the 850 $\mu$m continuum map with different color scales, highlighting the bright compact clumps 
in the top panel and the filamentary extended structures in the bottom panel. In 2264C, at least four dense clumps are found lined up along an east-west filamentary structure. A number of fainter streamer-like structures are revealed extending from this east-west filament to the northern and southern outskirts. In contrast to this, the bright clumps in 2264D are more scattered,
and they are likely connected with a fainter filamentary network. A number of fainter streamer-like structures can also be seen in 2264D extending from the filamentary network to the fainter outskirts. These filament/streamer-clump configurations are self-similar to the parsec-scale hub-filament configuration around NGC 2264 \citep{ku20}, appearing as typical hierarchical structures. We will further extract these filament-like structures in \autoref{sec:FIG}.

\subsection{Magnetic Field from 850 \texorpdfstring{$\mu$}m Continuum Polarization}
In order to ensure the robustness of polarization measurements, we select polarization detections based on the criteria $I/\sigma_{I}>20$ and $P/\sigma_{P} >3$. We further exclude detections with large uncertainties in polarization fraction ($\sigma_{P}>5\%$). As a result, a total of 622 polarization measurements are selected across NGC 2264.
For these selected data, the maximum and median uncertainty in the polarization position angle (PA) are 9.54\degr\ and 4.5\degr. In order to examine whether or not the observed polarization originates from magnetically-aligned dust grains, \autoref{sec:pp} shows the polarization fraction map and discusses how the polarization fraction correlates with the total intensity. Based on a Bayesian analysis we find that, indeed, dust grains are likely magnetically aligned in NGC 2264.

\autoref{fig:Bmap} shows the observed magnetic field. Its morphology appears organized over the entire region. In the northern 2264D, the overall magnetic field is oriented with a PA around 30\degr\ (measured counter-clockwise from north to east). In some of the dense regions, the magnetic field locally shows curved patterns, which seem to be associated with the elongated dense structures. In the southern 2264C, an overall magnetic field orientation around a similar PA $\sim$ 30\degr\ is observed.
Additionally, an hourglass-like pattern can be seen along the brightest filamentary structure (along a southeast-northwest direction, roughly perpendicular to the prevailing 30\degr\ field orientation). As a result, the prevailing magnetic field orientation has a wider range than in 2264D. 
At the northeastern end of 2264C, the magnetic field is flipped by almost 90\degr\ to an orientation around $\sim120\degr$, perpendicular to the overall prevailing orientation, becoming parallel to the central filamentary structure. This flipped magnetic field morphology is possibly influenced by  large-scale accreting filaments, or it might be part of the hourglass-like morphology but twisted by projection effects.

The larger-scale magnetic field, traced by the \textbf{\it Planck} 353 GHz data (beam size of 5\arcmin), reveals an organized nearly-uniform field with a mean orientation of 133\degr\ across the entire NGC 2264 region (overlaid in \autoref{fig:Bmap}). The clear difference in orientations between large- and small-scale magnetic field suggests that the sub-parsec scale magnetic field, traced by POL-2, has been decoupled from the larger-scale field traced by \textbf{\it Planck}.

\subsection{Velocity Dispersion traced by \texorpdfstring{$^{13}$}CO and C\texorpdfstring{$^{18}$}O lines}\label{sec:vd}
We extracted the velocity dispersion from our molecular
line data adopting the scousePy package \citep{henshaw16,henshaw19}.
This package is a Python implementation of the spectral line-fitting IDL code SCOUSE \citep{henshaw16b}, which can efficiently identify multiple velocity components in a large spectral cube. We applied scousePy on both the $^{13}$CO (3-2) and C$^{18}$O (3-2) data pixel-by-pixel, where the pixel size is 14\arcsec and the channel width is 0.11 \kms. The pixel rms noise was calculated using the line-free channels, and the identified velocity components were selected using the criteria that the peak brightness temperature be greater than 5$\sigma$, and the velocity dispersion less than 2 \kms. 

 The C$^{18}$O (3-2) line has a maximum brightness temperature of 6 K and is mostly below 4 K. This is much lower than the possible gas temperature in molecular clouds ($\sim10$--20 K), and it is thus likely optically thin. We further found that the intensity ratio of $^{13}$CO (3-2) and C$^{18}$O (3-2) is around a constant of 3.7 over the entire cloud (see details in \autoref{sec:13COvsC18O} in \autoref{fig:amp_C18Ovs13CO}), except for the few brightest pixels. This indicates that the $^{13}$CO (3-2) line is also optically thin except for the brightest pixels.

In the $^{13}$CO data cube, we commonly detected 2 to 3 velocity components across the source (for more information see \autoref{sec:13COvsC18O}). In contrast to this, only one major component was detected in the C$^{18}$O cube. Since both $^{13}$CO and C$^{18}$O  are optically thin, this difference possibly results from velocity components in the low-density outskirts, seen only in $^{13}$CO but not traced by C$^{18}$O. Since we aim at estimating the magnetic field strength from combining polarization and molecular line data, we mainly adopt the \C18O data for the following analyses. This is because they better match the POL-2 polarization data, in which the large-scale, low-intensity structures are filtered out. Nevertheless, we also perform the analysis using the $^{13}$CO data, with velocity dispersions higher by 30--50\% (\autoref{sec:CF_otherline}) to evaluate the possible impact of the choice of gas tracers.

\begin{figure*}
\includegraphics[width=\textwidth]{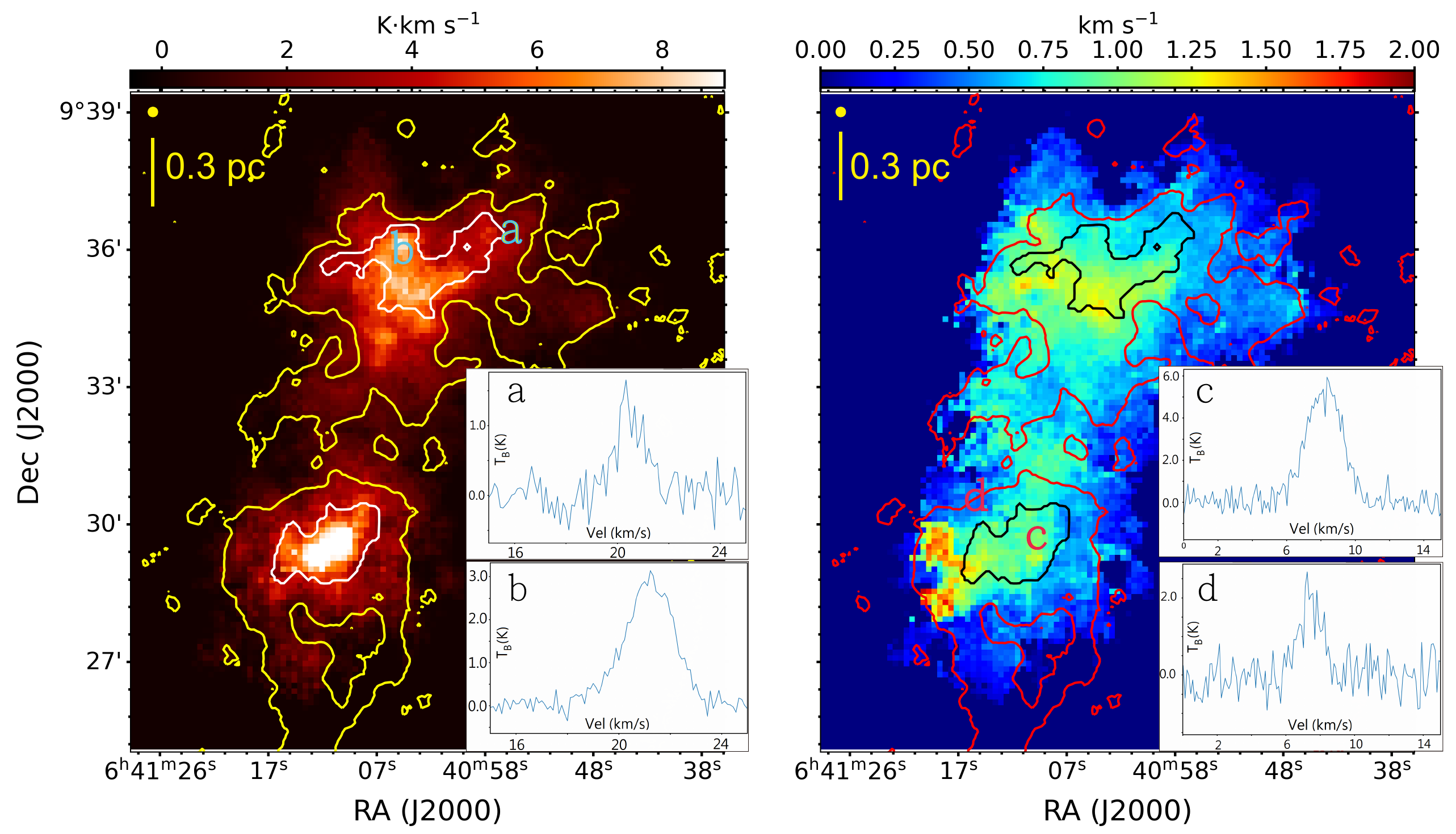}
\caption{(a) \C18O (3-2) integrated intensity map and (b) velocity dispersion map. The black and white contours mark the major clumps in 2264C and 2264D, extracted using a dendrogram algorithm. The 0.1 \mjya contour is shown in yellow and red, as in \autoref{fig:Bmap}. Examples of spectra are extracted at position a, b, c, and d.}\label{fig:C18O}
\end{figure*}

\autoref{fig:C18O} displays the \C18O integrated intensity and the extracted velocity dispersion maps. The moment maps are made using the moment-masking technique included in the \textsc{BTS} package (Clarke et al. 2018) with the masking threshold $T_C = 8 \sigma$ and $T_L = 3 \sigma$. Full details of the technique may be found in the code documentation\footnote{https://github.com/SeamusClarke/BTS}.The integrated intensity map shows structures visually similar to those in the continuum map. In order to identify the major components of 2264C and 2264D from the 850 $\mu$m continuum image, we performed a dendrogram analysis using the astrodendro package \citep{go09} with a minimum significance and threshold of 0.05 \mjya\ (5$\sigma$). 
The two branches matching the major clumps in 2264C and 2264D are selected and plotted in \autoref{fig:C18O}, with intensity levels of 2.0 and 1.3 \mjya, respectively. The estimated velocity dispersion seems to increase toward the intensity peaks, and thus a range of velocity dispersion values is present in each of the major clumps. 
The amplitude-weighted mean velocity dispersion in the 2264C and 2264D major clump is 0.94$\pm$0.24 and 1.04$\pm$0.23 \kms, where the uncertainties are the standard deviations of velocity dispersion within the two clumps. These values are much more turbulent than other low-mass filamentary clouds observed in the BISTRO survey (e.g., IC5146: 0.1--0.3 \kms\ \citep{wa19}, Ophiuchus C: 0.13 \kms\ \citep{liuJH19}, Serpens Main: 0.2--0.3 \kms\ \citep{kw22}), but similar to other massive regions (e.g.,Orion A: 1.33$\pm$0.31 \kms\ \citep{pa17} and Mon R2: 0.45--0.73 \kms\ \citep{hw22}).

\section{Analysis}\label{sec:ana}
\begin{figure*}
\includegraphics[width=\textwidth]{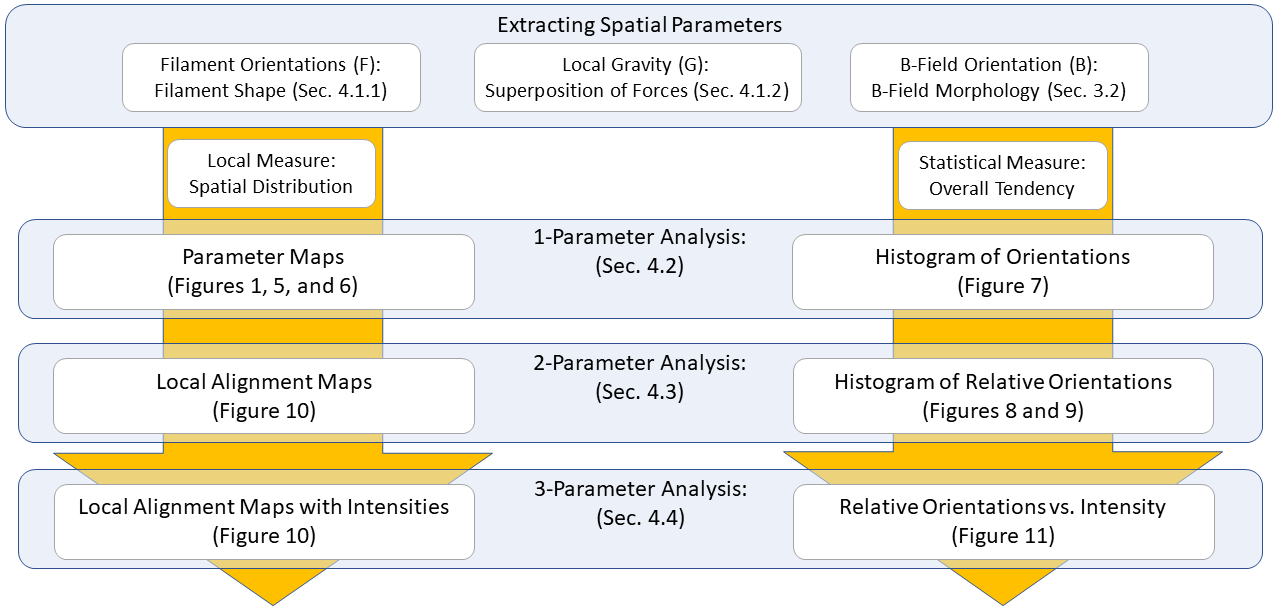}
\caption{Analysis flowchart. Our analysis consists of two aspects: local measures revealing the spatially localized information on maps, and statistical measures showing the overall trends from histograms. Our analysis focuses on individual parameters in \autoref{sec:1p_ana}, extends to the correlation between two parameters in \autoref{sec:2p_ana}, and further includes information of densities in \autoref{sec:3p_ana}. }\label{fig:flowchart}
\end{figure*}

One fundamental question in the formation of a HFS with its filamentary network is how the density structures are shaped by gravity and magnetic field. In order to answer this question, we aim to extract spatial features (\autoref{sec:FIGG}) from both the dust emission continuum and the polarization map, and investigate how these features possibly correlate. 
\autoref{fig:flowchart} shows a flowchart of our analysis scheme. In \autoref{sec:FIG}, we identify filaments (F) and estimate their orientations.
Additionally, we estimate the projected gravitational force (G) at every pixel to probe the gravitational vector field (\autoref{sec:G}), and we use the magnetic field (B) orientations as traced by the polarization data. With these spatially resolved parameters (F, G, B), we examine their spatial distributions on the maps (hereafter local measures) and their overall tendencies using histograms (hereafter statistical measures). \autoref{sec:1p_ana} focuses on the properties of these individual parameters, and \autoref{sec:2p_ana} investigates trends and possible correlations between these parameters. \autoref{sec:3p_ana} further addresses how these correlations evolve with the local densities. Finally, we conduct an analysis of the stability of 2264C and D in \autoref{sec:b_str}, focusing on their global properties.

\subsection{Introducing Local Measures}\label{sec:FIGG}

\begin{figure*}
\includegraphics[width=\textwidth]{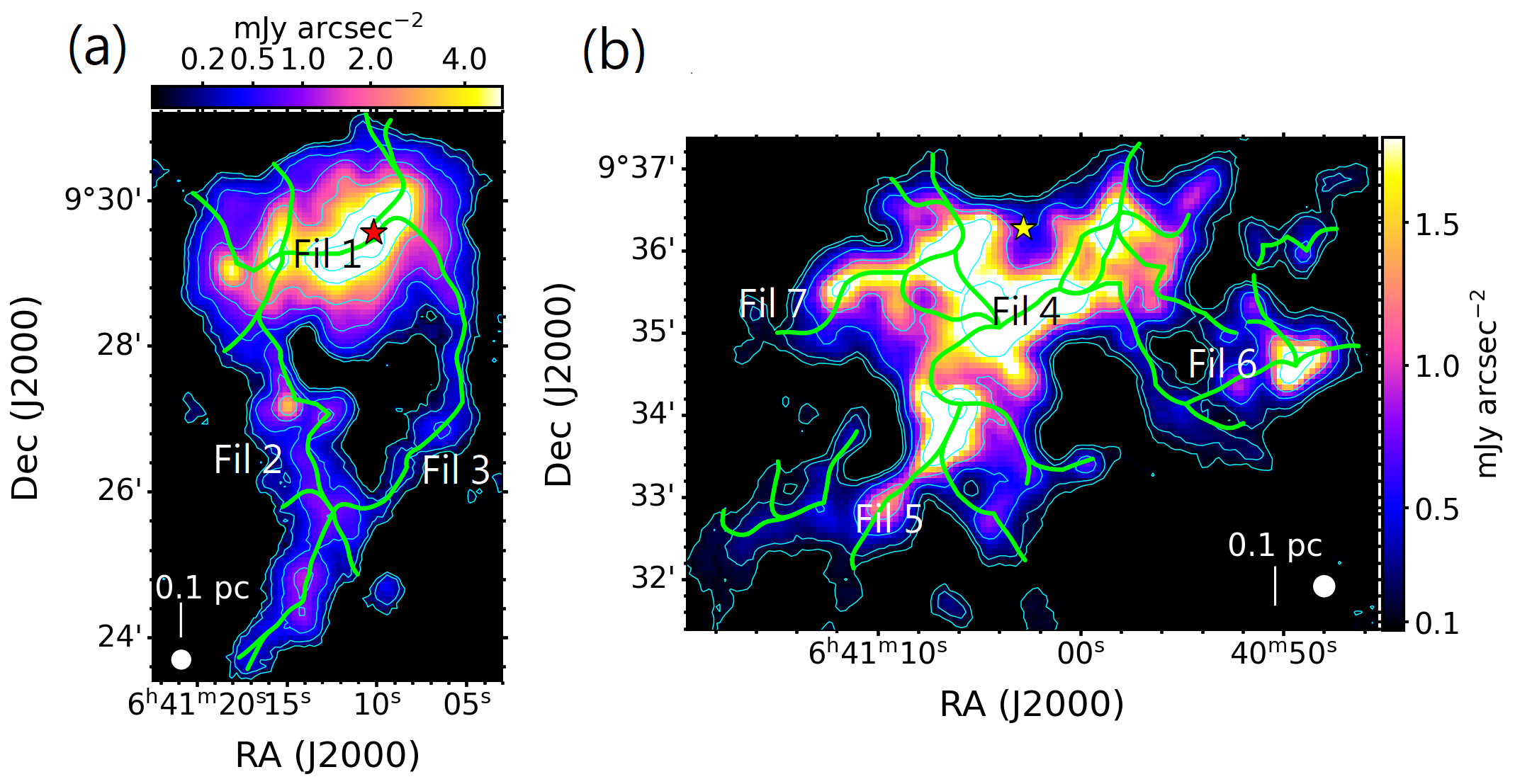}
\caption{Filaments (green lines) overlaid on 850$\mu$m continuum toward (a) 2264C and (b) 2264D. The cyan contours mark 0.1, 0.2, 0.5, 1.0, 2.0, 5.0, and 10 \mjya.
The red and yellow stars label IRS1 (zero-age main-sequence) and IRS2 (Class I), the dominating sources in 2264C and 2264D, respectively. We note that IRS2 is offset from any filament or intensity peak, while IRS1 is at the converging point of multiple filaments. The main filaments along the prevailing directions are labelled.
}\label{fig:FCmap}
\end{figure*}

\subsubsection{Filament Identification}\label{sec:FIG}
In order to identify the filamentary structures within NGC 2264, we apply the $DisPerSE$ algorithm \citep{so11} on the 4\arcsec-pixel JCMT 850 $\mu$m Stokes I image. We use a persistent threshold of 0.8 \mjya (8$\sigma$), which evaluates the intensity difference between two connecting critical points.
The identified filaments are smoothed by 28\arcsec (2$\times$ beam size). Filaments with lengths shorter than 3$\times$ beam size are excluded to ensure a minimum length.
We further examine the reliability and robustness of the 
extracted filaments
by performing Gaussian fits on the radial intensity profile at each pixel along the filaments.
This shows that at least 70\% of the radial profiles within each identified filament can be well described by a Gaussian, where the major peak can be identified and the fitted filament width is above 3$\sigma$. We note that those profiles that are not well fit are usually asymmetric profiles due to overlapping clumps or filaments. The identified filaments are plotted in \autoref{fig:FCmap}. The detailed filament properties are given in \autoref{sec:fi}. The identified filaments have typical lengths 
around 0.2--1.0 pc and widths around 0.1 pc. We note that this is a much smaller aspect ratio than seen for the typical interstellar filaments identified by \textbf{\it Herschel} \citep[e.g.,][]{an10,ar11}.

In 2264C, a bright 0.5-pc-long filament (Fil. 1) with an orientation around 120\degr\ connects most of the central bright structures.
Two long but fainter filaments  (Fil. 2 and 3) extend to the south from the two ends of the brightest filament (Fil. 1). In 2264D, the longest(0.4--0.7 pc) filaments (Fil. 4, 5, 6, and 7) seem nearly parallel to each other with orientations around 120\degr, with numerous shorter(0.2--0.3 pc) filaments extending roughly orthogonally from the long ones. 

To estimate the local orientations of the identified filament ridges, we extract the orientation of the tangent in each pixel along the filaments. For the $ith$ pixel along a filament, we fit the positions of the $(i-2)th$ to the $(i+2)th$ consecutive pixels along the filament with a straight line. With this we find a median fitting error in the local orientation of a filament of $\sim$2\degr.

\subsubsection{Local Gravitational Vector Field}\label{sec:G}
In order to evaluate whether the density structures in NGC 2264 are primarily driven by gravity, we estimate the projected gravitational vector field from the 4\arcsec-pixel JCMT 850 $\mu$m continuum data, following the polarization-intensity gradient technique in \citet{koch12a,koch12b}. 
This is done by calculating the vector sum ($\vec{F_{G,i}}$) of all gravitational forces from masses at all pixels $j$ over a map\footnote{We use the 4\arcsec -pixel continuum map to better sample the gravitational field. We note that while the over-sampled pixels may exhibit partial correlations, this does not impact the determined orientations of the gravitational field, due to the isotropic nature of the beam pattern.} \citep{wa21}, acting at a pixel location $i$ as
\begin{equation}
\vec{F_{G,i}} = kI_i\sum_{j=1}^{n}\frac{I_j}{r_{i,j}^2}\hat{r},
\end{equation}
where $k$ is a factor accounting for the gravitational constant and conversion from emission to total column density. $I_i$ and $I_j$ are the intensity at the pixel position $i$ and $j$, and $n$ is the total number of pixels within the area of relevant gravitational influence. $r_{i,j}$ is the plane-of-sky projected distance between pixel $i$ and $j$, and $\hat{r}$ is its unit vector. We only focus on the directions of the local gravitational forces and not their absolute magnitudes, 
and hence a constant factor $k=1$ is adopted. We apply a mask with a threshold of 0.1 \mjya (10$\sigma$) on the continuum map, because the gravitational force originating from the diffuse and extended surrounding material is both small and rather symmetrical, thus assumed to be mostly cancelled out. Assuming that the intensity distribution is a fair approximation for the distribution of the total mass, and that these mass components in NGC 2264 are roughly at the same distance, the calculated $\vec{F_{G,i}}$ can be used as a proxy for the projected gravitational vector field. 

\begin{figure*}
\includegraphics[width=\textwidth]{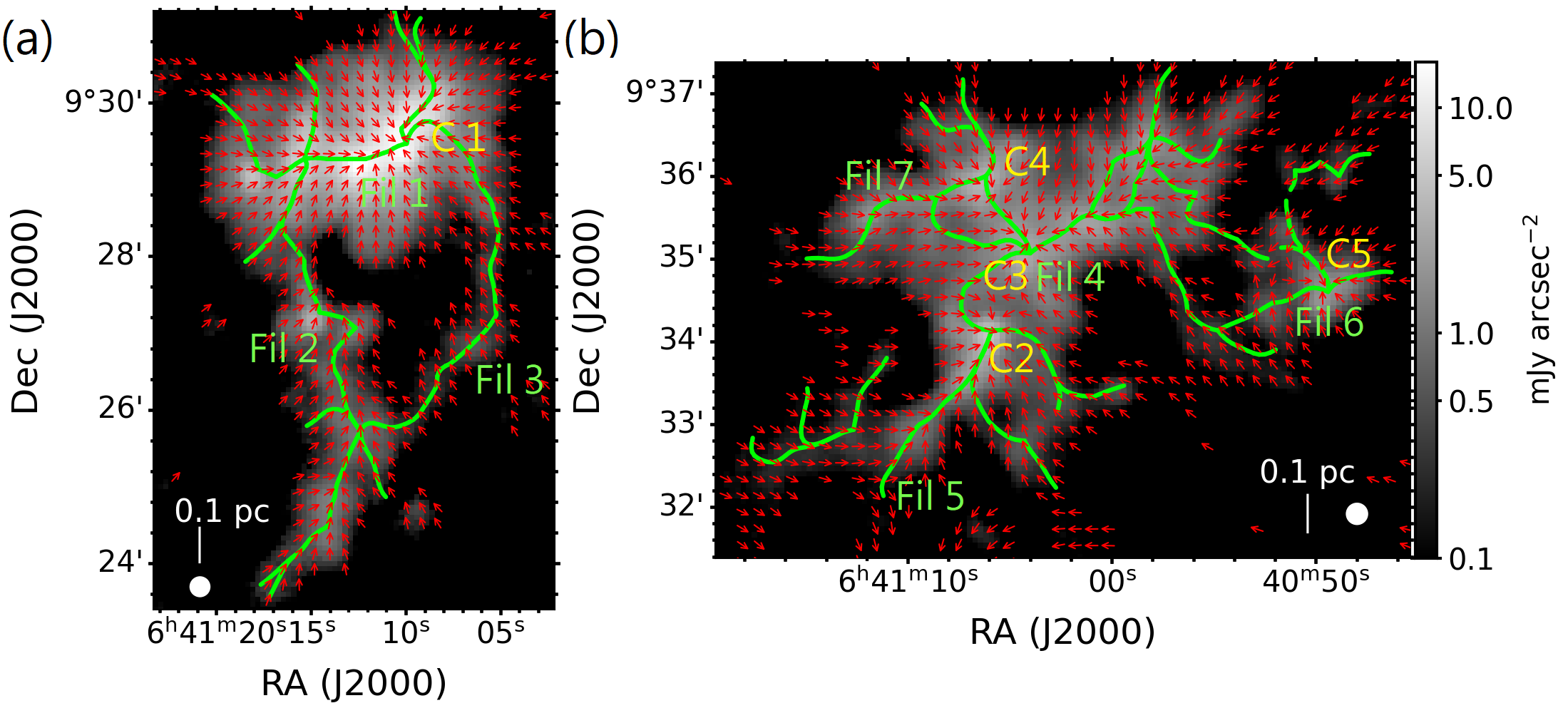}
\caption{Projected gravitational vector field (red) and filaments (green lines) overlaid on 850$\mu$m continuum toward (a) 2264C and (b) 2264D. 
The gravity vectors are calculated per pixel, but only shown per 3$\times$3 pixels for simplicity and with uniform length. The major gravitational converging centers are labelled as C1 to C5.
}\label{fig:Gmap}
\end{figure*}

\autoref{fig:Gmap} displays the local gravitational vector field in NGC 2264. In 2264C, the gravitational field is largely dominated by the densest elongated clump. Thus, the gravitational field visually tends to be either parallel or perpendicular to it. One clear gravitational converging point (C1) can be found at the massive major filament Fil 1.
The connection between gravitational vector field and the 
filaments Fil 2 and 3 extending from the major filament
appears more irregular. In 2264D, the gravitational field reveals four major converging points (C2 to C5) within four east-west filaments (Fil 4, 5, 6, and 7).
This suggests that 2264D is structured into 
four major filaments parallel to each other.

\subsection{1-Parameter Analysis -- Globally Prevailing Orientations}\label{sec:1p_ana}
\begin{figure*}
\includegraphics[width=\textwidth]{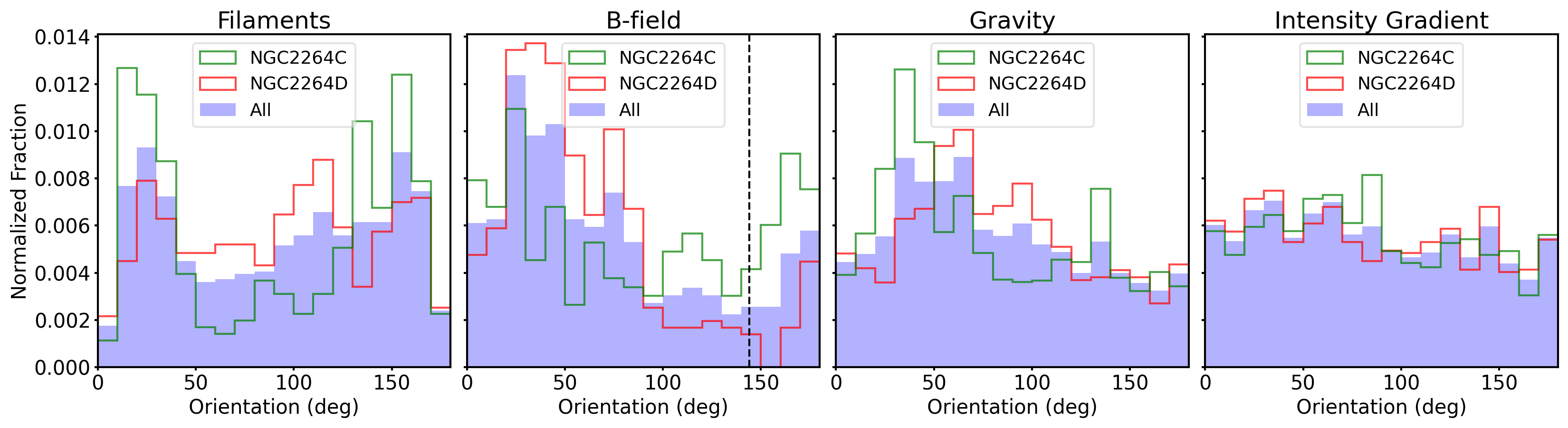}
\caption{Histograms of filament, magnetic field, local gravity, and local intensity gradient orientations. The red and green histograms represent 2264C and 2264D, respectively. The blue filled histograms are for the two combined. 
The vertical black dashed line labels the large-scale mean magnetic field orientation of 144\degr, as traced by \textbf{\it Planck} across the entire NGC 2264 region (\autoref{fig:Bmap}).
All four parameters show common peaks around 20\degr to 50\degr, both for 2264C and 2264D separately as well as for the combined data. Only the histograms of filament orientations reveal additional peaks around $\sim$150\degr.
The intensity gradient distribution in the rightmost panel is further discussed in \autoref{sec:IG_appendix}.
}\label{fig:hist}
\end{figure*}

We plot the histograms of the orientations of 
filament (F), B-field (B), and gravity (G)
in \autoref{fig:hist}, both individually for 2264C and 2264D, and also for them combined.
These histograms reveal the global tendencies of these parameters. None of the histograms is random. They either show a single-peaked or a bimodal distribution.


A common prevailing orientation, between about 20\degr to 50\degr, is found for all parameters.
This possibly implies that there is one underlying mechanism
that is regulating and driving these distributions.
Different from the single-peaked distributions found for the B-field and gravity, 
the filament orientations appear bimodal, peaking around 20--50\degr\ and additionally also around 150\degr. The two peaks are separated by almost 90\degr\ and hence indicate two prevailing orientations that are nearly perpendicular. The peak observed around 150\degr\ is specifically linked to the brightest filaments that are nearly parallel to each other (Fil 1, 4, 5, 6, and 7) as depicted in \autoref{fig:FCmap}. Therefore, we will henceforth refer to these filaments as the ``main filaments''.
We note the close similarity in the B-field and gravity distributions, with the B-field distribution being a little wider. Its wider peak is visible from the 0\degr/180\degr\ wrapping, i.e., an increase around 180\degr\ that is connecting to and continuing at 0\degr.

In summary, the distributions of the orientations of filament, B-field, and gravity
in 2264C and 2264D 
show similarities and a prominent approximate 90\degr\ spacing in the filament orientations.
This might suggest an evolution and connection between 
these parameters driven by the same process, as we further elaborate in 
\autoref{sec:sta_m} and \autoref{sec:local_m}.


\subsection{2-Parameter Analysis -- Relative Orientations among Local Parameters}\label{sec:2p_ana}

In this section we examine the relative orientations among the parameters F, G, and B. 
In other words, 
we assess how closely aligned or how clearly misaligned any two parameters are.
\autoref{sec:2p_statis} will first search for 
statistical trends, while the later \autoref{sec:GvsB} will look at the local alignment or misalignment on the maps, with a focus on B and G.

\subsubsection{Statistical Measure}\label{sec:2p_statis}
\begin{figure}
\includegraphics[width=\columnwidth]{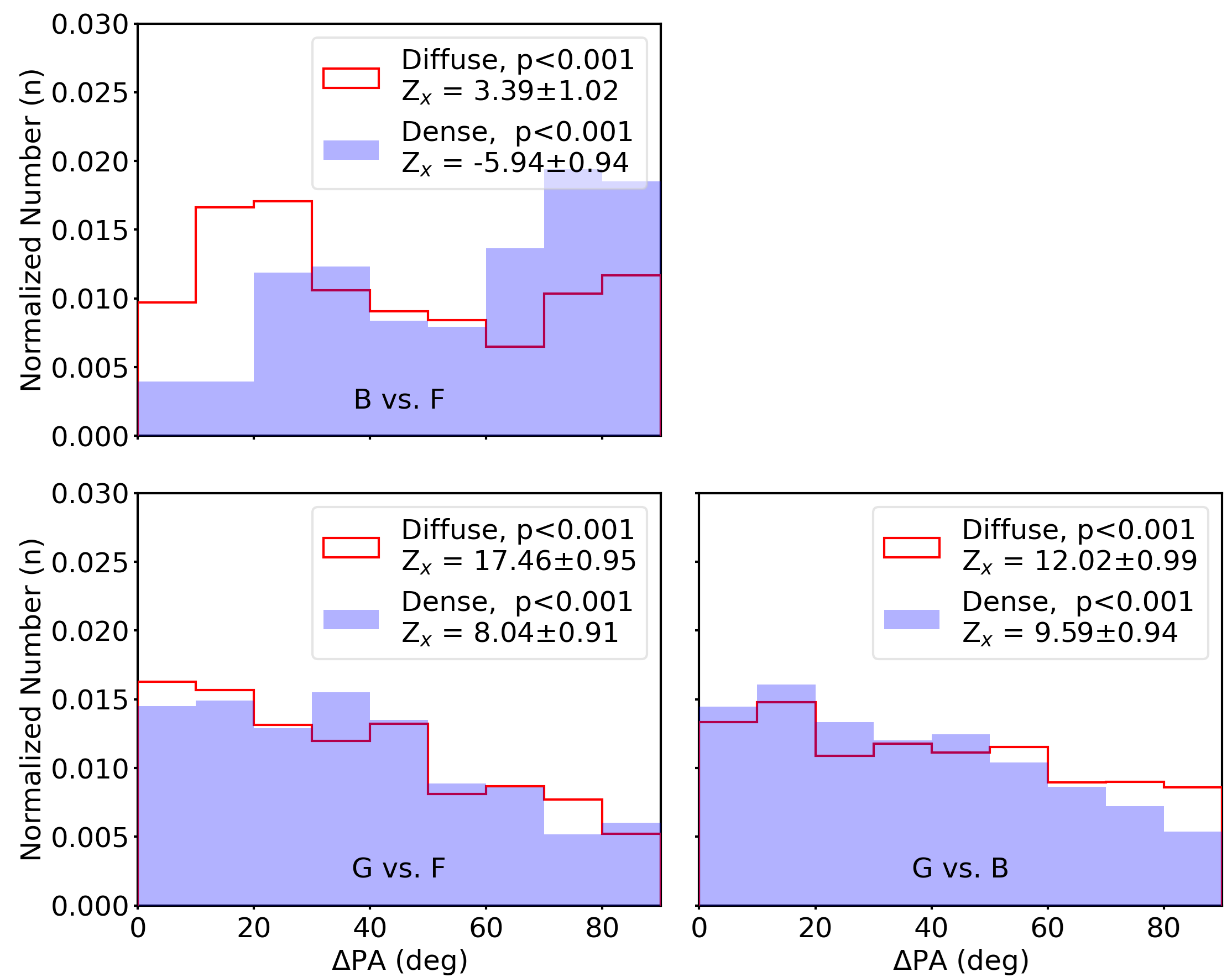}
\caption{Histograms of relative orientations between the spatial parameters F, G, and B for all associated pairs across the entire NGC 2264 region. The data are separated into low- and high-intensity regions with an intensity threshold of 2.0 \mjya. The p-value from the Rayleigh's test indicates the similarity between an observed and a uniform distribution, and p$<0.05$ favors a non-uniform distribution. PRS Z$_\textrm{x}$ indicates whether the relative orientations tend to be parallel ($Z_x > 0$) or perpendicular ($Z_x < 0$).}\label{fig:hist_dpa_2I}
\end{figure}

In order to define relative orientations, the different parameters are matched with each other
as follows.
For each estimated parameter, we select a nearest estimate of another parameter within a distance of 9\arcsec\ ($\sim\sqrt{2}/2\times$beam size). These two estimates are then defined as one associated pair. \autoref{fig:hist_dpa_2I} shows the relative orientations ($\Delta$PA) for all associated pairs for all three combinations among F, G, and B, for the entire NGC 2264 region. All combinations show significant non-uniform distributions, indicated by the Rayleigh's test (p$<$0.05)\footnote{The Rayleigh's test is defined in the domain of 0--$360\degr$. Since the range for our PA is 0--$90\degr$, we use 4$\Delta$PA as the input of this test. Note that the uniformity of 4$\Delta$PA remains unchanged compared to $\Delta$PA, and as a result, this change of variable preserves the validity of Rayleigh's test.
We note that we additionally also ran 
Kolmogorov-Smirnov tests on all distributions, 
and we found identical results, all pointing at 
non-uniform distributions.
}. 
The projected Rayleigh statistic (PRS, \citealt{jo18}) is a modification of Rayleigh's test to further examine the alternative hypothesis whether a distribution is rather parallel- ($Z_x > 0$) or perpendicular-like ($Z_x < 0$), with $Z_x$ defined as
\begin{equation}
    Z_x = \frac{\sum_{i}^{n} cos \theta_i}{\sqrt{n/2}},
\end{equation}
and the related uncertainties determined by
\begin{equation}
    \sigma_{Z_x}^2 = \frac{2\sum_{i}^{n} (cos \theta_i)^2 - (Z_x)^2}{n},
\end{equation}
where $\theta_i=2\Delta \textrm{PA}_i$, leading to $\theta_i\in[0,180\degr]$ for our measured $\textrm{PA}_i\in[0,90\degr]$. This change of variable is sufficient to match the domain of PRS (0--$360\degr$) as cosine is an even function, and allows us to examine the parallel ($\theta_i=0\degr$) and perpendicular cases ($\theta_i=180\degr$) simultaneously.

The PRS results suggest that most of the distributions are parallel-like, except for B vs. F in high-intensity regions.
This is consistent with the 1-parameter analysis (\autoref{sec:1p_ana}) which finds very similar prevailing orientations for F, G, and B, which implies that, at least statistically, one should also expect closely aligned orientations among these parameters. 
However, the 2-parameter analysis reveals 
differences once low- and high-density regions are separated. This is very obvious for B vs F -- with a clear change from rather aligned in the diffuse to rather orthogonal in the dense regions -- and possibly more subtle for G vs B -- with a possible hint of becoming more aligned in the dense regions (\autoref{fig:hist_dpa_2I}). While these differences start to show statistically in these histograms, this also raises the question whether their origin can be {\it localized} and understood in a comprehensive scenario. This will be further explored in the next section and also in \autoref{sec:local_m}.

\begin{figure}
\includegraphics[width=\columnwidth]{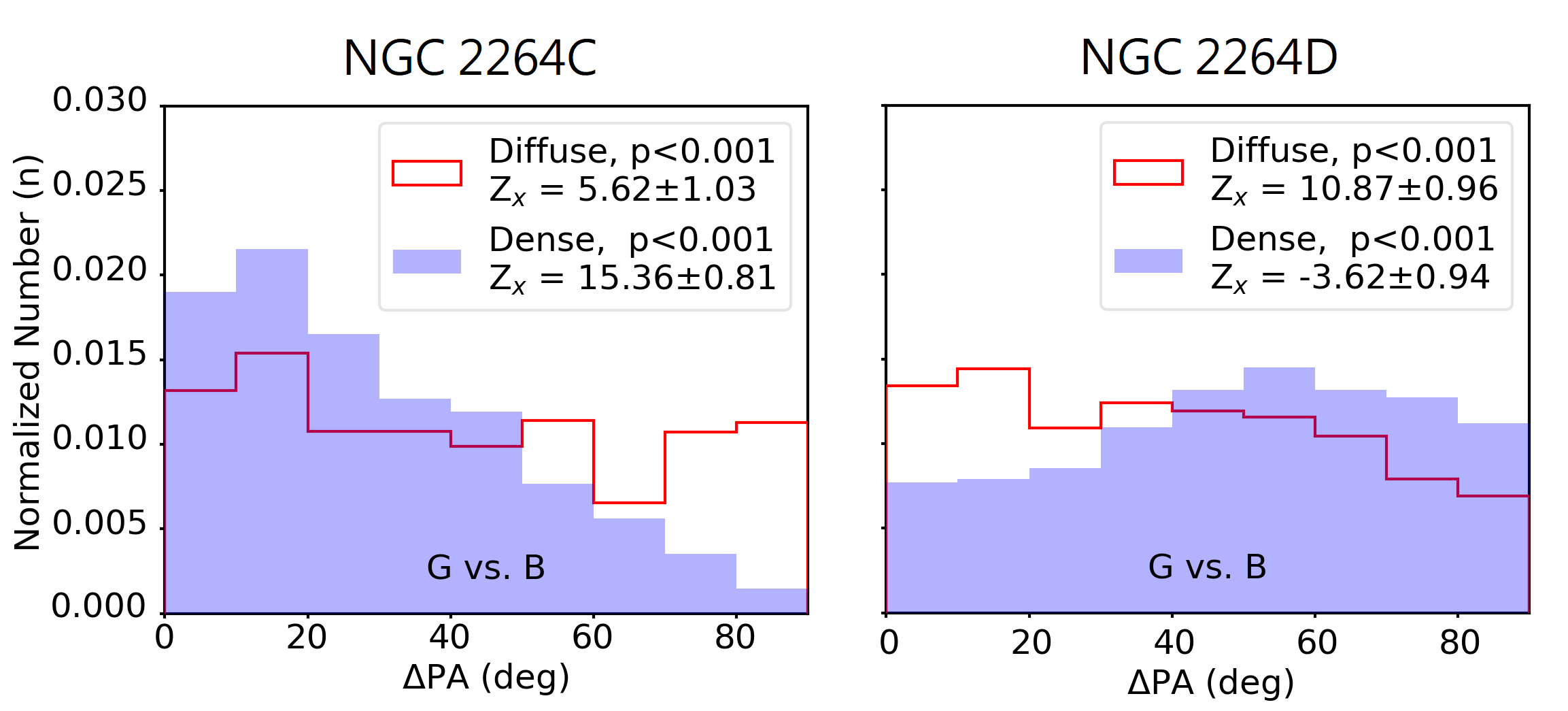}
\caption{Histograms of relative orientations between magnetic field and local gravity in 2264C and 2264D. The data are separated into low- and high-intensity regions with an intensity threshold of 2.0 \mjya. The p-value from the Rayleigh's test indicates the similarity between an observed and a uniform distribution, and p$<0.05$ favors a non-uniform distribution. PRS Z$_\textrm{x}$ indicates whether the relative orientations tend to be parallel ($Z_x > 0$) or perpendicular ($Z_x < 0$).}\label{fig:hist_BG_NS}
\end{figure}

If NGC 2264 is further separated into its northern and southern complex, further nuances can be seen (\autoref{fig:hist_BG_NS}). We focus on G vs B as this forms the basis for our analysis in the next sections and also \autoref{sec:local_m}.
While the diffuse regions in NGC 2264 C and D show rather flat distributions (similar to the entire region as seen in \autoref{fig:hist_dpa_2I}), the dense regions evolve differently.
$\Delta$PA peaks around 0--20\degr\ in 2264C -- indicating a statistically growing alignment between magnetic field and gravity -- and around 50--70\degr\ in 2264D -- indicating a statistically growing misalignment. This difference might be driven by the environment in the two regions, which will be examined with a global stability analysis in \autoref{sec:b_str}.
The additional combinations of relative orientations are 
presented in \autoref{sec:2p_hist_NS}.


\subsubsection{Local Alignment Maps}
\label{sec:GvsB}
\begin{figure*}
\includegraphics[width=\textwidth]{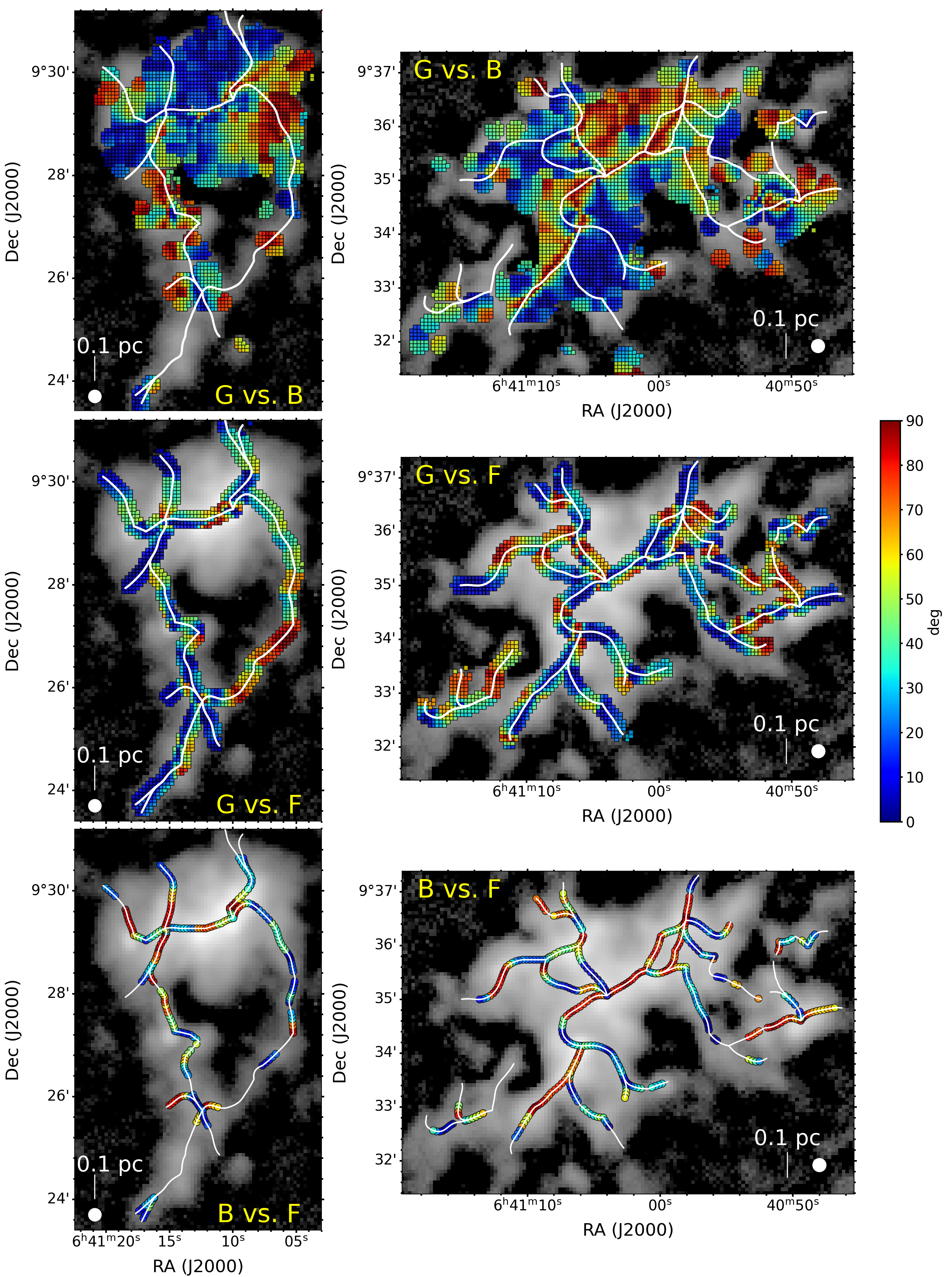}
\caption{Alignment maps between local gravity (G), magnetic field (B), and filament orientations (F) toward NGC 2264.}\label{fig:DPAmap}
\end{figure*}

\autoref{fig:DPAmap} displays maps of relative orientations for the three combinations among F, G, and B.
We first note that the areas of alignment or misalignment appear systematic and not random. They seem organized, sometimes confined in localized regions, sometimes transitioning towards or along filaments.

Motivated by the question of whether magnetic fields are sufficiently strong to support a system against gravitational collapse, we turn our attention to the $\Delta$PA maps for G vs B (the first row in \autoref{fig:DPAmap}). 
In 2264D, several stretches along filaments show relative orientations of nearly 90\degr (red). They appear to have transitioned from surrounding regions where $\Delta$PA is around 0 to 50\degr (blue to green). These regions coincide with those filaments along which the main gravitational converging points are located.
A similar transition occurs in the western part of 2264C around the dominating source IRS1.
The largest regions of close alignment between G and B are the northern end of 2264C and 
the nearby southern end of 2264D. Along the outer boundaries of the combined 2264C and 2264D complex tend to be larger regions of misalignment.

The 
bottom panels in \autoref{fig:DPAmap} are motivated by the question of how the observed filamentary structures are possibly shaped by gravity or magnetic field.
In both clouds, the $\Delta$PA maps for G vs F reveal that gravity is often parallel to filaments, especially towards the gravitational converging centers. The $\Delta$PA maps for B vs F
show certain filaments with B parallel to F while others show B perpendicular to F. 

This peculiar feature will be discussed more in 
\autoref{sec:local_m} where we will argue that the alignment of G vs F and B vs F reveals whether gravity and magnetic field tend to radially compress a filament ($\Delta PA\sim$90\degr) or pull/guide the gas flow along a filament ($\Delta PA\sim$0\degr). With this, the detailed inspection of the relative orientation, i.e., alignment or misalignment between F, G, and B in the maps can provide insight which is not accessible from the statistical measures presented in the previous section.
We finally note that many of the alignment trends seen in the northern end of 2264C appear to connect and continue in the southern part of 2264D.

\subsection{3-Parameter Analysis -- Dependence between Relative Orientations and Local Densities}\label{sec:3p_ana}

As noted in \autoref{sec:2p_statis}, the histograms 
reveal that the trends between F, G, and B appear to evolve with local intensity. To examine whether these trends are statistically significant, we adopt the histogram of relative orientation (HRO) technique \citep{so13,pl16}. This technique probes how the shape of a histogram of relative orientations $\Delta$PA 
changes with local density. The shape 
is described by the parameter 
\begin{equation}
    \xi = \frac{A_c-A_e}{A_c+A_e},
\end{equation}
where $A_c$ and $A_e$ are the areas of a histogram around its center ($0<|\Delta PA|<$22\fdg5) and at its extreme (67\fdg5$<|\Delta PA|<$90\degr). 
A positive $\xi$ indicates relative orientations close to 0, while a negative $\xi$ indicates perpendicular orientations.
The uncertainty $\sigma_{\xi}$ is obtained as
\begin{equation}
    \sigma_{\xi} = \frac{4(A^2_e\sigma^2_{A_c}-A^2_c\sigma^2_{A_e})}{(A_c+A_e)^4},
\end{equation}

with the variances $\sigma^2_{A_c}$ and $\sigma^2_{A_e}$ of $A_c$ and $A_e$ as
\begin{equation}
    \sigma_{A_{c,e}}^2 = h_k(1-h_k/h_{tot}),
\end{equation}
where $h_k$ is the number of data points in the central or extreme bins, and $h_{tot}$ is the total number of data points. The HRO analysis was originally used to study the correlation between magnetic field and intensity gradient \citep{so13}. Here, we further include local gravity and filament orientation in this analysis, in order to obtain a more complete picture of the physical parameters. 

\begin{figure}
\includegraphics[width=\columnwidth]{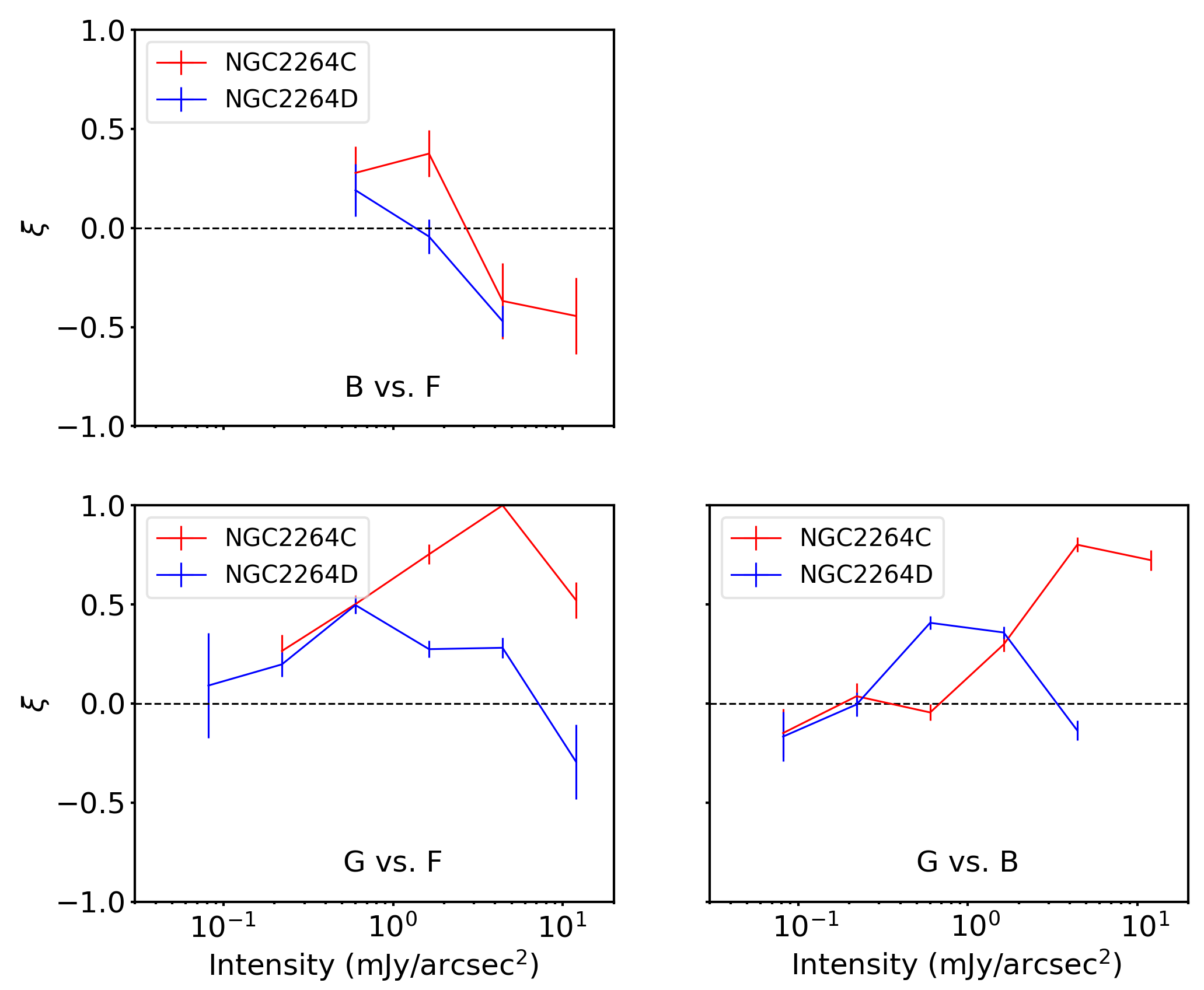}
\caption{Histogram shape parameter $\xi$ vs. local intensity for all parameter pairs toward 2264C (red) and D (blue). All pairs, except for B vs. F, show a clear tendency of increasing alignment 
as the local intensity increases, until certain turnover points. 
An opposite trend is seen in B vs. F which is more 
aligned in low-intensity and becoming more perpendicular in high-intensity regions. }\label{fig:HRO}
\end{figure}

\begin{figure}
\includegraphics[width=\columnwidth]{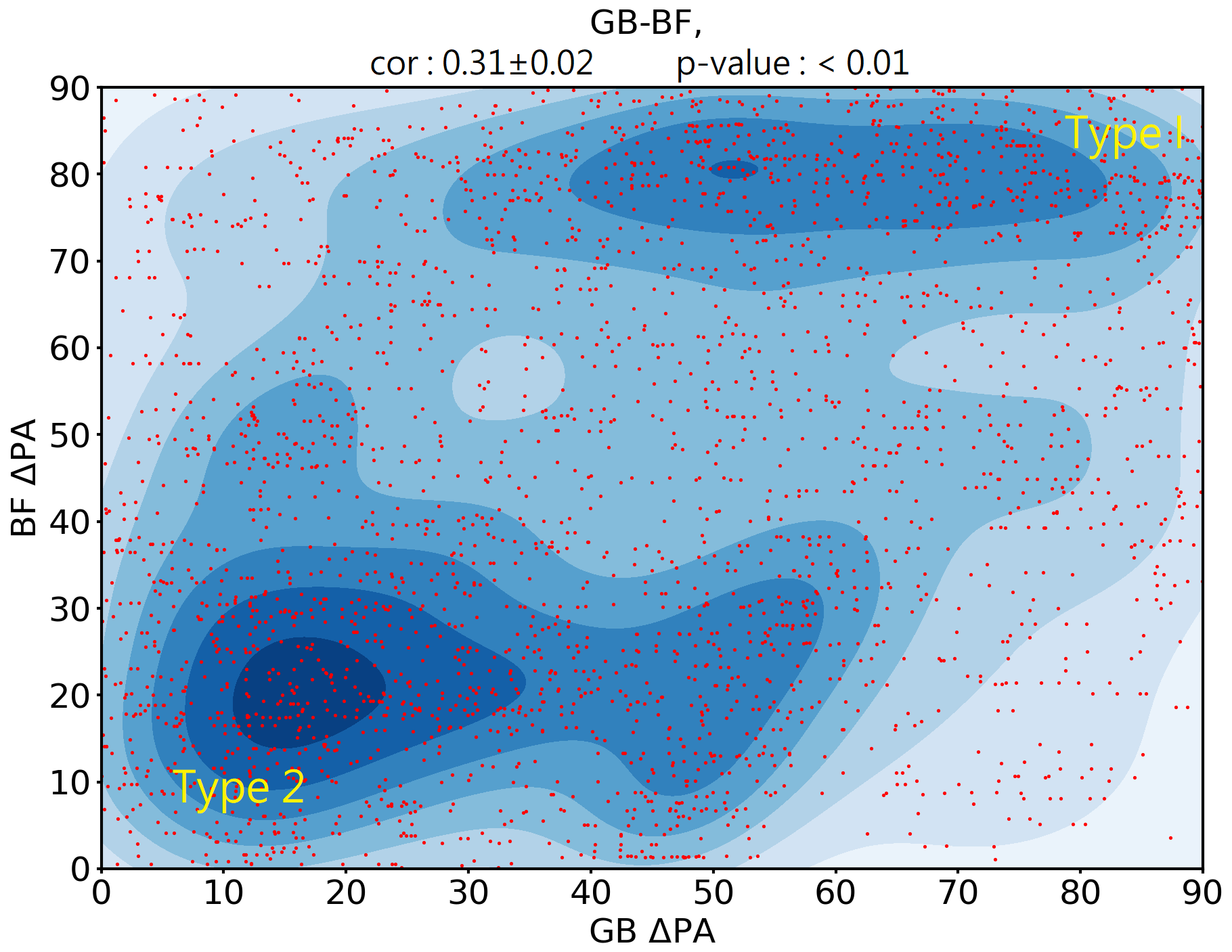}
\caption{2D KDE of relative orientations between gravity(G)-magnetic field(B) and magnetic field(B)-filament(F). The red dots are the samples estimated from individual pixels. The colored contours show the sample density (blue: largest; white: lowest). A Pearson correlation coefficient of 0.31$\pm$0.02 and a p-value $<0.01$ (against the null hypothesis of no correlation) indicate a positive correlation. The additional clustering in peak densities localizes the type I filament (upper right corner) and the type II filament (lower left corner), as discussed in Section \ref{sec:local_m}.
}\label{fig:GBvsBF}
\end{figure}

\autoref{fig:HRO} displays $\xi$ as a function of local intensity.
In both 2264C and 2264D, $\xi$ for these pairs becomes more positive with local intensity in low-intensity regions, but transitions to decrease in high-intensity regions. This suggests that these orientation pairs are getting more aligned with growing intensity, but this alignment breaks down in the densest areas. These breaking points vary, with a range of 1--5 \mjya\ ($\textrm{N}_{\textrm{H}_2}$ = $3\times10^{22}$--$10^{23}\ \textrm{cm}^{-2}$) for different pairs, and they occur at higher intensities in 2264C than in 2264D.
To examine whether the turn-over of $\xi$ in these pairs is occurring for the same filaments, 
we plot the 2D kernel density distributions (KDE) of $\Delta$PA for the G-B and B-F pairs in \autoref{fig:GBvsBF}, with a bandwidth of 2\degr\ determined by Scott's rule. These pairs are selected from filaments with an intensity threshold of 2.0 \mjya to exclude diffuse, outskirt regions. The $\Delta$PA for G-B and B-F show a positive correlation, seen with a Pearson correlation coefficient of 0.31$\pm$0.02, estimated using the bootstrap method to account for the observational uncertainty, and a p-value $<0.01$ against the null hypothesis of no correlation. This distribution further reveals two clustering regions: one in the upper right corner (B$\perp$F and G$\perp$B, which will later be introduced as type I configuration in \autoref{sec:local_m}) and one in the lower left corner (B$\parallel$F and G$\parallel$B, hereafter type II configuration). The possible implications and origins of these configurations are discussed in \autoref{sec:local_m} which will show
how local measures can reveal important features and 
differences that are unnoticed in global statistics.

\subsection{Stability of Major Clumps in NGC 2264}\label{sec:b_str}
In \autoref{sec:2p_ana}, we have shown that the magnetic field orientations can become perpendicular to local gravity within massive clumps and filaments. In such a configuration, magnetic fields might be important to support clumps and filaments against gravitational collapse, if the magnetic fields are sufficiently strong. To evaluate this further, we estimate the magnetic field strength using the Davis-Chandrasekhar-Fermi (DCF) method \citep{da51,ch53}. 
Based on a comparison between the observed polarization angular dispersion $\delta \phi$ and the line-of-sight non-thermal velocity dispersion $\sigma_{v,NT}$, and assuming equipartition of kinetic and magnetic energy, the DCF method yields a plane-of-sky magnetic field strength $B_{pos}$ as
\begin{equation}\label{eq:CF}
B_{pos}=Q~\sqrt[]{4\pi \rho}\frac{\sigma_{v,NT}}{\delta \phi},
\end{equation}
where $\rho$ is the gas volume density, and $Q=0.5$  is a factor accounting for complex magnetic field and inhomogeneous density structures \citep{os01}.

We select the polarization segments within the two major clumps in NGC 2264 C and D (\autoref{fig:C18O}), and calculate the angular dispersion using pairs from only adjacent segments as \begin{equation}\label{eq:dphi}
\delta \phi= \frac{\Sigma (\rm{PA}_i - \rm{PA}_j)}{N_{pairs}},
\end{equation}
where $\rm{PA}_i - \rm{PA}_j$ is the smaller angle between the two adjacent segments i and j, and $N_{pairs}$ is the total number of adjacent pairs in a clump. This is a simplified version of the angular dispersion function in \citet{hi09}, aimed at removing the angular dispersion contributed by large-scale magnetic field structures, 
but only focused on the shortest resolved scale.
We estimate the total mass of the two major clumps, assuming a constant dust temperature of 15 K \citep{ward2000} and a dust opacity $\kappa$ of 0.0125 cm$^2$/g at 850 $\mu$m \citep{hi83}. The total mass of a clump is then obtained by integrating the column density over the selected region, and the clump volume is estimated using the spherical volume with the effective radius $\sqrt{A/\pi}$, where A is the clump area.
From the observed \C18O velocity dispersion (\autoref{sec:vd}), we remove the thermal velocity component and derive the non-thermal velocity dispersion as
\begin{equation}\label{eq:vdisp}
\sigma^2_{v,NT}=\sigma^2_{obs} - \frac{k_B T_\mathrm{kin}}{m_{\mathrm{C}^{18}\mathrm{O}}},
\end{equation}
where the kinematic energy $T_\mathrm{kin}$ is assumed to be 15 K \citep{pe06}.

With the above estimates, the derived magnetic field strengths are 0.24$\pm$0.06 and 0.22$\pm$0.05 mG for 2264C and 2264D (\autoref{tab:CF_main}).
The mass-to-flux criticality $\lambda_{obs}$, commonly used to evaluate the relative importance between magnetic field and gravity \citep{na78}, is estimated as
\begin{equation}
\lambda_{obs}=2\pi \sqrt{G}\frac{\mu m_{H}N_{H_2}}{B_{pos}}.
\end{equation}
where $\mu$=2.8 is the mean molecular weight per H$_2$ molecule, $G$ is the gravitational constant, and $N_{H_2}$ is the molecular hydrogen column density. We adopt a statistical average factor of $1/3$ \citep{cr04} to account for the projection effect for oblate structures, flattened perpendicular to the magnetic field. The corrected mass-to-flux ratio $\lambda$ becomes 
\begin{equation}\label{eq:mf}
\lambda=\frac{\lambda_{obs}}{3}.
\end{equation} Since the large uncertainties in $\sigma_{v,NT}$ can cause unequal upper and lower uncertainties in $\lambda$, we use a Monte-Carlo approach with 10,000 samplings to represent the distribution of the input uncertainties and handle the error propagation. An uncertainty of 7\% in the intensity measurements, due to the uncertainty in the flux conversion factor \citep{ma21}, is included. The resulting upper and lower uncertainties (\autoref{tab:CF_main}) are defined as the difference between the mean value, the 16th, and the 84th percentile. The estimated $\lambda$ are $1.4\substack{+0.4 \\ -0.4}$ and $0.6\substack{+0.1 \\ -0.2}$ for cloud C and D.

In order to evaluate the stability of the major clumps,
we perform a Virial analysis using the above derived quantities. The Virial theorem can be written as
\begin{equation}\label{eq:virial}
\frac{1}{2}\ddot{I}=2(\mathcal{T}-\mathcal{T}_s)+\mathcal{M}+\mathcal{W},
\end{equation}
\citep[e.g.,][]{me56,mc07} where $I$ is a quantity proportional to the trace of the inertia tensor of a cloud. The sign of $\ddot{I}$ determines the acceleration of the expansion or contraction of the spherical clump. The term
\begin{equation}
\mathcal{T}=\frac{3}{2}M\sigma^2_{tot}
\end{equation}
is the total kinetic energy, where M is the total mass and $\sigma_{tot}$ is the total velocity dispersion in a clump, estimated as $\sigma_{tot}^2$ = $\sigma_{v,NT}^2$ +$\sigma_{thermal}^2$. The thermal velocity dispersion $\sigma_{thermal}^2$ for mean free particles is calculated assuming a temperature of 2 K and a mass of mean free particle = 2.33. We neglect the surface kinetic term $\mathcal{T}_s$, because it cannot be determined from the current observations. We note that the presence of substantial external pressure might still influence a cloud's instability. The magnetic energy term is 
\begin{equation}
\mathcal{M}=\frac{1}{2}Mv^2_A,
\end{equation}
where $v_A=B/\sqrt{4\pi\rho} $ is the Alfv\'{e}n velocity and $\rho$ is the mean density as in \autoref{eq:CF}. Since we estimate the magnetic field strengths using the DCF method (\autoref{eq:CF}), the Alfv\'{e}n velocity is determined by the polarization angular dispersion $\delta \phi$ and the line-of-sight non-thermal velocity dispersion $\sigma_{v,NT}$.
As the DCF method only constrains the plane-of-sky magnetic field component, we use the statistical average to correct and estimate the total magnetic field strength as $B = (4/\pi)B_{pos}$ \citep{cr04}. 
The term
\begin{equation}
\mathcal{W}=-\frac{3}{5}\frac{GM^2}{R}
\end{equation}
is the gravitational potential of a sphere with a uniform density $\rho$ and a radius $R$. The derived $\mathcal{T}/\mathcal{W}$ and $\mathcal{M}/\mathcal{W}$ are listed in \autoref{tab:CF_main}.

Our mass-to-flux criticality and Virial analysis both show that the magnetic energy is smaller than the gravitational and kinematic energy in 2264C  ($\mathcal{T}$:$\mathcal{M}$:$\mathcal{W}$ = 0.3$^{\pm0.2}$:0.2$^{\pm0.1}$:1, normalized to $\mathcal{W}$). In contrast to this, the constituents are comparably strong within uncertainties in 2264D ($\mathcal{T}$:$\mathcal{M}$:$\mathcal{W}$ = 0.9$^{\pm0.4}$:0.8$^{\pm0.4}$:1). This could explain the difference in distributions of $\Delta$PA for G-B in 2264C and 2264D (\autoref{fig:hist_BG_NS}): since 2264C is dominated by gravity, the magnetic field is expected to be more aligned (parallel) with gravity toward higher densities. On the other hand, since none of the constituents dominates in 2264D, magnetic field and gravity do not simply appear parallel or perpendicular. As the stability analysis here is based on global and averaged properties (\autoref{tab:CF_main}), a different analysis (\autoref{sec:local_m}) is still necessary to shed light on the more detailed and local role of gravity and magnetic field. 

We note that a major source of uncertainty in our estimates is the possible complexity of velocity structures. 
Multiple velocity components, possibly tracing the low-density outskirts, are identified in $^{13}$CO. C$^{18}$O on the other hand appears to trace the major high-density components which are more likely associated with the regions traced by POL-2.
In \autoref{sec:13COvsC18O}, we show that the velocity dispersion from $^{13}$CO is systematically larger than the one from C$^{18}$O by 30--50$\%$. Additionally, the velocity dispersion estimated from N$_{2}$H$^{+}$ (1-0) lines in \citet{pe06}, tracing higher densities, is systematically lower than our estimates by $\sim$30$\%$
This suggests that the lower a density is traced, the larger the velocity dispersion can be. 
This is possible if the diffuse gas is distributed over a larger volume and thus mixed with more complex velocity structures. 
Given that also the polarization efficiency, 
determined by unknown dust properties and radiation
field together with the SCUBA2/POL-2 large-scale filtering effect \citep{sa13}, is rather unclear, it is not obvious which gas tracer is the best to use together with continuum polarization in the DCF method. 
To illustrate these caveats, we present tables of the estimated magnetic field strengths, mass-to-flux ratios, and energy ratios for all structures identified by the dendrogram and for the three different gas tracers (C$^{18}$O, $^{13}$CO, N$_{2}$H$^{+}$) in \autoref{sec:CF_otherline}. Nevertheless, regardless of whether the $^{13}$CO or C$^{18}$O line is chosen, our conclusion regarding the relative importance of energy sources remains valid.

Finally, we note that outflow activities were extensively observed in NGC 2264 \citep[e.g.,][]{cu16}. \citet{ma09} conducted an analysis focusing on the equilibrium between the force exerted by outflows in 2264C and the gravitational force. Their findings indicate that these outflows fall short by a factor of $\sim$25 in terms of providing significant support against the gravitational collapse of the 2264C protocluster. Consequently, it is unlikely that the outflows will have a substantial influence on the stability analysis presented here.

\begin{deluxetable*}{cccccccc}
\tablecaption{Magnetic field strengths and mass-to-flux ratios in NGC 2264. \label{tab:CF_main}}
\renewcommand{\thetable}{\arabic{table}}
\tablenum{1}
\tablehead{\colhead{Region} & \colhead{$n_{H_2}$} & \colhead{$\sigma_{v,NT}$} & \colhead{$\delta \phi$} & \colhead{$B_{pos}$} & \colhead{$\lambda$}  & \colhead{$T/W$} & \colhead{$M/W$}\\
\colhead{} & \colhead{$\textrm{cm}^{-3}$} &\colhead{(\kms)} &\colhead{(deg)} &\colhead{(mG)} & \colhead{} & \colhead{} & \colhead{} }
\startdata
NGC 2264 C & $(1.5\pm0.1)\times10^5$ & $0.94\pm0.24$ & $30.2\pm0.3$ & $0.24\pm0.06$ & $1.4\substack{+0.4 \\ -0.4}$ & $0.3\substack{+0.2 \\ -0.2}$ & $0.2\substack{+0.1 \\ -0.1}$ \\
NGC 2264 D & $(5.1\pm0.4)\times10^4$ & $1.04\pm0.23$ & $21.5\pm0.4$ & $0.22\pm0.05$ & $0.6\substack{+0.1 \\ -0.2}$ & $0.9\substack{+0.4 \\ -0.4}$ & $0.8\substack{+0.4 \\ -0.4}$ \\
\enddata
\tablecomments{Magnetic field strengths and mass-to-flux ratios are derived from the DCF method. The uncertainties listed here are obtained from propagating the observational uncertainties through the corresponding equations using a Monte Carlo approach. Possible additional systematic uncertainties due to, e.g., the unknown dust opacity, are not included. }
{\addtocounter{table}{-1}}
\end{deluxetable*}

\section{Discussion}\label{sec:dis}
\subsection{Statistical Measures of the Filamentary Network 
}\label{sec:sta_m}
The filamentary network in NGC 2264 is embedded in the hub of a 10-pc-scale hub-filament system \citep{ku20}. This network
is therefore denser ($\textrm{N}_{\textrm{H}_2} = 10^{22}$--$10^{24} \textrm{cm}^{-2}$) and more compact (lengths = 0.2--1.0 pc) than the typical filamentary networks discovered by \textbf{\it Herschel}, i.e., $\textrm{N}_{\textrm{H}_2} = 10^{21}$--$10^{22} \textrm{cm}^{-2}$ and lengths = 1--10 pc in IC5146 \citep{ar11}. A major scientific question that we are addressing in the following is how the network in NGC 2264 is formed and shaped.

\subsubsection{1-Parameter Statistics: Global Trends from Prevailing Orientations}

\begin{figure*}
\includegraphics[width=\textwidth]{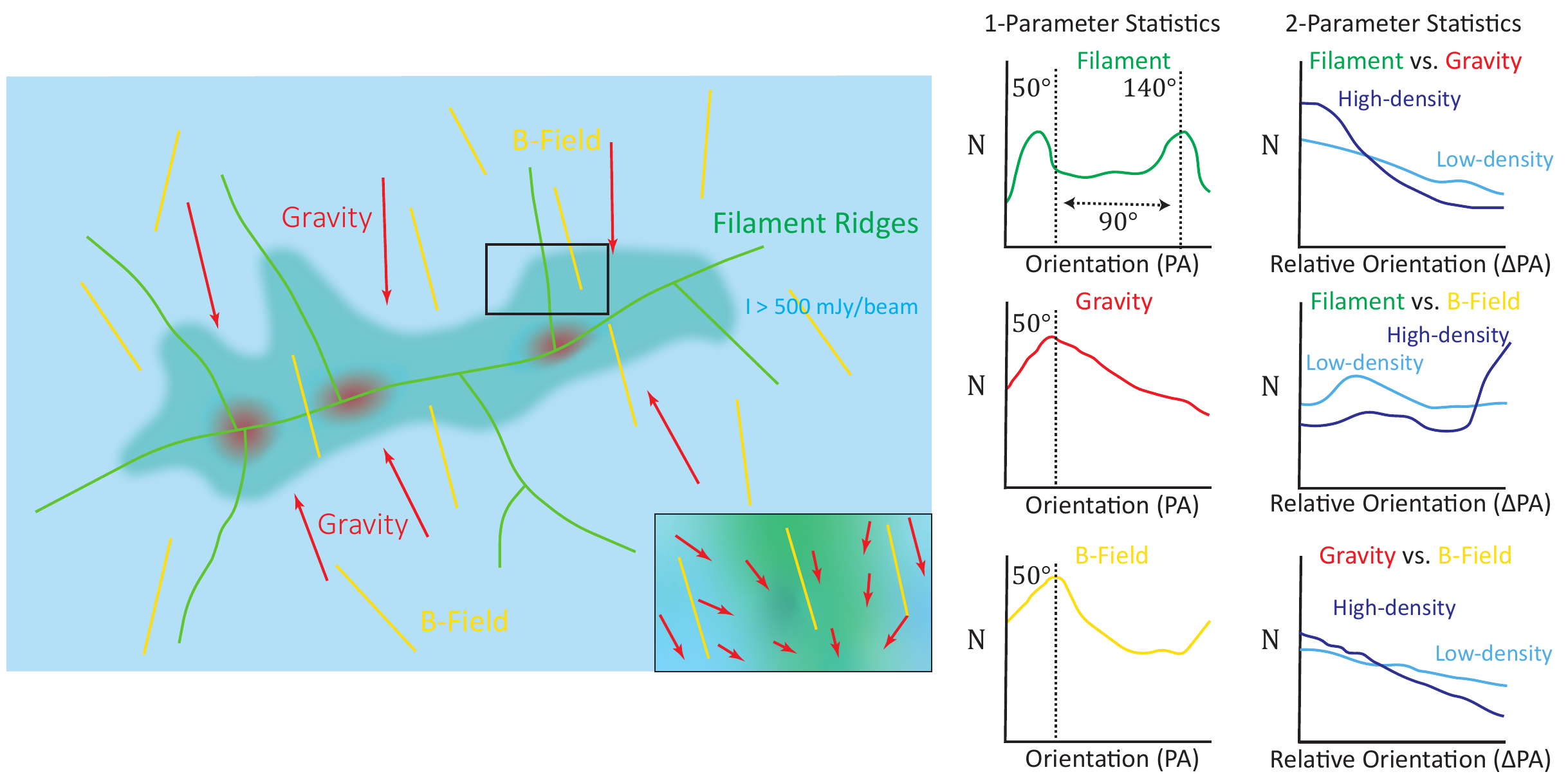}
\caption{Cartoon illustrating the correlation between filaments, gravity, and magnetic field in NGC 2264. The green lines, red arrows, and yellow segments represent the filament ridges, direction of gravity, and magnetic field orientation. The right bottom corner shows a zoom-in to the filament ridge in the black rectangle. The right panels are the expected histograms of the 1-parameter and 2-parameter statistics for this illustrated picture. All three parameters share a common prevailing orientation around $\sim$50\degr, while filaments show an additional prevailing orientation around $\sim$140\degr. Gravity displays a broader distribution due to the feature in the zoom-in. The magnetic field distribution is even wider due to the 0\degr/180\degr wrapping.}\label{fig:Statistical_Measure}
\end{figure*}

The 1-parameter statistics reveal that filaments, gravity, and magnetic fields share a common prevailing orientation (\autoref{sec:1p_ana}, \autoref{fig:hist}). It appears that the filamentary network primarily consists of filaments with orientations approximately perpendicular to each other ($\sim$30\degr\ and $\sim$150\degr). 
This arrangement is similar to the so-called main-filament and sub-filament (or striation) structures found in other filamentary clouds, such as Serpens South \citep{su11} and Taurus B211/213 \citep{pal13}.
The one prevailing orientation in gravity (around $\sim$40\degr\ to 50\degr) indicates {\it one dominating converging direction for gravity} along a $\sim$40\degr\ to 50\degr\ orientation. The finding that one of the two prevailing filament orientations, namely the one around $\sim$150\degr, is approximately perpendicular to the converging gravity direction, is suggestive of a picture where gravity (on a {\it global} scale across both 2264C and 2264D) is predominantly driving mass accretion along this direction (\autoref{fig:Statistical_Measure}). The 
second prevailing filament orientation around $\sim$30\degr\ aligns roughly with the prevailing gravity orientation, though gravity shows a broader distribution in orientations. This broader distribution (also seen in the intensity gradient) likely is the result of small (local) deviations of gravity from its prevailing global direction towards the filaments.

The B-field orientations are again around a prevailing orientation of $\sim$40\degr\ to 50\degr, similar to gravity, yet with an even broader distribution. 
This broadening with respect to gravity
is likely capturing a lag in alignment between B-field and gravity. The broader wings in the distribution likely result from the more diffuse regions where the B-field orientation is less aligned with gravity. This is also seen from the 2-parameter statistics discussed in the next section.
Since the polarization measurements are independent from the intensity maps (which are used for the filament extraction and the derivation of gravity), these similar orientations around $\sim50\degr$ imply that the magnetic field morphology is strongly connected to the density structures. This observed filament-magnetic field configuration can initially be generated by either a cloud collapsing/fragmenting along the magnetic field \citep{na08,li13,va14} or a bow-shaped magnetic field associated with filaments formed via the compression of ISM bubbles \citep{in15,in18,tahani19,ta22a,ta22b}.
The cartoon in \autoref{fig:Statistical_Measure} summarizes the suggested scenario based on these global trends.

\subsubsection{2-Parameter Statistics: Global Alignment Statistics}\label{sec:2p_dis}

The relative orientation analysis (\autoref{fig:hist_dpa_2I}) suggests that filament orientation and local gravity
are all correlated with the magnetic field orientation,
i.e., their relative alignment is not random but systematic. Moreover, low- and high-intensity regions appear different, with a trend that denser regions display a closer alignment, i.e., a shift in the distributions towards smaller $\Delta$PA. The only exception is the alignment between magnetic field and filaments which shows an opposite trend, the magnetic field orientation becoming more orthogonal to the filaments in denser regions. Finally, all these trends among all parameters also remain when binned to finer intensity steps, except for the turn-over trends only seen in regions with the highest densities (\autoref{fig:HRO}). This indicates that these alignment correlations are a function of the local density.

Previous statistical analyses \citep[e.g.,][]{pl16,kw22} based on polarization observations toward a number of filamentary systems have also shown that magnetic fields tend to be parallel to less-dense filamentary structures, but become perpendicular towards denser filaments. As the filaments identified in our work are distinct from filaments identified using \textbf{\it Planck} and \textbf{\it Herschel} data, our results suggest that the alignment in NGC 2264 transitions in 0.1-pc-scale regions with much higher densities ($\sim10^4$--$10^5$ cm$^{-3}$ or $10^{22}$--$10^{23}$ cm$^{-2}$) than the transition density of $10^{21}$--$10^{21.5}$ cm$^{-2}$ reported in \citet{pl16}.

In order to explain the observed transition in alignment, recent numerical simulations have proposed a scenario where such a transition 
originates from changing physical processes in the filament formation, namely from the phase dominated by magnetically regulated MHD turbulence in low-density regions to the phase dominated by gravitational collapse or converging flows compression in high-density regions \citep[e.g.,][]{so17,se20,Ib22,ga23}. However, these simulations predict a wide range of transitional density thresholds, from 10$^2$ -- $10^3$ cm$^{-3}$ \citep{se20, ga23}, 10$^3$ -- $10^4$ cm$^{-3}$ \citep{Ib22} to 10$^4$ -- $10^5$ cm$^{-3}$ \citep{so17}. Our results favor a higher threshold of $\sim10^4$--$10^5$ cm$^{-3}$.



\subsection{Local Measures of the Filamentary Network: Two Different Types of Filaments?}\label{sec:local_m}

The analysis presented in \autoref{sec:sta_m} has revealed statistical trends, globally across the entire NGC 2264 region. It is evident from the alignment maps in \autoref{fig:DPAmap} that the relative orientations $\Delta$PA between any two parameters are not random. They seem to appear in certain spatial locations, possibly in organized structures and possibly even with characteristic length scales. The detailed {\it local} connections among magnetic field, gravity, and filament are largely unexplored. In this section we focus on a particular aspect, namely the combined magnetic field--gravity connection in shaping filaments. 
This is unseen in global statistics, but apparent in local structures.

\begin{figure*}
\includegraphics[width=\textwidth]{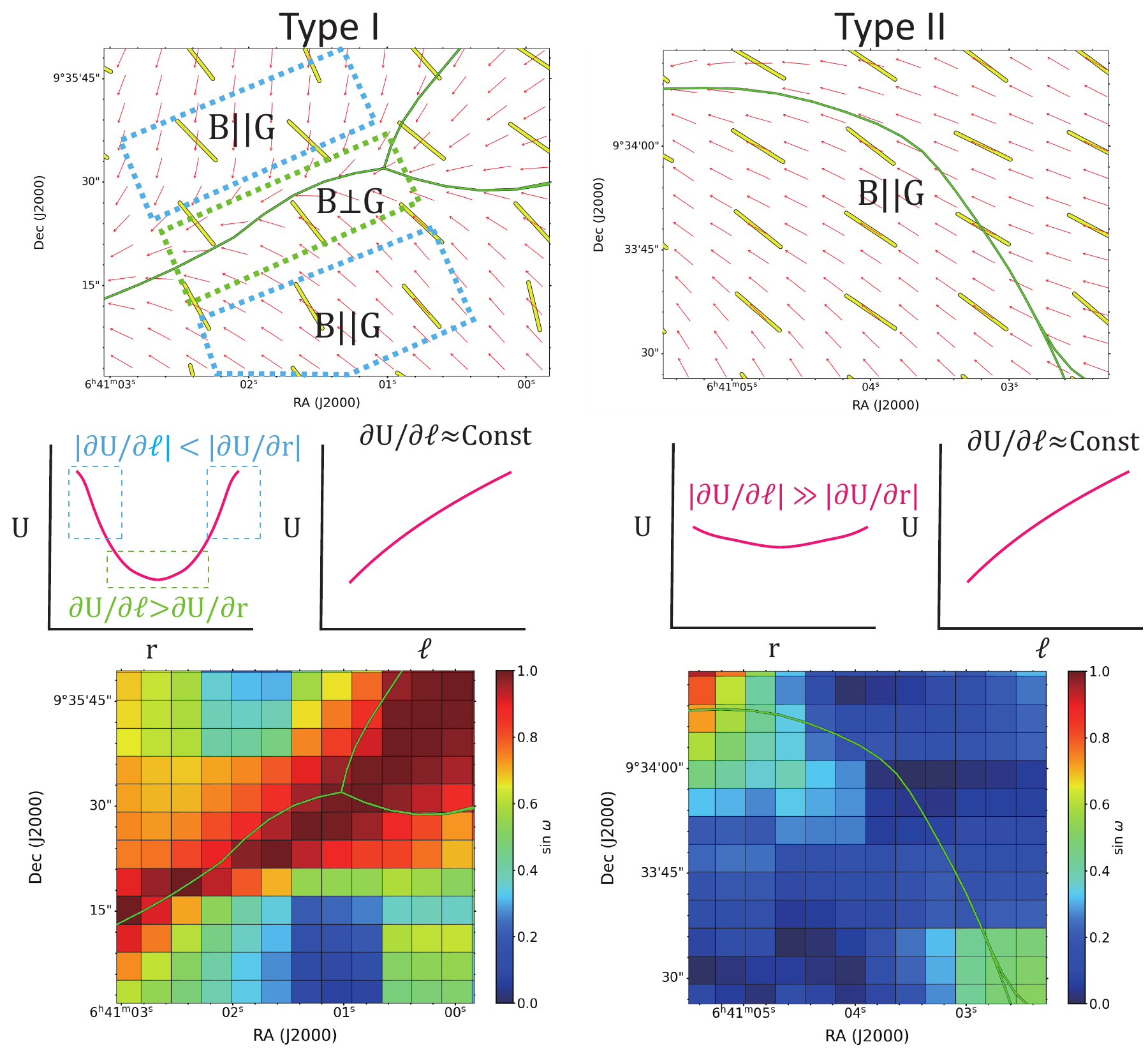}
\caption{Illustration of the role of gravity and magnetic field in two types of filament (type I, left panels; type II, right panels). The two maps in the top row are two representative regions as observed in NGC 2264 C and D, where the green lines, red arrows, and yellow segments are the filament, gravity, and magnetic field orientations. Middle row: for type I filaments, the radial component of the gravitational field ($\frac{\partial U}{\partial r}$) is greater or comparable to the longitudinal component ($\frac{\partial U}{\partial \ell}$) in the outer regions, but the longitudinal component starts to  dominate at the filament ridges. The magnetic field is therefore parallel to gravity in the outer regions, but becomes more perpendicular towards the ridges. In contrast to this, the gravitational field in type II filaments is always dominated by the longitudinal component. Bottom row: sin $\omega$ maps for the two selected regions in the top row.
}\label{fig:Local_Measure}
\end{figure*}

Motivated by the two types of configurations revealed in the $\Delta$PA distribution of B-F and B-G (\autoref{fig:GBvsBF}), we extract two representative regions in NGC 2264, associated with these two types (\autoref{fig:Local_Measure}).
Configuration I (left panel) shows local gravity (G) converging from the lower-density surrounding toward the filament (F), and then once close enough, G aligns with the filament ridge. On the resolved scale in these data, the magnetic field (B) remains perpendicular to the filament. {\it The magneto-gravitational configuration shows G parallel to B at larger distances from F, but then transitioning to G becoming increasingly perpendicular to B closer to F. } This configuration is seen both in lower- and higher-density areas. 
Configuration II (right panel) shows local gravity 
being mostly along the filament and not yet converging toward the filament ridge (at the resolved scale in these data). Additionally, the magnetic field is also predominantly aligned with both local gravity and the filament. Here, {\it the magneto-gravitational configuration shows G parallel to B both at larger distances and also close to F.} This configuration is also seen both in lower- and higher-density areas.

In the following we discuss possible implications and origins of these two different magneto-gravitational configurations. These two systematically different configurations
might be suggestive of two different types of filament. 
A quantitative measure to assess differences in these constellations is the $\sin\omega$ measure \citep{ko18,ko22}. 
With the angle $\omega$ between B and G, $\sin\omega$ quantifies a fraction (between 0 and 1) of the magnetic field tension force that can work against a local gravitational pull. Small values in $\omega$, i.e., close alignment between gravity and magnetic field, allow gravity to maximally accelerate inflowing material. 
Large values in $\omega$, i.e., gravity and magnetic field close to orthogonal, indicate maximal obstruction by the magnetic field and hence a reduced acceleration. 
Consequently, the type I filament (left panel in \autoref{fig:Local_Measure}) will initially experience fast accretion from the outside (small $\omega$; middle panels in \autoref{fig:Local_Measure}), hence short radial timescales towards the filament but then have a longer longitudinal timescale along the filament ridge (larger $\omega$). The type II filament (right panel) will favour a short longitudinal timescale for any motion or collapse along the filament (small $\omega$). On a more speculative note, these findings might further imply that the type I filament is able to accrete more material.

The two different types of filament and magneto-gravitational configurations can also be explained in terms of their gravitational potentials (middle panels in \autoref{fig:Local_Measure}). The gravitational potential $U$ of a filament within a HFS is expected to be a valley. Radially, within a filament ($r <$ filament width) $U$ becomes shallower toward its bottom at the ridge  (e.g., a filament with a Plummer-like density profile \citep{ar11}). Longitudinally, driven by the global gravitational field influenced by the nearest massive hub, 
$U$ smoothly decreases toward the major potential well. Whether material is accreted onto a filament is determined by the competition between the radial and longitudinal components of the gradient of the gravitational potential. The type I filament (left panels in \autoref{fig:Local_Measure}) will have a radial gravitational potential profile ($\frac{\partial U}{\partial r}$) which is steeper (as a result of a higher density or a steeper density profile) than its longitudinal gravitational potential profile ($\frac{\partial U}{\partial \ell}$) (as a result of mass and distance to the major potential well). Thus, this configuration favors accretion from the ambient material. 
Once approaching the filament ridge, where the radial gravitational potential approaches its bottom and $\frac{\partial U}{\partial r}$ becomes zero, the longitudinal motion along the filament will start to dominate. This then causes a transition from small to larger $\omega$ at the observed resolution. In contrast to this, type II filaments (right panels in \autoref{fig:Local_Measure}) have a radial gravitational potential profile that is shallower than the longitudinal profile.
Therefore, the pull from the global gravitational field is always more efficient. This picture has also been a subject of discussion in theoretical studies on hierarchical collapse within molecular clouds \citep[e.g.,][]{go14, va19}, where both longitudinal and radial gravitational accretion modes of filaments, linked to the global and local cloud collapse, can occur simultaneously and impact the filaments' stationary, growing, or dissipating states \citep{va19, na22}.


The role of magnetic fields can easily be incorporated into the above picture. Since the magnetic field tension force is perpendicular to field lines, a magnetic field component perpendicular to a filament will suppress the longitudinal potential, while a component parallel to a filament will suppress the radial potential. In other words, a magnetic field more perpendicular to a filament will favor the type I filament, whereas a field more parallel to a filament will favor the type II filament. 
This reveals itself as the positive correlation we found in the $\Delta$PA of B-F and G-B pairs in \autoref{fig:GBvsBF} which identifies the type I and type II filaments.

In short, {\it the two different magneto-gravitational configurations have different effective ways of driving  motions, either more effectively perpendicular to a filament (type I) or more effectively parallel to a filament (type II).
Based on all of the above, we propose two different types of filament resulting from two different magneto-gravitational configurations.}

\subsection{Star Formation in NGC 2264}\label{sec:YSO}
\begin{figure*}
\includegraphics[width=\textwidth]{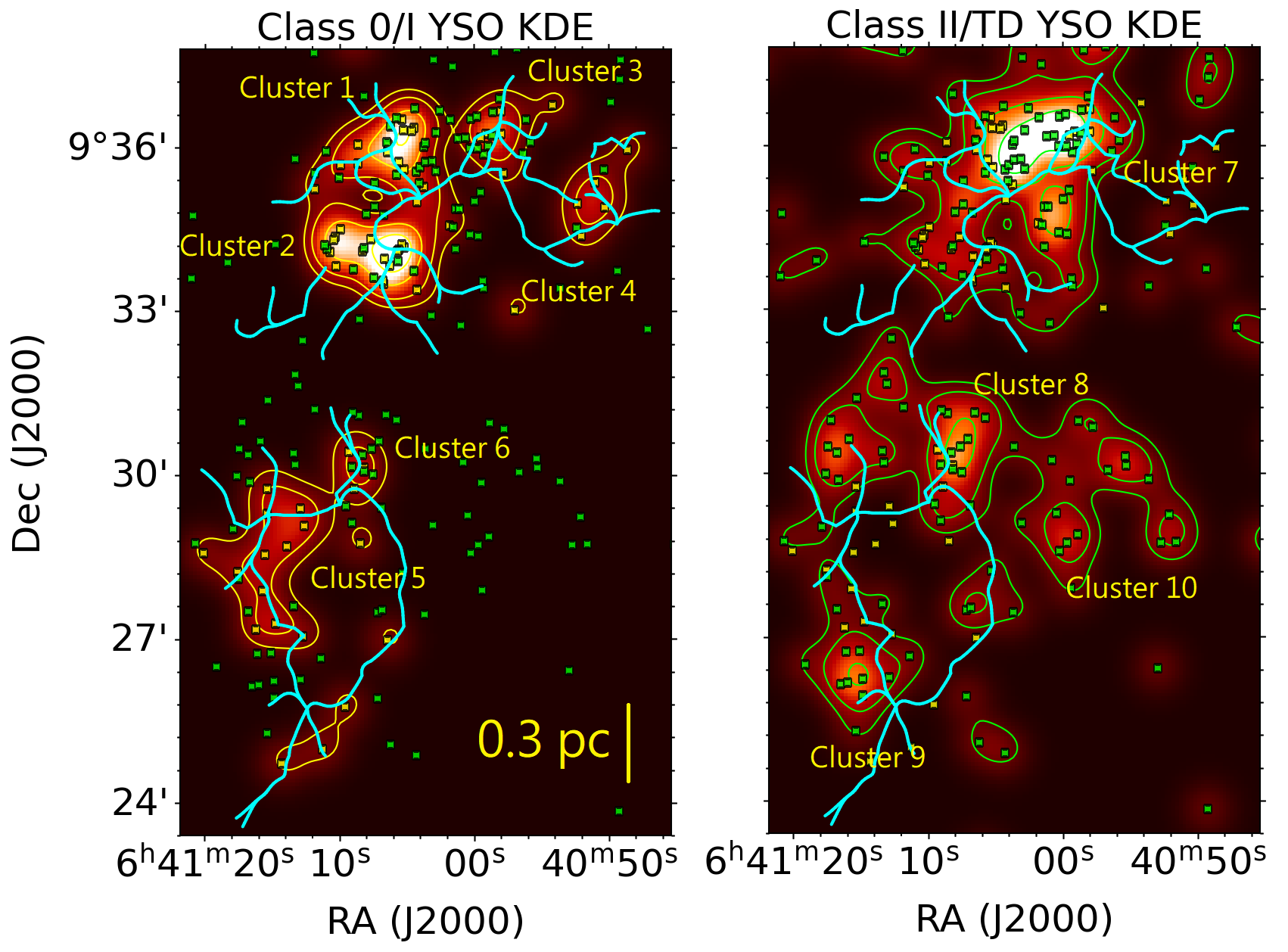}
\caption{YSO kernel density distribution maps for Class 0/I YSOs (left panel) and Class II/TD sources (right panel). The cyan lines are the identified filaments.
The yellow and green squares label the positions of Class 0/I and Class II/TD sources. The obvious YSO clusters are labelled as Cluster 1 to 10.
}\label{fig:yso}
\end{figure*}

YSOs are observed to form within filaments \citep{an14,pi22}. It is thus a question which type of filament tends to form stars. 
In order to trace the star formation activity in NGC 2264, we adopt the YSO catalog in \citet{ra14}, which is based on the $Spitzer$ and Two Micron All Sky Survey data. We select the YSOs classified as Class 0/I, II, and transitional disks (TD) from this catalog, and plot the YSO KDE for the young YSOs (Class 0/I) and evolved YSOs (Class II and TD) in \autoref{fig:yso}. We do not include the Class III YSOs, because those disk-less YSOs are difficult to distinguish from field stars using only infrared data \citep{ra14}. Since this YSO catalog covers an area greater than NGC 2264, we only select YSOs within RA (6$^{h}$40$^{m}$36$^{s}$) to (6$^{h}$41$^{m}$37$^{s}$) and Dec (9\degr22\arcmin23$^{\prime\prime}$) to (9\degr42\arcmin11$^{\prime\prime}$).

The kernel density distribution uses a Gaussian kernel with a FWHM width of 200\arcsec. This is chosen because it is comparable to the median proper motion velocity of 1.1 mas/yr \citep{bu20} times the free-fall dynamical timescale of 1.7$\times10^5$ yr for the NGC 2264 clumps \citep{pe06}. In this way, this represents the spatial probability distribution of each YSO, accounting for its possible migration. We note that this FWHM width is merely an order of magnitude estimate. A more accurate YSO migration would need to be described with an N-body model also considering the interaction between YSOs.

The derived YSO KDE maps show that the distributions of young and evolved YSOs are clearly different (\autoref{fig:yso}). The young YSOs are more tightly clustered.
In both 2264C and 2264D, the Class 0/I YSOs are clustered closely either on filaments or filament converging points.
In contrast to this, the overall distribution of the evolved YSOs is more scattered. The few evolved YSO clusters that can still be identified have their density peaks offset from filaments. In 2264C, these clusters are located to the north of the major filaments (Cluster 8), and only the two clusters (Cluster 9 and 10) in the southern 2264D are likely associated with filaments. In 2264D, evolved YSOs (Cluster 7) are concentrated in a filament-surrounded cavity, where no filaments are identified at the center.

\begin{figure}
\includegraphics[width=\columnwidth]{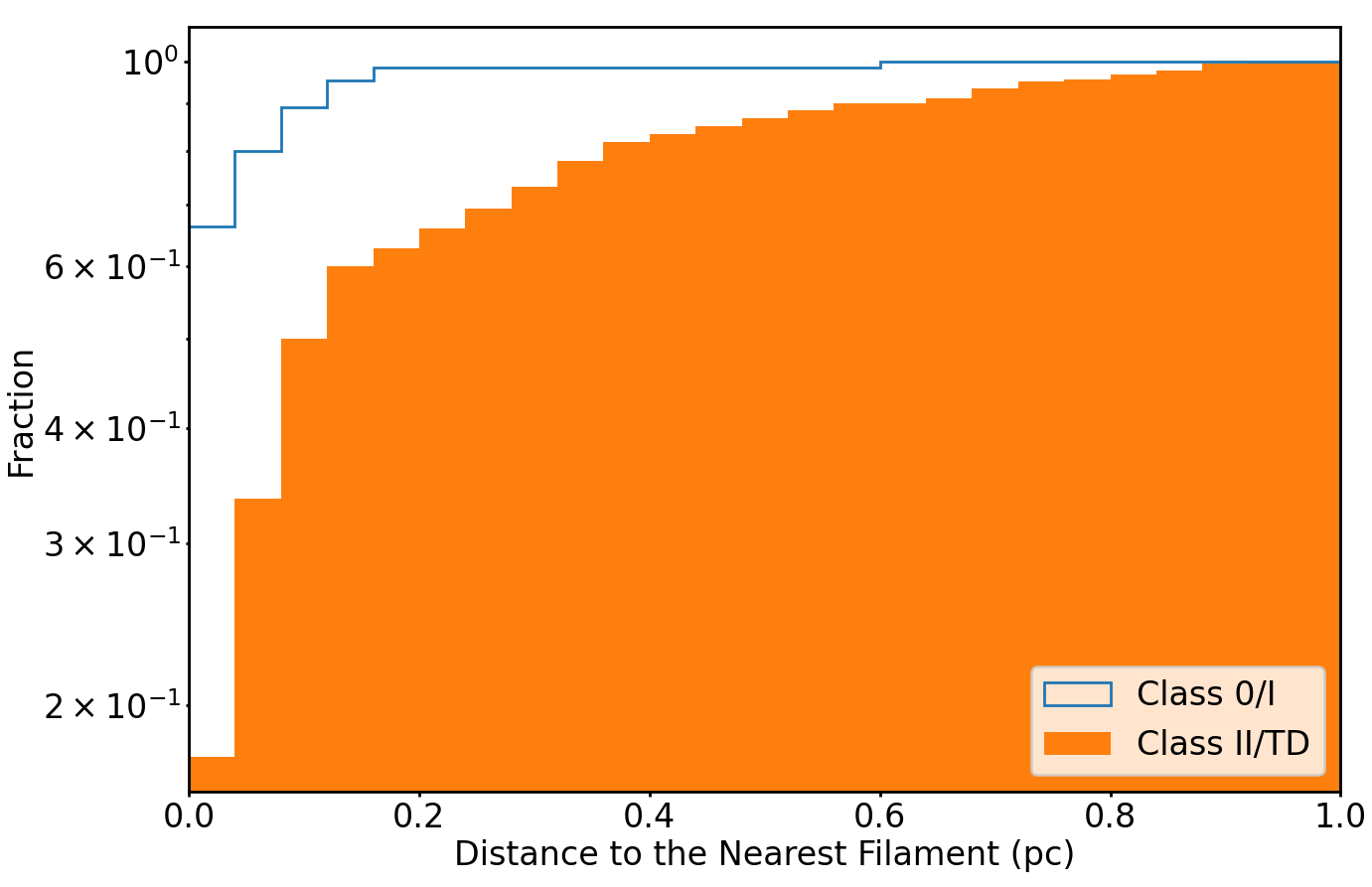}
\caption{Cumulative histogram of distances between YSOs and the nearest filament. The blue and orange histograms are for Class 0/I and Class II/TD, respectively. The 90 percentile of the nearest distance is 0.07 pc for Class 0/I and 0.45 pc for Class II/TD YSOs.}\label{fig:yso_dist}
\end{figure}

In order to further confirm that the evolved YSOs are more distant from filaments, we show cumulative histograms of the nearest distance between YSOs and filaments in \autoref{fig:yso_dist}. The 90th percentile of the nearest distance is 0.07 pc for Class 0/I YSOs and 0.45 pc for class II/TD YSOs. This suggests that nearly all Class 0/I YSOs are located very close to filaments, while evolved YSOs tend to be more distant from filaments, up to 0.5--1 pc in some cases.

In the following we discuss two possibilities that can cause offsets between evolved YSOs and filaments: (1) YSO migration and (2) change of filamentary structures.

(1) Due to their older age, evolved YSOs can migrate further from their primordial birth sites than younger YSOs, leading to a spatially more scattered distribution.  This possibility has been used to explain the different spatial distribution between protostars and disk sources in Southern Orion A \citep{ma16}. \citet{bu20} found a correlation between the evolutionary stage and source concentration in NGC 2264: 69.4\% of Class 0/I, 27.9\% of Class II, and 7.7\% of Class III objects are found to be clustered.
Based on this, migration of evolved YSOs was used to 
explain the differences in concentration in NGC 2264
\citep{bu20}. Nevertheless, no information of cloud structure was included in their study. In addition, the observed offset direction is not fully consistent with the direction of proper motions of the evolved YSOs. For example, the proper motion direction of the evolved YSOs is $\sim$240--270\degr\ in cluster 8 and $\sim$270--290\degr\ in cluster 7 \citep{bu20}, which does not likely trace back to existing filaments.

(2) The evolved YSOs might have formed from filaments that have been dissipated. \citet{in16} argue that present YSOs can have partially formed within filaments that existed in the past. This is suggested because the total mass of the observed YSOs often exceeds the product of the total filament mass and the star-formation efficiency. Since the free-fall dynamical timescale ($\sim2\times 10^5$ yr, \citet{pe06}) of the clumps in NGC 2264 protoclusters is shorter than the typical age of Class II YSOs (1--2 Myr, \citep{ev09}), those filaments or clumps forming the evolved YSOs might have dissipated in between. In addition, the star formation feedback, such as outflows or expanding shells, might also clear the primordial local density structures. 
This possibility is supported by the presence of cavities associated with evolved YSO clusters (e.g., Cluster 7 and 10), and it also aligns with the scenario proposed by \citet{ku20}, which suggests that NGC 2264 originated from a collision between filaments on a parsec-scale. In this scenario, the first-generation massive stars are formed at the filament collision center (Cluster 8). The feedback from the massive star subsequently clears the material in the gap between 2264C and 2264D, while 2264C and 2264D are the remnants of the filaments that collided.
Finally, it is worth mentioning that according to \citet{nony21}, the weakly embedded Class II sources (such as Cluster 10) have the potential to migrate $\sim$6 pc if they move freely without being bound to any structures. Thus, those Class II sources that are significantly separated from filaments could also originate from distant regions.

\onecolumngrid 
\section{Conclusions}\label{sec:con}
This study utilizes JCMT SCUBA-2/POL-2 850 $\mu$m continuum polarization observations to investigate the hub-filament system NGC 2264. An organized magnetic field morphology is uncovered with the following main outcome.

\begin{itemize}
    \item Orientations of filaments (F), local gravity (G), and magnetic field (B)
    are estimated. These spatial parameters are found to share common prevailing orientations around 20 to 50\degr, with only the filaments showing a second prevailing orientation around 150\degr. This suggests that the evolution of these four parameters is possibly regulated by the same physical processes.
    
    \item The pairwise analysis of the relative orientations between these parameters indicates that they are {\it locally} correlated. Most of the parameter pairs tend to become more tightly aligned toward higher intensities, suggesting that these parameters are likely aligned by the increased gravity. A transition of the relative orientation becoming more perpendicular at very high densities is found in the HRO diagram. This is likely associated with the transition between two types of filaments. 

    \item A significant number of the filaments in NGC 2264 can be categorized into two distinct groups.
    This is based on their alignment properties as revealed in the 2D KDE map for B-F vs B-G: B$\perp$F and B$\perp$G (Type I filament) and B$\parallel$F and B$\parallel$G (Type II filament). 

    \item  We examine the spatial arrangements of the magneto-gravitational configuration {\it locally} on the maps associated with the two types of filaments. These maps reveal that the magneto-gravitational configuration in Type I filaments shows G parallel to B at larger distances from F, but then transitions to G becoming increasingly perpendicular to B closer to F. In contrast to that, the Type II filaments display G parallel to B both at larger distances and also close to F. 
    
    \item We propose a picture to explain the observed features of the two types of filaments. This involves the competition between the radial and longitudinal collapsing time scale. We interpret Type I filaments as potentially being able to accumulate more nearby material, while Type II filaments are rather shaped by a dominating longitudinal pull from the larger-scale global gravitational field.

    \item Magnetic field strengths are estimated for the major clumps in 2264C and 2264D together with a Virial analysis. The estimated ratios of kinematic energy ($\mathcal{T}$): magnetic energy ($\mathcal{M}$): gravitational energy ($\mathcal{W}$) are 0.3$^{\pm0.2}$:0.2$^{\pm0.1}$:1 in 2264C and 0.9$^{\pm0.4}$:0.8$^{\pm0.4}$:1 in 2264D. This indicates that the {\it global} evolution of 2264C is dominated by gravity, while the three factors appear comparably important in 2264D. 

    \item The Class 0/I YSOs are found to be closer in distance to the identified filaments or filament converging points. Most of the Class II and transition disk YSOs are located at systematically 
    larger distances from filaments.
    
\end{itemize}

\acknowledgments
The James Clerk Maxwell Telescope is operated by the East Asian Observatory on behalf of The National Astronomical Observatory of Japan; Academia Sinica Institute of Astronomy and Astrophysics; the Korea Astronomy and Space Science Institute; the National Astronomical Research Institute of Thailand; Center for Astronomical Mega-Science (as well as the National Key R{\&}D Program of China with No. 2017YFA0402700). Additional funding support is provided by the Science and Technology Facilities Council of the United Kingdom and participating universities and organizations in the United Kingdom and Canada. Additional funds for the construction of SCUBA-2 and POL-2 were provided by the Canada Foundation for Innovation. The authors wish to recognize and acknowledge the very significant cultural role and reverence that the summit of Maunakea has always had within the indigenous Hawaiian community. 
We are most fortunate to have the opportunity to conduct observations from this mountain. J.-W.W and P.M.K. acknowledge support from the National Science and Technology Council (NSTC) in Taiwan through grants
NSTC 111-2112-M-001-039 and NSTC 112-2112-M-001-049.
K.Q. is partially supported by the National Key R\&D Program of China No. 2022YFA1603100, No. 2017YFA0402604. K.Q. acknowledges the National Natural Science Foundation of China (NSFC) grant U1731237 and the science research grant from the China Manned Space Project with No. CMS-CSST-2021-B06. F.P. acknowledges support from the Agencia Canaria de Investigaci\'{o}n, Innovaci\'{o}n y Sociedad de la Informaci\'{o}n (ACIISI) under the European FEDER (FONDO EUROPEO DE DESARROLLO REGIONAL) de Canarias 2014-2020 grant No. PROID2021010078. The work of MGR is supported by NOIRLab, which is managed by the Association of Universities for Research in Astronomy (AURA) under a cooperative agreement with the National Science Foundation. L.F. acknowledges support from the Ministry of Science and Technology of Taiwan, under grants No. 111-2811-M-005-007 and 109-2112-M-005-003-MY3. D.J.\ is supported by NRC Canada and by an NSERC Discovery Grant. K.P. is a Royal Society University Research Fellow, supported by Grant No. URF\textbackslash R1\textbackslash 211322. CE acknowledges the financial support from grant RJF/2020/000071 as a part of the Ramanujan Fellowship awarded by the Science and Engineering Research Board (SERB), Department of Science and Technology (DST), Government of India. C. W.L. was supported by the Basic Science Research Program through the National Research Foundation of Korea (NRF) funded by the Ministry of Education, Science and Technology (NRF-2019R1A2C1010851), and by the Korea Astronomy and Space Science Institute grant funded by the Korea government (MSIT; project No. 2023-1-84000). W.K. was supported by the National Research Foundation of Korea (NRF) grant funded by the Korea government (MSIT) (NRF-2021R1F1A1061794). M.T. is supported by JSPS KAKENHI grant No.18H05442. 
\twocolumngrid

\appendix
\section{Polarization Properties}\label{sec:pp}
\begin{figure*}
\includegraphics[width=\textwidth]{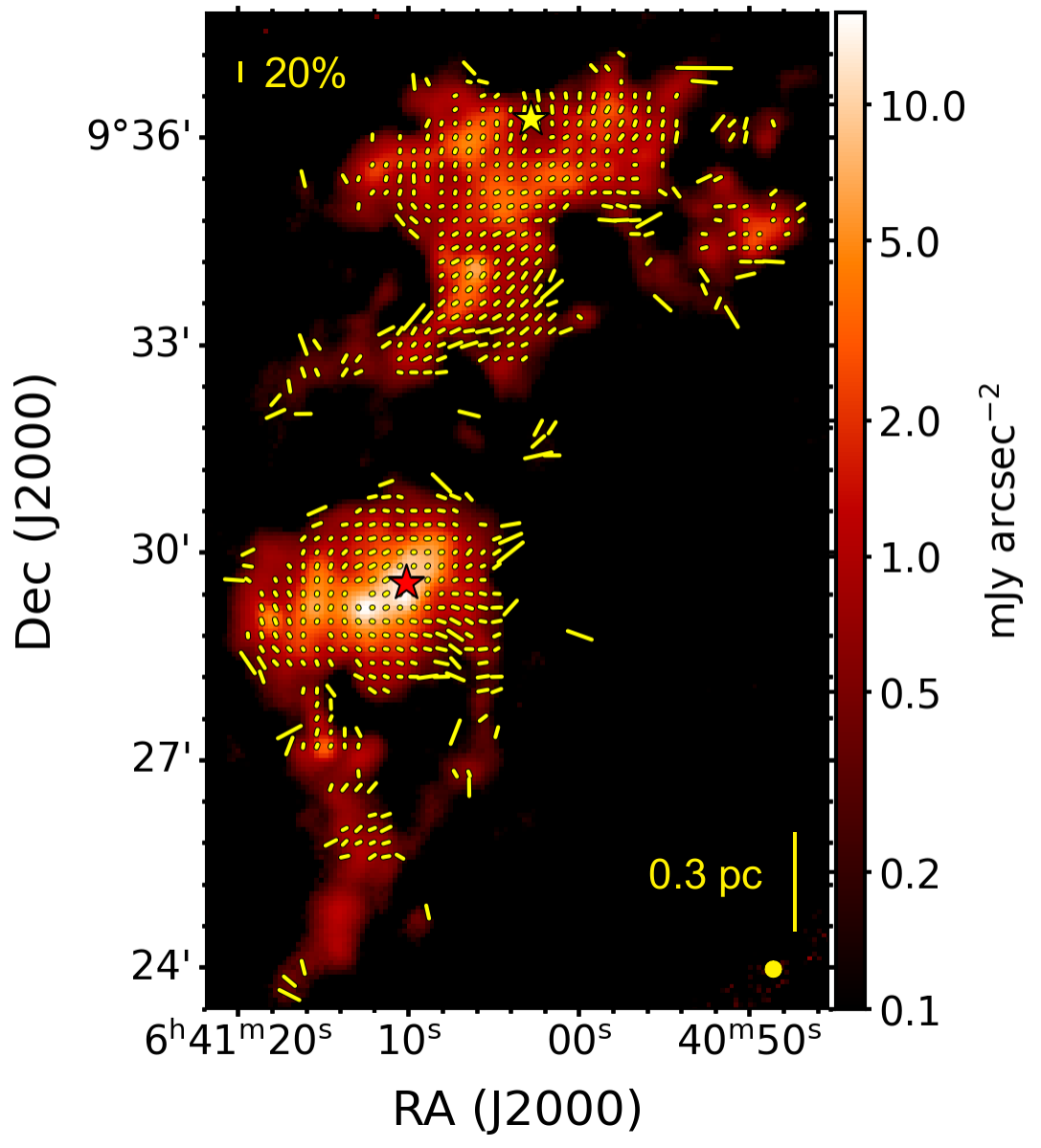}
\caption{POL-2 polarization segments overlaid on 850 $\mu$m dust continuum toward NGC 2264. The yellow segments show polarization detections selected by $I/\sigma_{I}>20$ and $P/\sigma_{P} >3$.
The lengths of the segments are scaled with the polarization fractions. The yellow scale bar at the top left shows a 20 \% polarization fraction. The red and yellow stars label IRS1 (zero-age main-sequence) and IRS2 (Class I), the dominating sources in 2264C and 2264D.
}\label{fig:pfmap}
\end{figure*}

\begin{figure}
\includegraphics[width=\columnwidth]{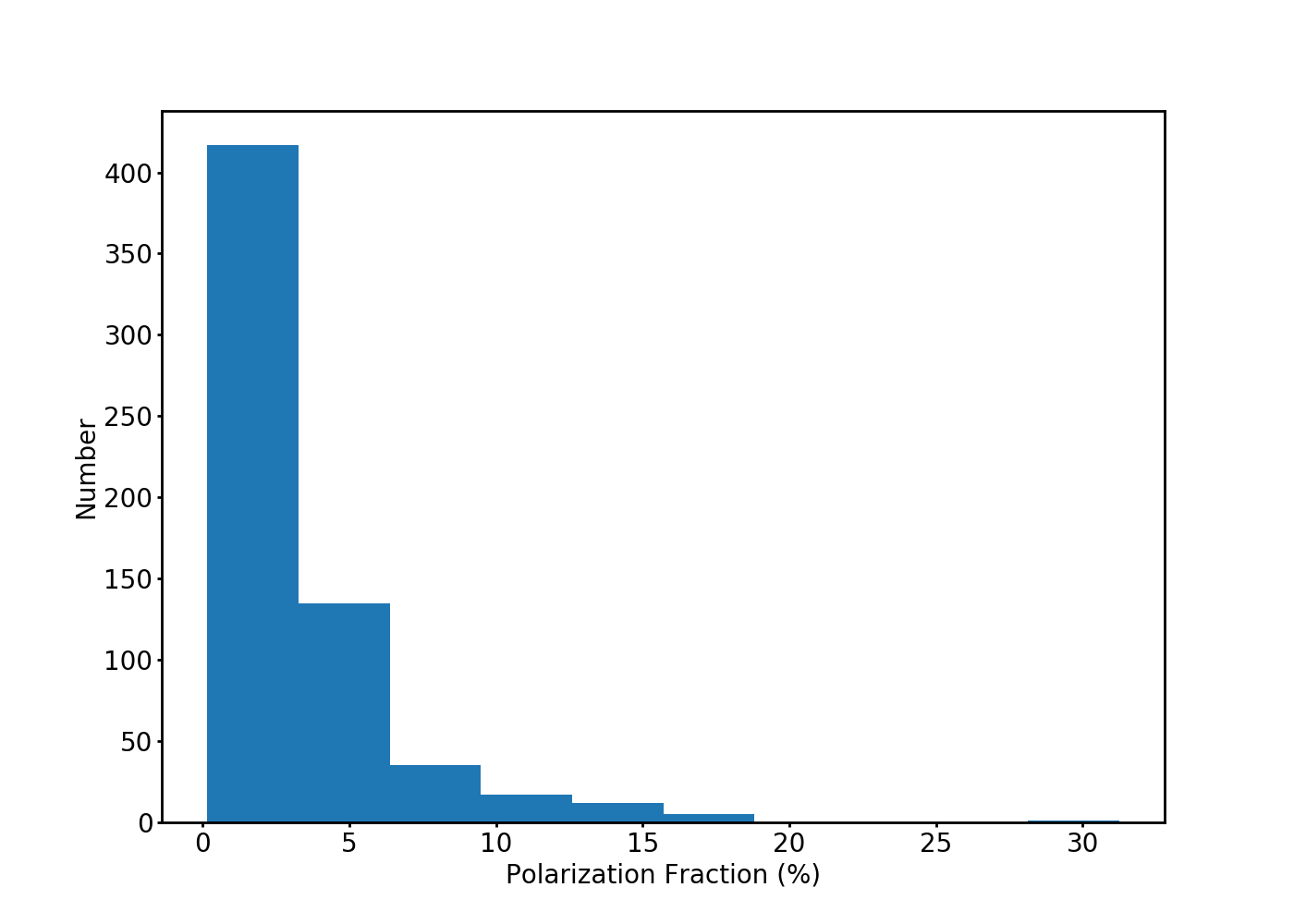}
\caption{Histogram of debiased polarization fraction of the selected samples.
Most of the segments show a polarization fraction smaller than 10\% with a median value of 2.4\%.}\label{fig:P_hist}
\end{figure}

\autoref{fig:pfmap} shows the 850 $\mu$m POL-2 polarization map, where the lengths of the segments are scaled with the polarization fraction. The resulting histogram is reported in \autoref{fig:P_hist}. The polarization fraction, with a median value of 2.4\%, is mostly below 10\%. Values of 10--20\% are rare, typically occuring near the edge of the clouds.

The correlation between Stokes $I$ and polarization fraction $P$ is commonly used to examine whether the observed polarization is originated from aligned dust grains. This is essential to confirm that the polarization measurements can trace the magnetic fields within molecular clouds \citep[e.g.,][]{an15,jo15,pl15,jo16,wa19,pa19,pl20}. 
The $I$--$P$ relation can be described by a power law, $P=PI/I\propto I^{-\alpha}$, where $PI$ is the polarized intensity. If the dust grains are not aligned with the magnetic field, $PI$ is independent of Stokes $I$ and $\alpha$ is 1. $\alpha$ smaller than 1 indicates that the polarization fraction results from magnetically aligned dust grains.

\begin{figure}
\includegraphics[width=\columnwidth]{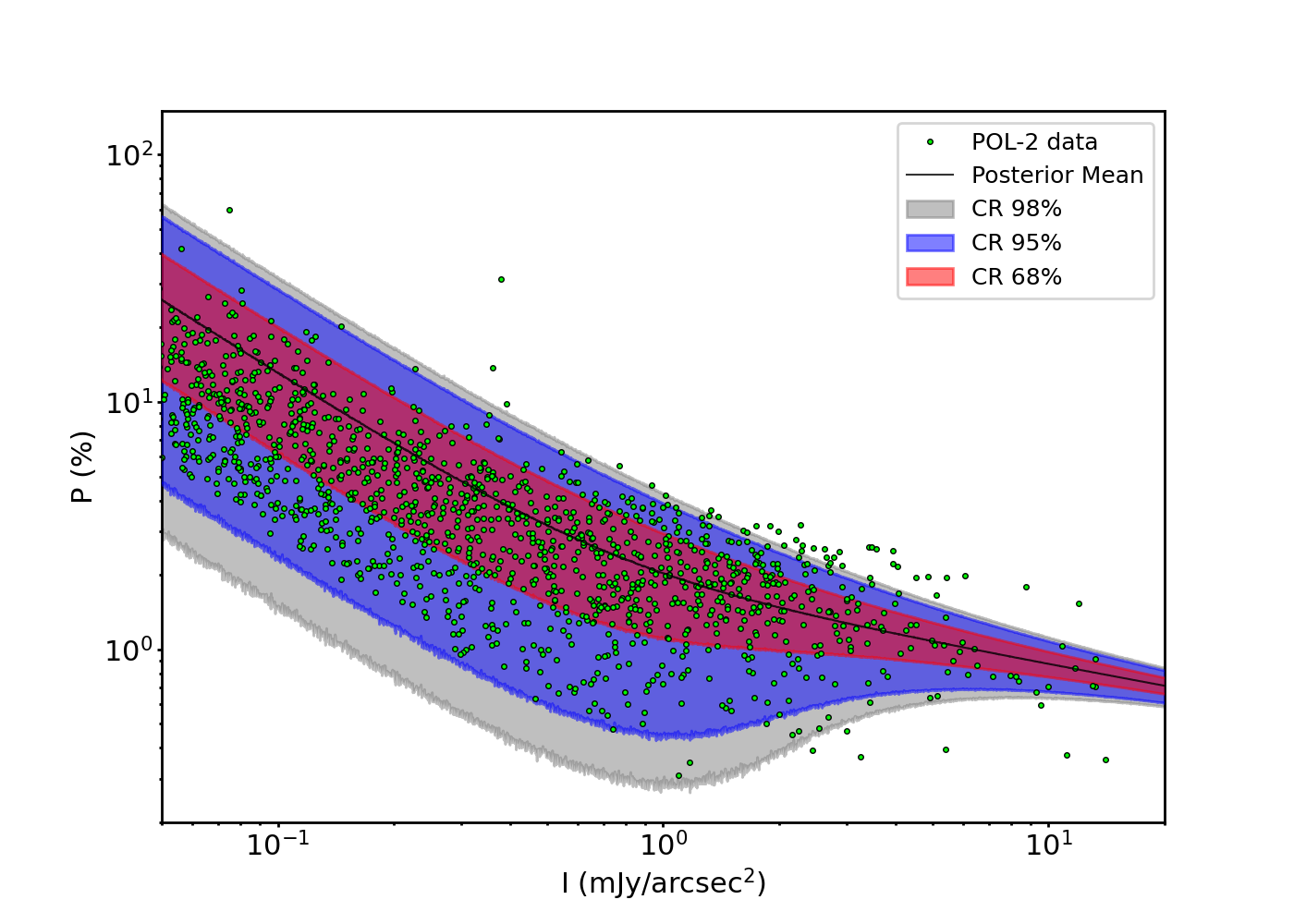}
\caption{850 $\mu$m total intensity $I$ vs polarization fraction $P$. The green points are the non-debiased POL-2 polarization measurements, selected by $I/\sigma_{I}>$10 and $\sigma_I<$0.02 \mjya\ . 
The colored regions label the 68\%, 95\%, and 98\% confidence regions (CR), predicted by the Bayesian model. The black line indicates the posterior mean. Most of the data points are within the 98\% confidence region of our prediction.}\label{fig:IP}
\end{figure}

The observed $I$--$P$ relation for NGC 2264 is shown in \autoref{fig:IP}. We follow the Bayesian analysis in \citet{wa19} to determine the power-law index $\alpha$ using the model 
\begin{equation}\label{eq:prior}
P =\beta I^{-\alpha}.
\end{equation}
The Bayesian model allows us to directly model the Ricean distribution \citep{wa74} as the error function of the observed $P$
\begin{equation}\label{eq:rice}
F(P|P_0)=\frac{P}{\sigma_P^{2}}\exp\left[-\frac{P^2+P_0^2}{2{\sigma_P}^2}\right]I_0\left(\frac{PP_0}{\sigma_P^2}\right),
\end{equation}
where $P$ is the observed polarization fraction, $P_0$ is the real polarization fraction, $\sigma_P$ is the Ricean dispersion in the polarization fraction, and $I_0$ is the zeroth-order modified Bessel function. We further assume that the uncertainty in the polarization fraction is dominated by the uncertainties in Stokes Q and U
\begin{equation}
\sigma_P=\sigma_Q/I,
\end{equation}
where $\sigma_Q$ is the dispersion in Stokes Q, assuming the same noise in Stokes Q and U.

\begin{figure}
\includegraphics[width=\columnwidth]{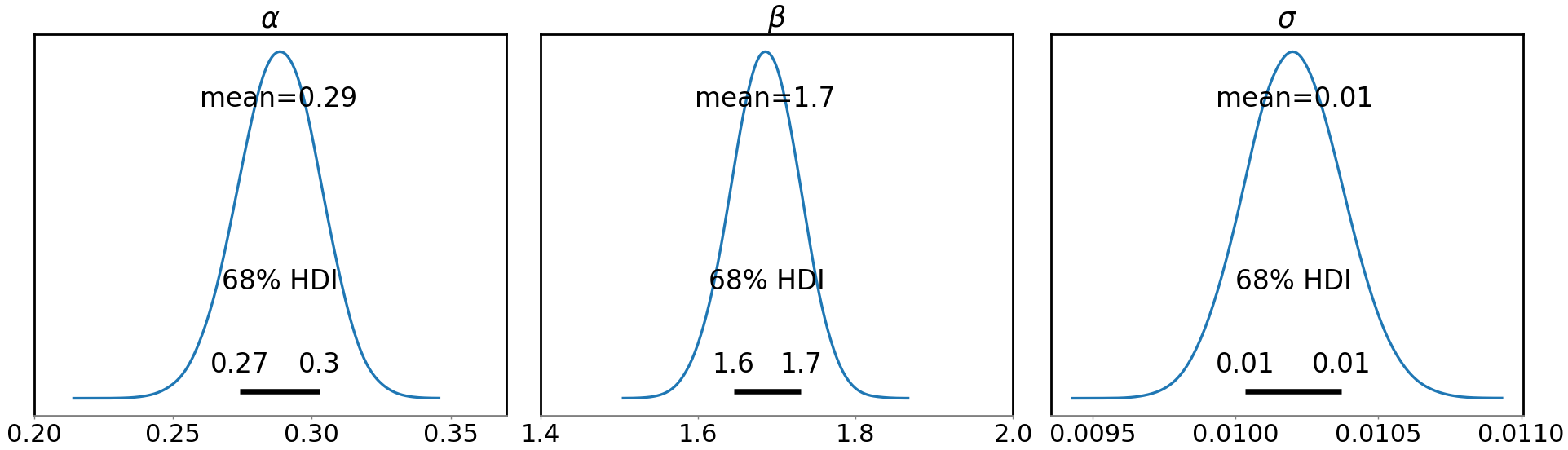}
\caption{Posterior distributions of our Bayesian model for $I-P$ distribution. The black lines delineate the 68\% ($1\sigma$) highest density interval (HDI). The most probable $\alpha$ of $0.29\pm0.01$ is significantly less than unity, suggesting that the observed polarization likely originates from aligned dust grains in NGC 2264. }\label{fig:IPpost}
\end{figure}

We re-select the polarization data based on the criteria $\sigma_I < 0.02$ \mjya\ and $I/\sigma_I>10$, in order to have sufficient data points to model the noise-dominated regime. Additionally, the non-debiased polarization fractions are used because the Ricean noise is well accounted for in this model. The $I$--$P$ relation predicted by our Bayesian model is shown in \autoref{fig:IP}. Most samples are well covered by the predicted 98\% confidence region (CR). Nevertheless, some data points with polarization fractions higher than predicted by the model can be seen in the high-intensity regions ($I >$ 2.0 \mjya). This can be due to different dust properties or additional radiative alignment torques caused by the embedded sources. The computed posterior distributions of the Bayesian model are reported in \autoref{fig:IPpost}. $\alpha$ is constrained to $0.29\pm0.01$ which is significantly smaller than 1. Consequently, this suggests that the observed polarization likely originates from magnetically aligned dust grains, and hence traces the magnetic field morphology within NGC 2264. 

\section{\texorpdfstring{\13CO} (3-2) and \texorpdfstring{\C18O (3-2)} Molecular Lines}\label{sec:13COvsC18O}
\begin{figure}
\includegraphics[width=\columnwidth]{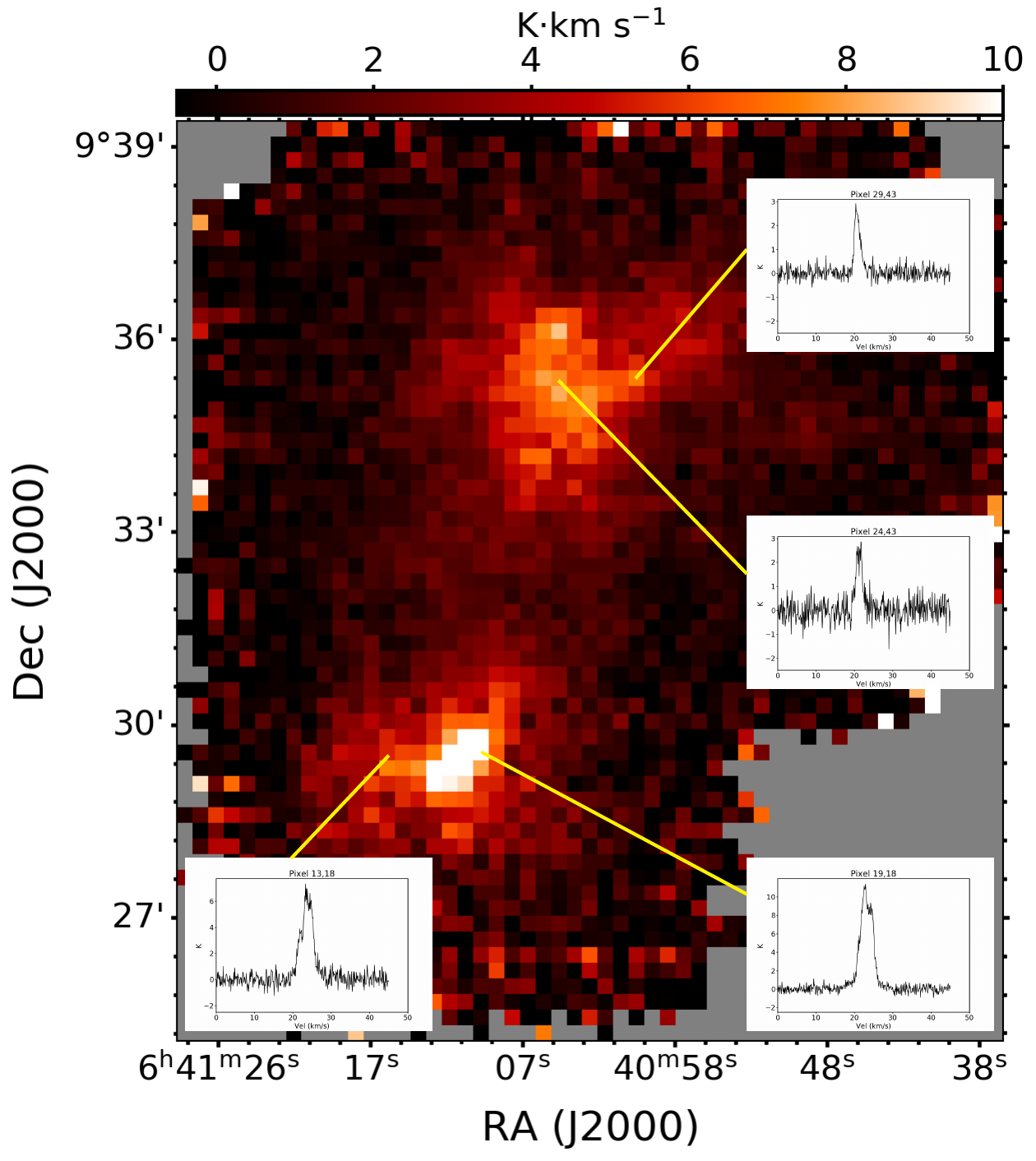}
\caption{\13CO (3-2) integrated intensity map with spectra extracted at different pixels. Double or triple velocity components separated by only a few \kms are seen.}\label{fig:13CO}
\end{figure}

In this section, we discuss the possible influence of the choice of the \13CO (3-2) or \C18O (3-2) line to estimate the velocity dispersion.
\autoref{fig:13CO} shows example spectra of \13CO (3-2) in NGC 2264 at different positions. In most of the locations in NGC 2264, we identified multiple velocity components or asymmetric line profiles in the spectra. The separations of those multiple components are usually less than or comparable to their linewidths. 

\begin{figure}
\includegraphics[width=\columnwidth]{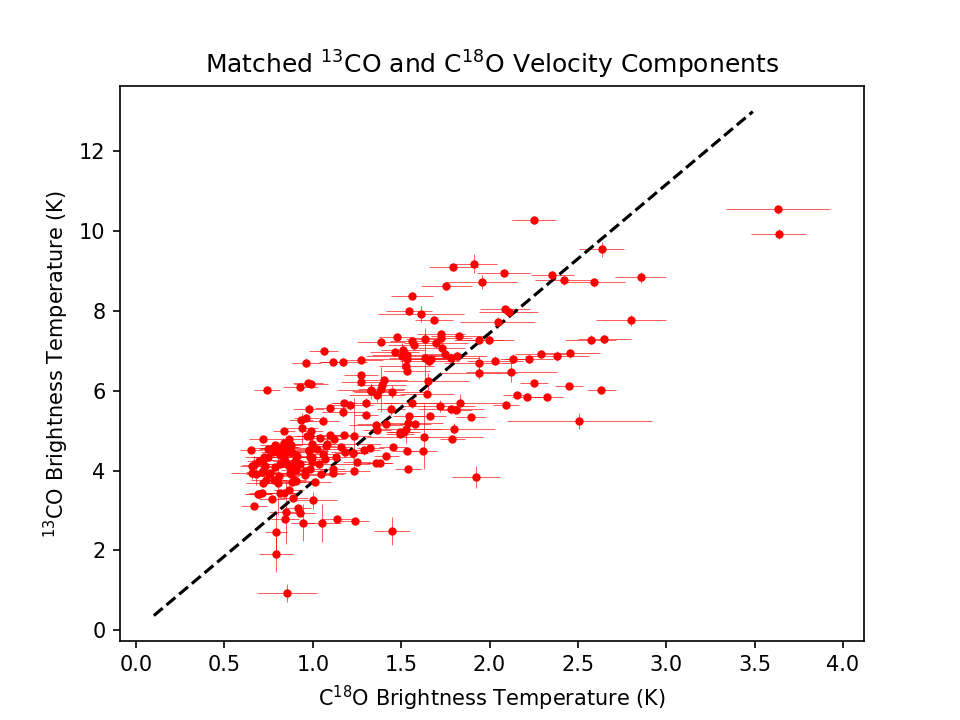}
\caption{Comparison of the fitted peak amplitudes of the matched velocity components in \13CO (3-2) and \C18O (3-2). \C18O (3-2) is likely optically thin, because the brightness temperature is much lower than the possible gas temperature in molecular clouds. The dashed line marks the best-fit intensity ratio of 3.7. 
Most points follow the fitted intensity ratio, except for a few pixels where the \13CO amplitude is above 10 K. This suggests that \13CO (3-2) 
is likely optically thin in most of the regions.
}\label{fig:amp_C18Ovs13CO}
\end{figure}

The fitted amplitude of the velocity components in \13CO (3-2) and \C18O (3-2) are reported in \autoref{fig:amp_C18Ovs13CO}. These components are matched if their centroid velocity is close within 2 times the fitted uncertainties in the velocity component identification. The black dashed line represents the best-fit T(\13CO)/T(\C18O) ratio of 3.7. Most points follow well the fitted line, suggesting that \13CO (3-2) and \C18O (3-2) are both optically thin.

\begin{figure}
\includegraphics[width=\columnwidth]{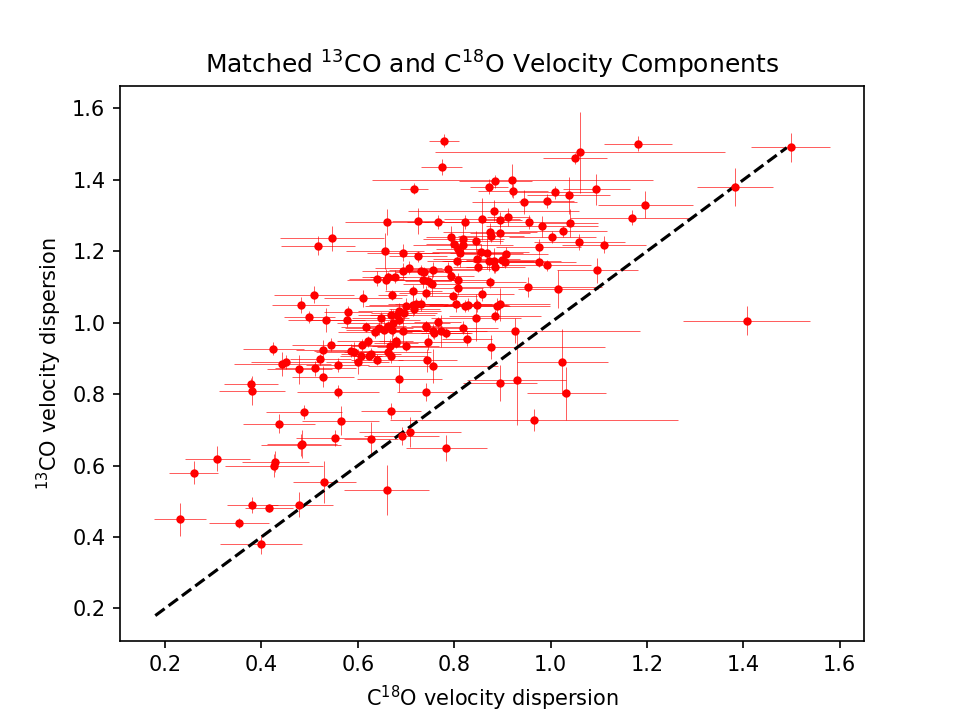}
\caption{Comparison of the fitted velocity dispersion of the matched velocity components in \13CO (3-2) and \C18O (3-2). The dashed line marks where the two dispersion values are equal. 
The velocity dispersion values extracted from \13CO 
are systematically wider than the ones from \C18O by 30-50\%.
}\label{fig:vdisp_C18Ovs13CO}
\end{figure}

\autoref{fig:vdisp_C18Ovs13CO} shows a comparison between the velocity dispersion estimated from the two tracers \13CO (3-2) and \C18O (3-2). These velocity components are matched as \autoref{fig:amp_C18Ovs13CO}, but with an additional criterion 2.5 $<$ T(\13CO)/T(\C18O) $<$ 5.5 to ensures that those components are really associated. The \13CO velocity dispersion is systematically larger by 30--50\%. We speculate that \13CO might trace more of the relatively low-density outskirts of NGC 2264, which could be affected by the large-scale accreting filaments, hence being more turbulent and leading to larger velocity dispersion. Therefore, we conclude that \C18O is more reliable to trace the velocity dispersion within the NGC 2264 clumps. The five outliers with $\sigma_V$(\C18O)/$\sigma_V$(\13CO) $>$ 1.2 are from the faint pixels, where the velocity components are not well-constrained by the \C18O data, and thus we flagged the five pixels from calculating the mean velocity dispersion in the DCF analysis.

\section{Filament Properties -- Width, Line Mass, and Filament-to-Cloud Fraction}\label{sec:fi}
\begin{figure}
\includegraphics[width=\columnwidth]{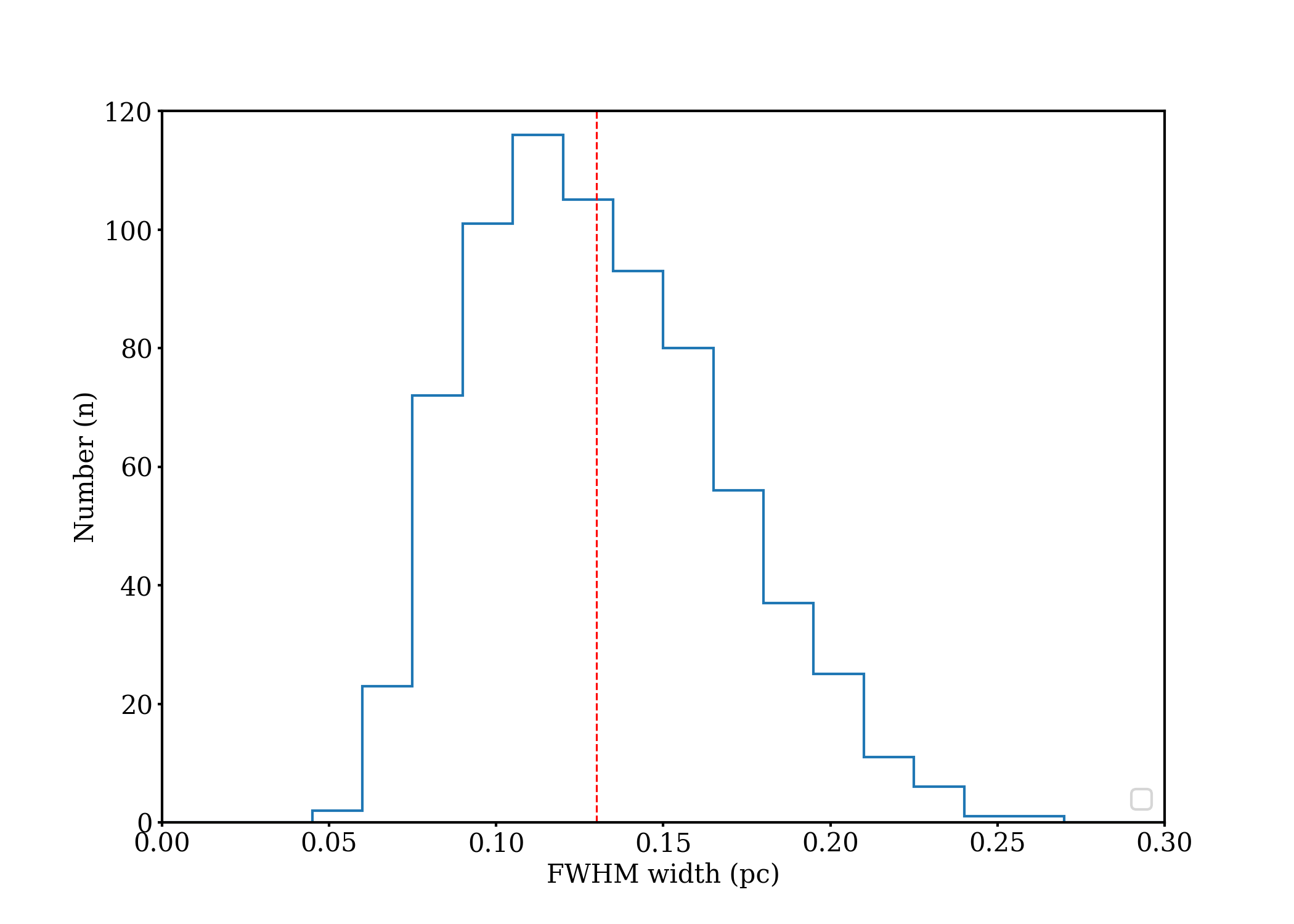}
\caption{Histogram of FWHM filament widths estimated from individual intensity radial profiles. The red dashed line is the median width of 0.13 pc.}\label{fig:Fwidth}
\end{figure}

\begin{figure}
\includegraphics[width=\columnwidth]{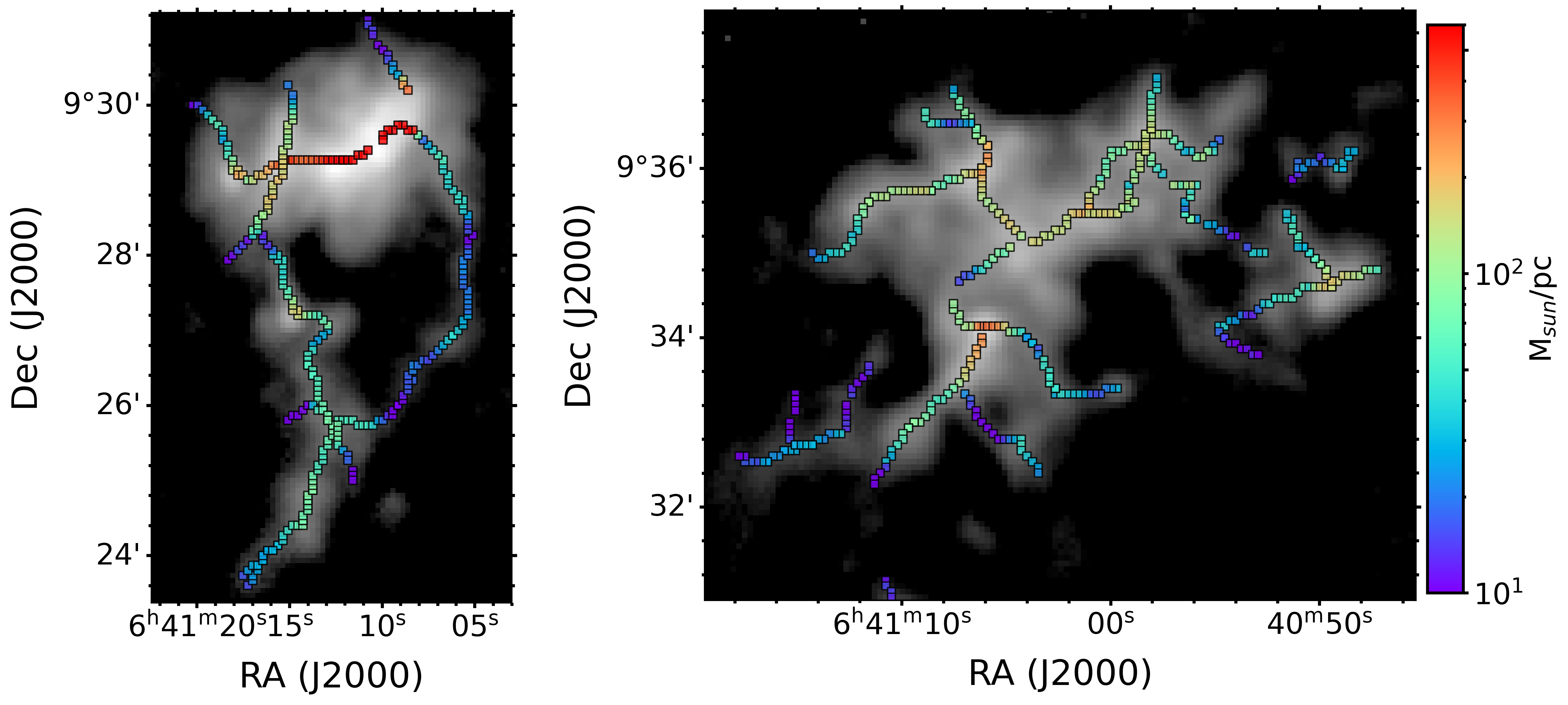}
\caption{Maps of filament line mass. The color indicates the line mass at each position, estimated from the radial intensity profile. The filament line mass gradually increases from a range of [10,50] M$_{\sun}$/pc in the outskirts to above 100 M$_{\sun}$/pc in the converging points. The main filament in 2264C has the highest line mass of $\sim$600 M$_{\sun}$/pc.}\label{fig:Flmass}
\end{figure}

We derive the properties of the identified filaments by fitting the radial intensity profile at each pixel along the filaments using the bootstrap method. A fitting result is selected if (1) one major peak is found, and (2) the fitted width is above 3$\sigma$. 
We further derive the central column densities based on the fitted amplitude (assuming a constant dust temperature of 15 K \citep{ward2000} and a dust opacity $\kappa$ of 0.0125 cm$^2$/g at 850 $\mu$m \citep{hi83}) and a line mass (M$_{\textrm{line}}$; also known as mass-per-unit-length) by multiplying the central column density with the filament width. \autoref{fig:Fwidth} shows a histogram of the FWHM filament width. The mean and median filament widths are both 0.13 pc which is very similar to the characteristic width of 0.1 pc found in \citet{ar11}. The shape of the filament width distribution with a rapid decline to larger widths is consistent with the filament formation via supersonic turbulence \citep{pr22}. 
\autoref{fig:Flmass} plots the line mass along each filament. A gradual increase is seen, ranging from 10 to 50 M${\sun}$/pc at the outskirts, to over 100 M${\sun}$/pc at the converging points. Notably, the main filament within 2264C has the highest line mass, reaching approximately 600 M$_{\sun}$/pc. This suggests that these filaments are actively accumulating ambient material, and possibly transport the accumulated mass toward the converging points.

The critical filament line mass \citep{fi20} can be estimated by
\begin{equation}
   M_{line,critical}=2(c_s^2+\sigma_{v,NT}^2)/G,
\end{equation}
where $c_s$ is the sound speed and $\sigma_{v,NT}$ is the non-thermal velocity dispersion. Assuming a non-thermal velocity dispersion of 1.0 km/s and a gas temperature of 15 K, the $M_{line,critical}$ is about 400 M$_{\sun}$/pc. This is higher than the derived line mass in most of the regions. However, it is important to note that the critical line mass is determined for filaments assumed to be in hydrostatic balance. This might not be the case 
for most of the observed filaments with accretion onto or along them as indicated by velocity gradients \citep[e.g.,][]{pe13,st16,wi18,zh22}. These motions could then 
change the energy balance. In addition, HFSs, with the presence of dense hubs, behave as major gravitational potential wells.
Thus, the gravitational potential in a filament is a superposition of both the filament's self-gravity and the global mass distribution. This is much more complex than the case of an isolated, static filament. Consequently, a dynamic filament model is desirable to more accurately describe the stability of this system.

Assessing the proportion of mass within a cloud that resides in filaments is essential to evaluate the final star formation efficiency \citep{in12,in16}. To accomplish this, we employ two different methods to calculate the total mass of the identified filaments. 
Firstly, we integrate the products of $M_{line}$ 
and $\Delta$L, where $\Delta$L represents the length of each filament segment. According to this calculation, the total filament mass amounts to 757 M${\sun}$, which corresponds to 26\% of the total cloud mass of 2960 M${\sun}$ \citep{pe06}.
Secondly, we calculate the total mass associated with the pixels belonging to filaments, selected if pixels are within the fitted 2$\sigma$ width. To estimate the large-scale background, we use a Gaussian kernel with a $\sigma$ of 5 pixels to interpolate the intensity map of the unselected pixels. After subtracting the contribution of the large-scale background, the total filament mass is determined to be 567 M${\sun}$. This is equivalent to 19\% of the total cloud mass.
While the first method might be an upper limit 
(because filaments partially overlap near the converging points), the second method might 
be a lower limit (due to a possible over-subtraction of the background). 
We note that the major uncertainty in this estimation is the large-scale filtering effect. This is because the total mass recovered by our POL-2 continuum map is 1150 M${\sun}$, which is only 32\% of the total mass estimated from the unfiltered 1.2 mm continuum data in \citet{pe06}.

%

\section{Intensity Gradient}\label{sec:IG_appendix}
In addition to filament extraction algorithms (Section \ref{sec:FIG}), the intensity gradient can also characterize features in density structures. \citet{koch12a,koch12b} show that the local intensity gradient is a measure for the resulting direction of motion in the magneto-hydrodynamic (MHD) force equation, and hence it can be used to evaluate the relative importance of the magnetic field together with other constituents (such as local gravity, \autoref{sec:G}). 
One outcome of their polarization -- intensity gradient technique is the magnetic field significance $\Sigma_B$=$\frac{sin \Phi}{cos\delta}$ which is a ratio that describes the relative importance between magnetic field and gravity solely based on measurable angles. $\Phi$ is the angle between intensity gradient and local gravity (\autoref{sec:G}) and $\delta$ the angle between intensity gradient and magnetic field. 
This method has been applied to a 50-source sample of star-forming regions to quantify the importance of the magnetic field in these sources \citep{koch14}. An additional method to characterize the magnetic field and density morphologies are histograms of relative orientations \citep{so13} which have been used for a variety of systems, from larger-scale to filamentary systems \citep[e.g.,][]{pl16,kw22}, finding that the relative orientation (which is identical to the angle $\delta$) can be either parallel or perpendicular or transitioning, depending on the local density.

The local intensity gradients in our 4\arcsec-pixel 850 $\mu$m continuum image are derived. We note that even though the over-sampled pixels may be partially correlated, the derived orientations of the intensity gradients remain unaffected. This is because of the isotropic beam pattern. The gradients are calculated by fitting at each grid point over 3$\times$3 pixels in the continuum image \citep{go93} with
\begin{equation}\label{eq:ig}
I(x,y)= c_{x}x + c_{y}y + c_0,
\end{equation}
where x and y are the coordinates of each pixel, and $c_0$, $c_{x}$, and $c_{y}$ are free parameters representing the first-order expansion of the intensity field. 

\begin{figure*}
\includegraphics[width=\textwidth]{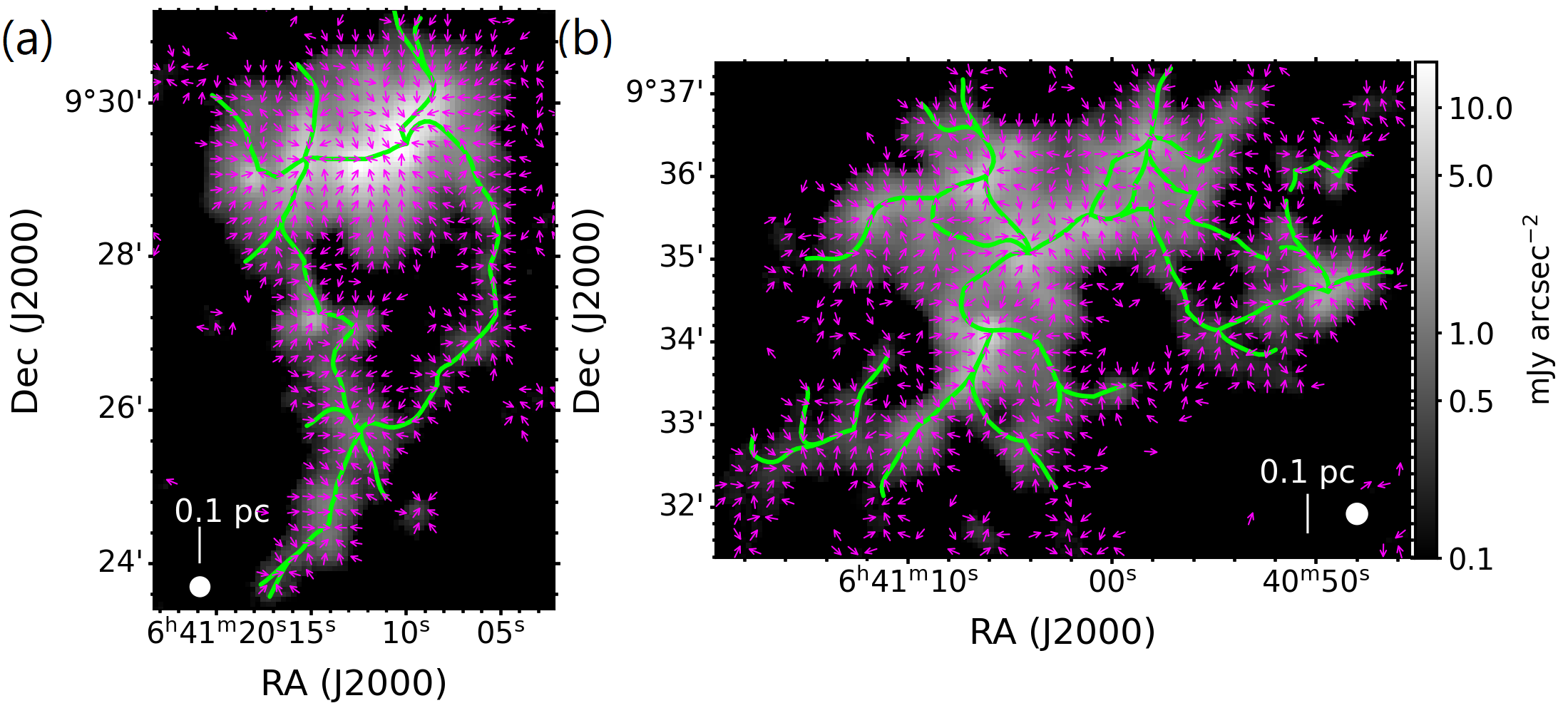}
\caption{Filaments (green lines) and intensity gradients (magenta arrows) overlaid on 850$\mu$m continuum toward (a) 2264C and (b) 2264D. 
The intensity gradients are calculated per pixel, but only shown per 3$\times$3 pixels and with uniform length for simplicity.}\label{fig:Fmap}
\end{figure*}

The resulting intensity gradients are shown as arrows in \autoref{fig:Fmap}, with a median  orientation uncertainty of 6.3\degr.
Overall, the estimated intensity gradients are pointing toward both the massive clumps and filament ridges. This suggests that the local mass is pulled by not only the filaments themselves but also by global gravity toward the cloud's main mass centers. As a result, the intensity gradients often show offset angles of $\sim$30\degr--60\degr\ to the filament ridges, instead of simply being parallel or perpendicular. 
The histogram of intensity gradients (right panel in \autoref{fig:hist}) displays a less pronounced but still visible broad peak
around 30--60\degr. This is a similar prevailing orientation as seen for the other parameters (filament, magnetic field, and gravity), but flatter 
with a very broad shoulder. 

\begin{figure}
\includegraphics[width=\columnwidth]{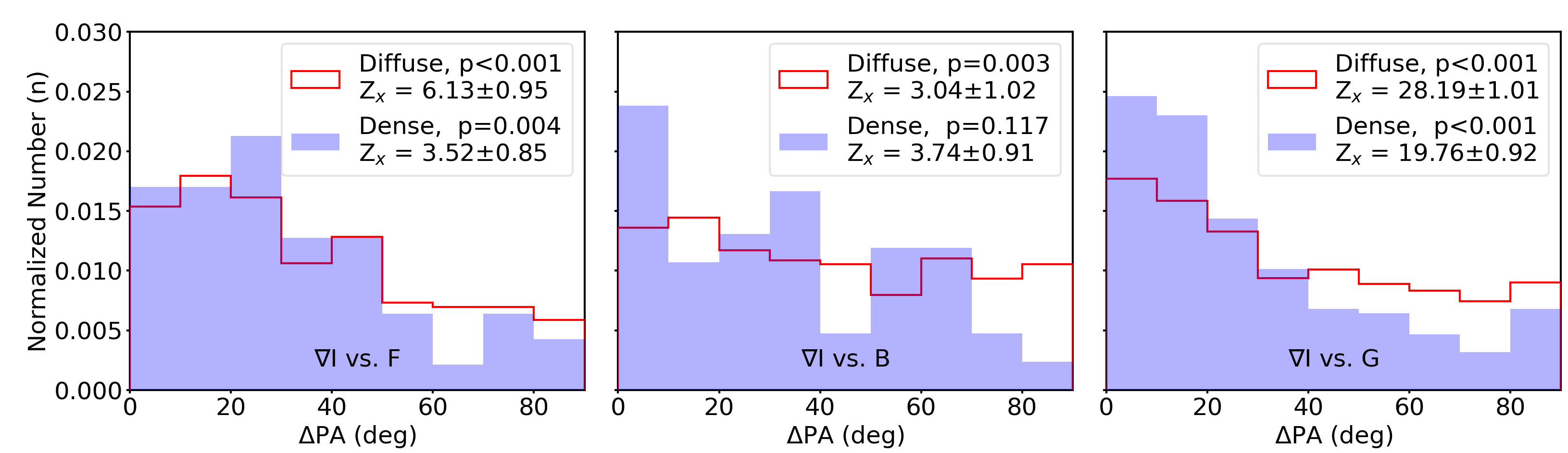}
\caption{Histograms of relative orientations between \IG\ and other spatial parameters over all samples. The samples are separated into low- and high-intensity regions with an intensity threshold of 2.0 \mjya. The p-value from the Rayleigh's test indicates the similarity between an observed and a uniform distribution, and p$<0.05$ favors a non-uniform distribution. PRS Z$_\textrm{x}$ indicates whether the relative orientations tend to be parallel ($Z_x > 0$) or perpendicular ($Z_x < 0$).}\label{fig:hist_dpa_2I_IG}
\end{figure}

\begin{figure*}
\includegraphics[width=\textwidth]{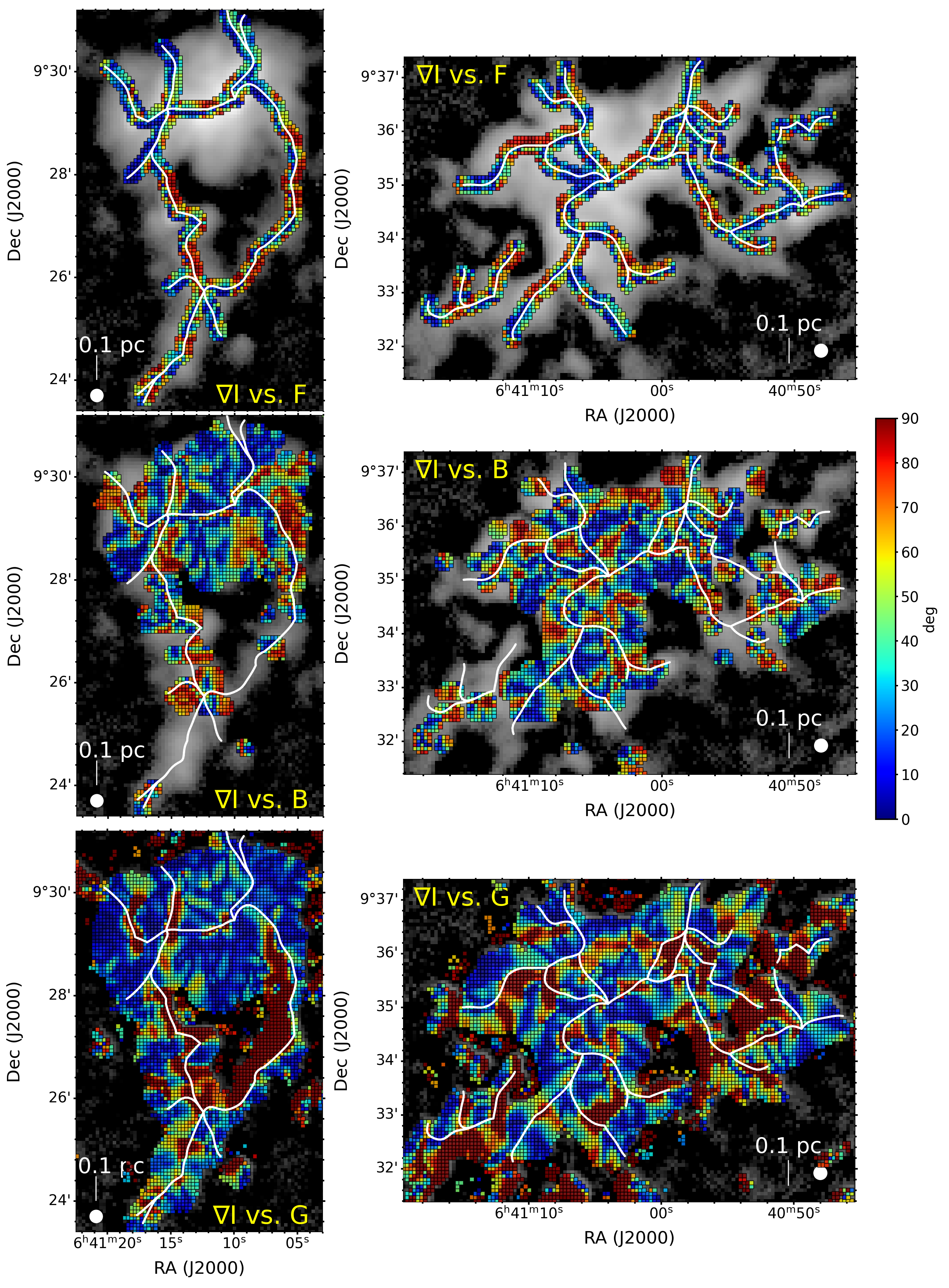}
\caption{Maps of relative orientations between intensity gradient (\IG), filament (F), local magnetic field (B), and local gravity (G) orientations toward NGC 2264.}\label{fig:DPAmap_IG}
\end{figure*}

\begin{figure*}
\includegraphics[width=\textwidth]{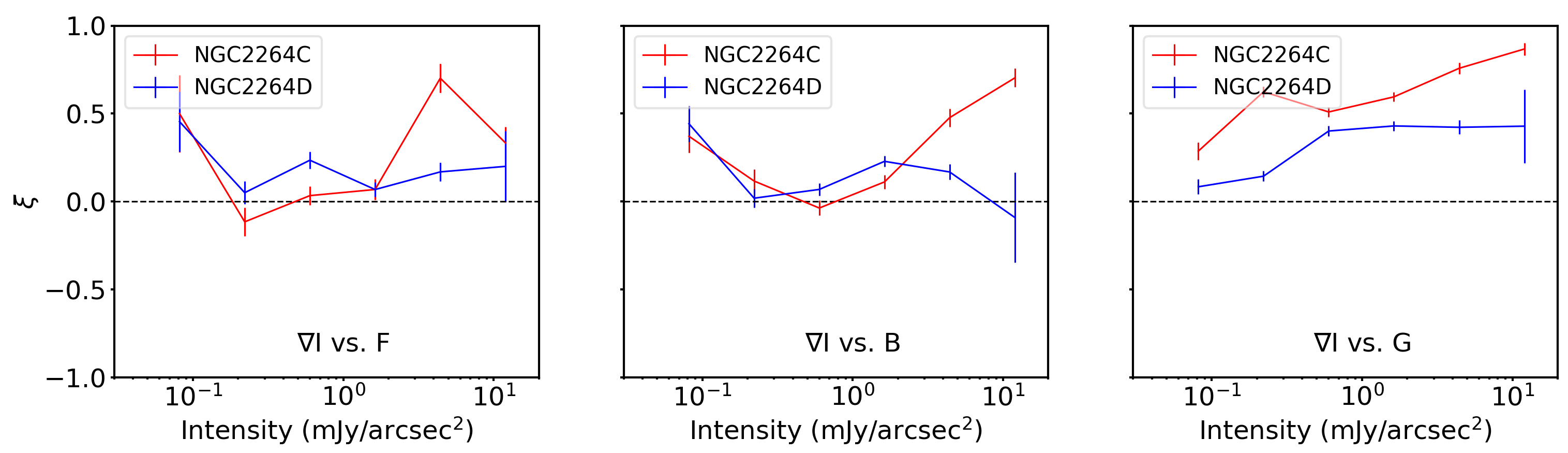}
\caption{
Histogram shape parameter $\xi$ vs. local intensity for \IG\ and other parameters toward 2264C (red) and D (blue). All pairs tend to be aligned, and show a tendency of increasing alignment as the local intensity increases, except for \IG\ vs. B in NGC 2264 D. }\label{fig:HRO_IG}
\end{figure*}

This is possibly because our resolved physical scale captures both filamentary structures and clumps, which are both delineated by closed intensity contours. These closed contours typically lead to a broad distribution in intensity gradient orientations with a broad peak that is reflecting a prevailing shape.
\autoref{fig:hist_dpa_2I_IG} shows histograms of the relative orientations between the intensity gradient \IG\ and other spatial parameters, with the corresponding visualization maps in \autoref{fig:DPAmap_IG}. All the plots reveal a trend that \IG\ is more aligned with filaments, magnetic field, and local gravity toward high-intensity regions. This trend is most obvious with local gravity, suggesting that the density structures are shaped by an increasingly dominating gravitational force. \autoref{fig:HRO_IG} presents how the histogram shape changes with local intensity for the relative alignments between intensity gradient and the other spatial parameters. \IG\ is preferentially aligned with all parameters.

\section{Histograms of Relative Orientations in 2264C and 2264D}\label{sec:2p_hist_NS}
\begin{figure}
\includegraphics[width=\columnwidth]{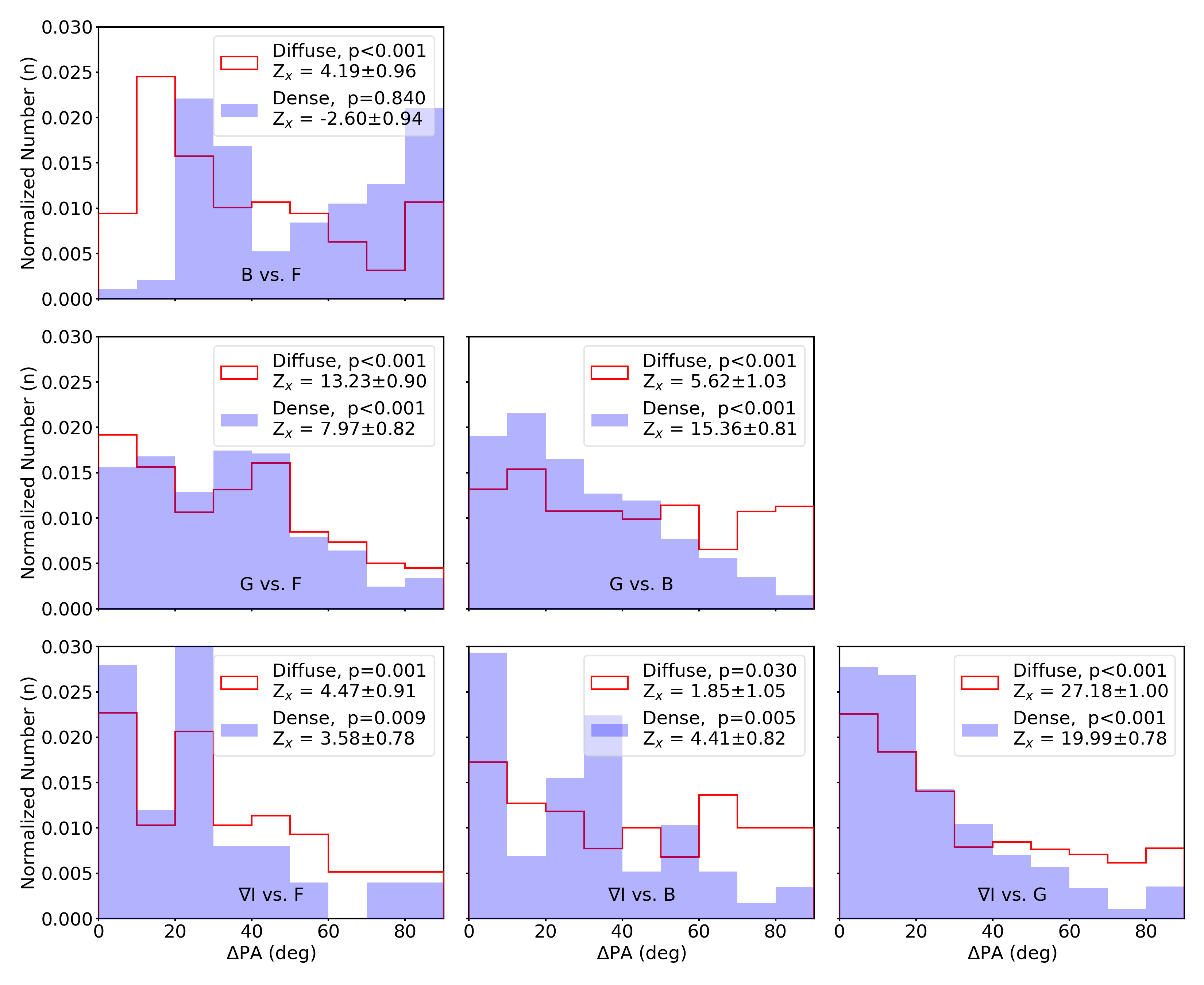}
\caption{Histograms of relative orientations between all spatial parameters selected in 2264C. The samples are separated into low-density and high-intensity regions with an intensity threshold of 2.0 \mjya. The p-value from the Rayleigh's test indicates the similarity between an observed and a uniform distribution, and p$<0.05$ favors a non-uniform distribution. PRS Z$_\textrm{x}$ indicates whether the relative orientations tend to be parallel ($Z_x > 0$) or perpendicular ($Z_x < 0$).}\label{fig:hist_dpa_2I_S}
\end{figure}

\begin{figure}
\includegraphics[width=\columnwidth]{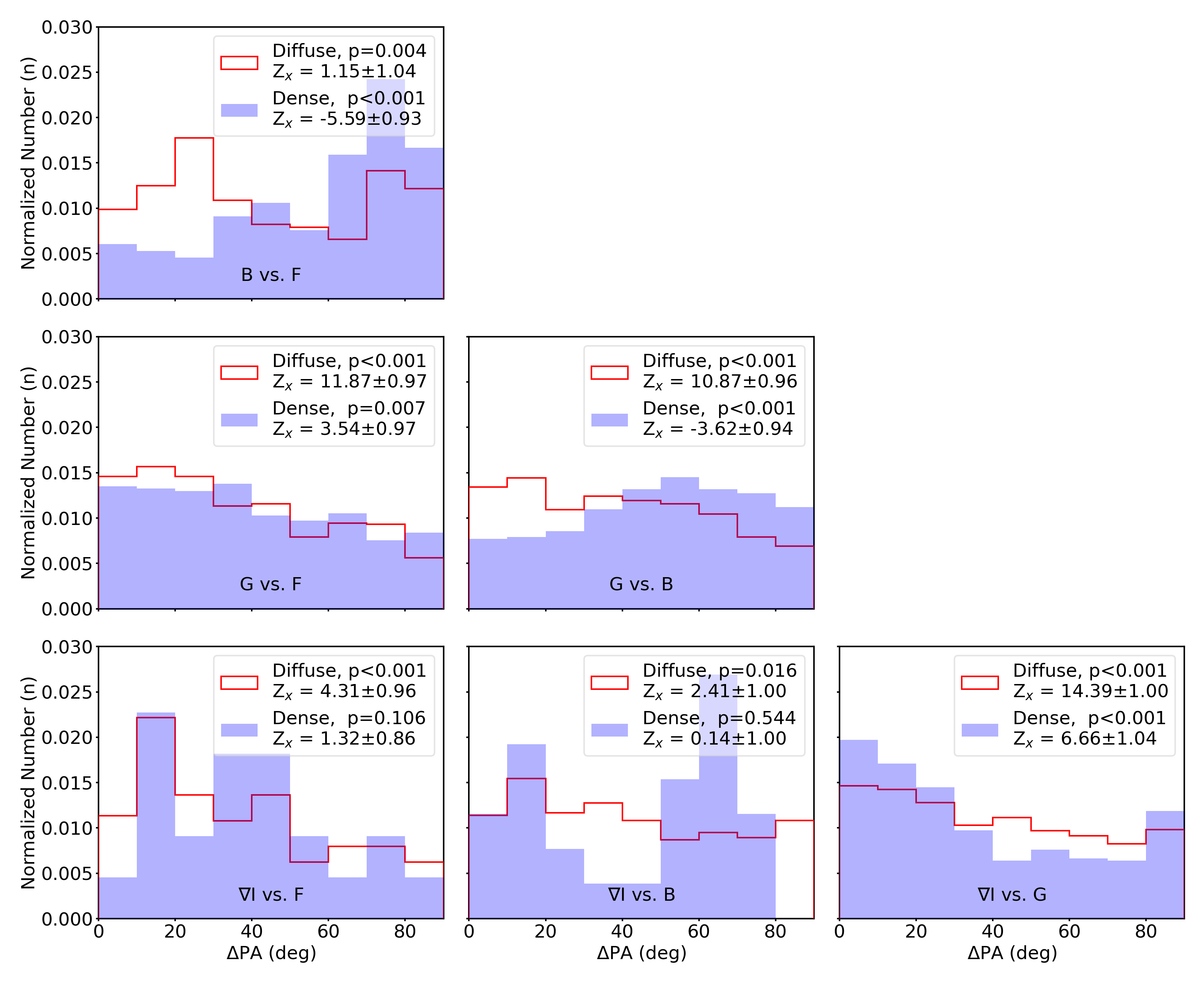}
\caption{Identical to \autoref{fig:hist_dpa_2I_S} but for 2264D.
}\label{fig:hist_dpa_2I_N}
\end{figure}

\autoref{fig:hist_dpa_2I_S} and \autoref{fig:hist_dpa_2I_N} present the histograms of relative orientations between all spatial parameters selected in 2264C and 2264D separately. For most of the pairs, the overall trends in 2264C and 2264D are not significantly different, and the only pair showing an obvious difference is G vs B, which is further discussed in \autoref{sec:2p_statis} and \autoref{sec:2p_dis}.

\section{Magnetic Field Strength Estimates Using Different Molecular Lines}\label{sec:CF_otherline}
\begin{figure*}
\includegraphics[width=\textwidth]{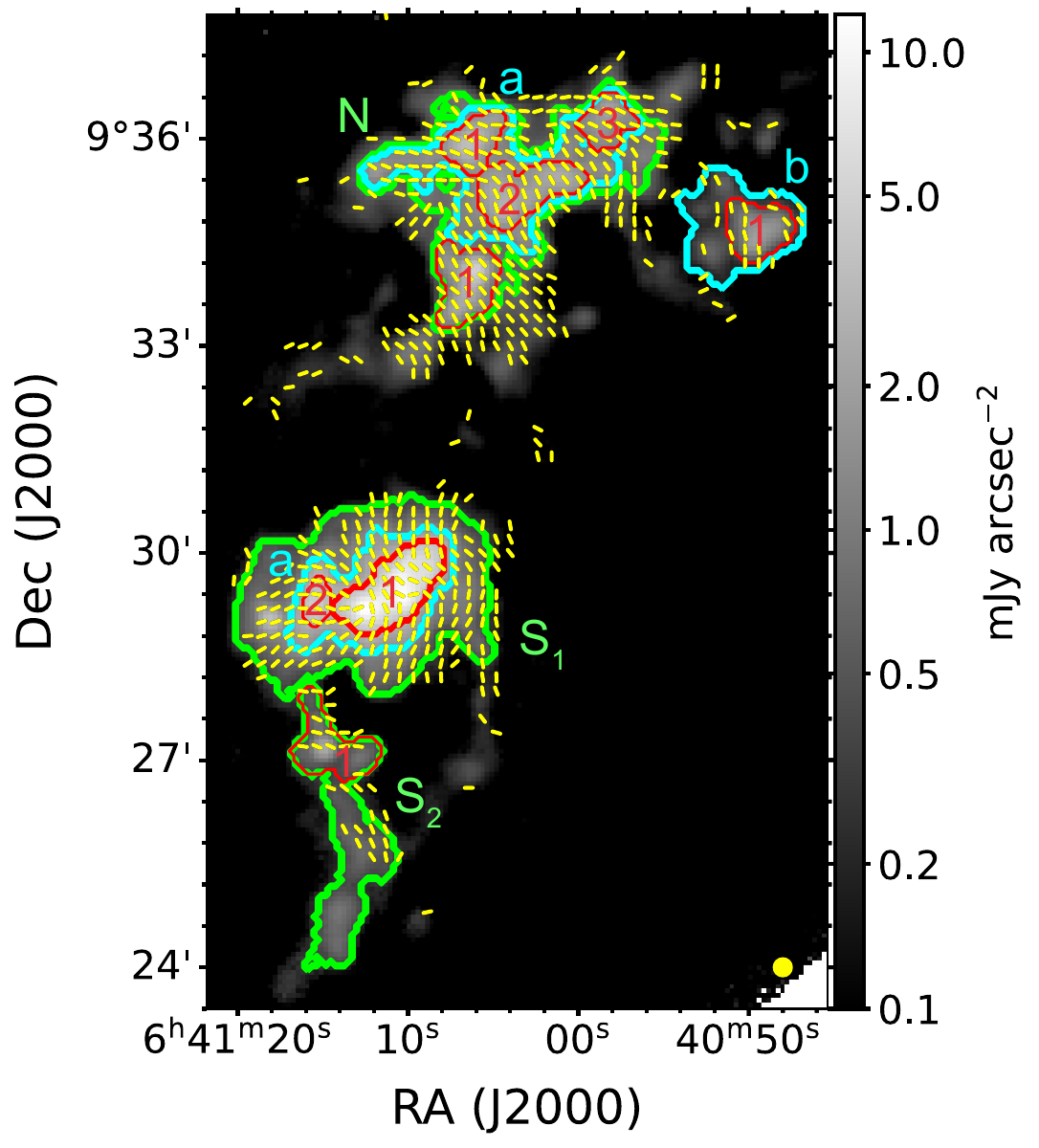}
\caption{Identified clumps in NGC 2264. Contours are the boundaries of the clumps identified using Dendrogram. The green, cyan, and red labels indicate whether the clumps correspond to trunks, branches, or leaves. }\label{fig:clump}
\end{figure*}

We estimate magnetic field strengths, mass-to-flux ratios, and energy ratios for all structures identified in NGC 2264 with different gas tracers. The structures identified from the Dendrogram algorithm are labelled in \autoref{fig:clump}. They are named by three symbols: trunk (N, S$_1$, or S$_2$) + branch (a or b) + leaf (1, 2, or 3). If a level is not identified, it is omitted.  

In addition to the \13CO and \C18O data, we further adopt the N$_2$H$^+$ data in \citet{pe06} toward 2264D. The estimates using \13CO, \C18O, and N$_2$H$^+$ are listed in \autoref{tab:CF_13CO}, \autoref{tab:CF_C18O}, and \autoref{tab:CF_N2H+}, respectively. In general, the estimated magnetic field strengths increase by 10-50\% if \13CO is used instead of \C18O, and they decrease by 10--70\% if N$_2$H$^+$ is used instead of \C18O.

\begin{deluxetable*}{cccccccc}
\tablecaption{Magnetic field strengths and mass-to-flux ratios in NGC 2264, using $^{13}$CO.\label{tab:CF_13CO}}
\renewcommand{\thetable}{\arabic{table}}
\tablenum{2}
\tablehead{\colhead{Leaves} & \colhead{$n_{H_2}$} & \colhead{$\sigma_{v,NT}$} & \colhead{$\delta \phi$} & \colhead{$B_{pos}$} & \colhead{$\lambda$}  & \colhead{$T/W$} & \colhead{$M/W$}\\
\colhead{} & \colhead{$\textrm{cm}^{-3}$} &\colhead{(\kms)} &\colhead{(deg)} &\colhead{(mG)} & \colhead{} & \colhead{} & \colhead{} }
\startdata
\hline
S${_1}$a1 & $(3.7\pm0.2)\times10^5$ & $1.21\pm0.43$ & $34.0\pm0.3$ & $0.43\pm0.16$ & $1.5\substack{+0.3 \\ -0.5}$ & $0.6\substack{+0.4 \\ -0.4}$ & $0.2\substack{+0.1 \\ -0.1}$\\
S${_1}$a2 & $(6.2\pm0.4)\times10^5$ & $0.92\pm0.28$ & $9.5\pm1.4$ & $1.56\pm0.54$ & $0.2\substack{+0.1 \\ -0.1}$ & $1.4\substack{+0.7 \\ -0.7}$ & $7.0\substack{+4.3 \\ -4.3}$ \\
S${_2}$1 & $(5.6\pm0.3)\times10^4$ & $0.95\pm0.26$ & $14.3\pm2.1$ & $0.32\pm0.10$ & $0.2\substack{+0.1 \\ -0.1}$ & $3.9\substack{+1.8 \\ -1.8}$ & $8.3\substack{+4.7 \\ -4.6}$ \\
N1 & $(1.5\pm0.1)\times10^5$ & $0.90\pm0.20$ & $8.5\pm0.4$ & $0.82\pm0.18$ & $0.2\substack{+0.1 \\ -0.1}$ & $1.3\substack{+0.5 \\ -0.5}$ & $7.3\substack{+3.1 \\ -3.0}$ \\
Na1 & $(2.2\pm0.2)\times10^5$ & $0.93\pm0.46$ & $16.4\pm2.0$ & $0.53\pm0.27$ & $0.5\substack{+0.1 \\ -0.3}$ & $1.8\substack{+1.3 \\ -1.3}$ & $2.8\substack{+2.3 \\ -2.3}$ \\
Na2 & $(1.4\pm0.1)\times10^5$ & $1.01\pm0.21$ & $11.4\pm0.7$ & $0.67\pm0.15$ & $0.3\substack{+0.1 \\ -0.1}$ & $1.4\substack{+0.6 \\ -0.6}$ & $4.5\substack{+1.9 \\ -1.9}$ \\
Na3 & $(1.6\pm0.1)\times10^5$ & $1.10\pm0.36$ & $9.3\pm0.9$ & $0.93\pm0.32$ & $0.2\substack{+0.1 \\ -0.1}$ & $3.4\substack{+1.9 \\ -1.9}$ & $17.1\substack{+10.3 \\ -10.2}$ \\
Nb1 & $(8.7\pm0.6)\times10^4$ & $1.04\pm0.08$ & $28.0\pm2.7$ & $0.22\pm0.03$ & $0.4\substack{+0.1 \\ -0.1}$ & $3.4\substack{+0.5 \\ -0.5}$ & $1.9\substack{+0.5 \\ -0.5}$ \\
\hline
S${_1}$a & $(1.5\pm0.1)\times10^5$ & $1.09\pm0.39$ & $30.2\pm0.3$ & $0.28\pm0.10$ & $1.3\substack{+0.5 \\ -0.5}$ & $0.50\substack{+0.3 \\ -0.3}$ & $0.2\substack{+0.1 \\ -0.1}$ \\
Na & $(5.1\pm0.4)\times10^4$ & $1.01\pm0.32$ & $21.5\pm0.4$ & $0.21\pm0.07$ & $0.7\substack{+0.2 \\ -0.2}$ & $0.9\substack{+0.5 \\ -0.5}$ & $0.8\substack{+0.5 \\ -0.5}$ \\
Nb & $(2.5\pm0.2)\times10^4$ & $1.09\pm0.12$ & $23.0\pm1.8$ & $0.15\pm0.02$ & $0.3\substack{+0.1 \\ -0.1}$ & $3.9\substack{+0.8 \\ -0.8}$ & $3.1\substack{+0.8 \\ -0.8}$ \\
\hline
S${_1}$ & $(4.7\pm0.3)\times10^4$ & $1.06\pm0.37$ & $31.1\pm0.3$ & $0.14\pm0.05$ & $1.4\substack{+0.3 \\ -0.6}$ & $0.5\substack{+0.3 \\ -0.3}$ & $0.2\substack{+0.1 \\ -0.1}$ \\
S${_2}$ & $(2.4\pm0.2)\times10^4$ & $0.74\pm0.29$ & $31.4\pm1.3$ & $0.07\pm0.03$ & $0.7\substack{+0.3 \\ -0.3}$ & $1.8\substack{+1.1 \\ -1.2}$ & $0.8\substack{+0.5 \\ -0.5}$ \\
N & $(3.4\pm0.2)\times10^4$ & $1.00\pm0.31$ & $22.6\pm0.4$ & $0.16\pm0.05$ & $0.8\substack{+0.2 \\ -0.3}$ & $0.8\substack{+0.4 \\ -0.4}$ & $0.6\substack{+0.4 \\ -0.4}$ \\
\enddata
\tablecomments{Magnetic field strengths and mass-to-flux ratios derived from DCF method. The uncertainties listed here are obtained from propagating the observational uncertainties through the corresponding equations using a Monte Carlo approach. Possible additional systematic uncertainties due to, e.g., the unknown dust opacity, are not included. }
{\addtocounter{table}{-1}}
\end{deluxetable*}

\begin{deluxetable*}{cccccccc}
\tablecaption{Magnetic field strengths and mass-to-flux ratios in NGC 2264, using C$^{18}$O.\label{tab:CF_C18O}}
\renewcommand{\thetable}{\arabic{table}}
\tablenum{3}
\tablehead{\colhead{Leaves} & \colhead{$n_{H_2}$} & \colhead{$\sigma_{v,NT}$} & \colhead{$\delta \phi$} & \colhead{$B_{pos}$} & \colhead{$\lambda$}  & \colhead{$T/W$} & \colhead{$M/W$}\\
\colhead{} & \colhead{$\textrm{cm}^{-3}$} &\colhead{(\kms)} &\colhead{(deg)} &\colhead{(mG)} & \colhead{} & \colhead{} & \colhead{} }
\startdata
\hline
S${_1}$a1 & $(3.7\pm0.3)\times10^5$ & $0.97\pm0.25$ & $34.0\pm0.4$ & $0.35\pm0.09$ & $1.6\substack{+0.4 \\ -0.5}$ & $0.4\substack{+0.2 \\ -0.2}$ & $0.1\substack{+0.1 \\ -0.1}$\\
S${_1}$a2 & $(6.2\pm0.4)\times10^5$ & $0.69\pm0.18$ & $9.5\pm1.4$ & $1.16\pm0.35$ & $0.3\substack{+0.1 \\ -0.1}$ & $0.8\substack{+0.4 \\ -0.3}$ & $3.8\substack{+2.1 \\ -2.1}$ \\
S${_2}$1 & $(5.6\pm0.4)\times10^4$ & $0.58\pm0.09$ & $14.3\pm2.1$ & $0.19\pm0.04$ & $0.3\substack{+0.1 \\ -0.1}$ & $1.5\substack{+0.4 \\ -0.4}$ & $2.9\substack{+1.2 \\ -1.2}$ 
\\
N1 & $(1.5\pm0.1)\times10^5$ & $0.90\pm0.07$ & $8.5\pm0.4$ & $0.82\pm0.08$ & $0.2\substack{+0.1 \\ -0.1}$ & $1.2\substack{+0.2 \\ -0.2}$ & $7.0\substack{+1.3 \\ -1.3}$ \\
Na1 & $(2.2\pm0.2)\times10^5$ & $0.92\pm0.11$ & $16.4\pm2.0$ & $0.55\pm0.09$ & $0.4\substack{+0.1 \\ -0.1}$ & $1.4\substack{+0.3 \\ -0.3}$ & $2.3\substack{+0.7 \\ -0.7}$  \\
Na2 & $(1.4\pm0.1)\times10^5$ & $1.22\pm0.15$ & $11.4\pm0.7$ & $0.80\pm0.11$ & $0.2\substack{+0.1 \\ -0.1}$ & $2.0\substack{+0.5 \\ -0.5}$ & $6.4\substack{+1.7 \\ -1.7}$ \\
Na3 & $(1.6\pm0.1)\times10^5$ & $0.78\pm0.07$ & $9.3\pm0.9$ & $0.67\pm0.09$ & $0.2\substack{+0.1 \\ -0.1}$ & $1.7\substack{+0.3 \\ -0.3}$ & $8.0\substack{+2.0 \\ -2.0}$ \\
Nb1 & $(8.7\pm0.6)\times10^4$ & $0.70\pm0.08$ & $28.0\pm2.7$ & $0.15\pm0.02$ & $0.6\substack{+0.1 \\ -0.1}$ & $1.6\substack{+0.3 \\ -0.3}$ & $0.8\substack{+0.2 \\ -0.2}$ \\
\hline
S${_1}$a & $(1.5\pm0.1)\times10^5$ & $0.94\pm0.24$ & $30.2\pm0.3$ & $0.24\pm0.06$ & $1.4\substack{+0.4 \\ -0.4}$ & $0.3\substack{+0.2 \\ -0.2}$ & $0.2\substack{+0.1 \\ -0.1}$ \\
Na & $(5.1\pm0.4)\times10^4$ & $1.04\pm0.23$ & $21.5\pm0.4$ & $0.22\pm0.05$ & $0.6\substack{+0.1 \\ -0.2}$ & $0.9\substack{+0.4 \\ -0.4}$ & $0.8\substack{+0.4 \\ -0.4}$ \\
Nb & $(2.5\pm0.2)\times10^4$ & $0.72\pm0.09$ & $23.0\pm1.8$ & $0.10\pm0.01$ & $0.5\substack{+0.1 \\ -0.1}$ & $1.8\substack{+0.4 \\ -0.4}$ & $1.3\substack{+0.4 \\ -0.4}$ \\
\hline
S${_1}$ & $(4.7\pm0.3)\times10^4$ & $0.91\pm0.25$ & $31.1\pm0.3$ & $0.13\pm0.04$ & $1.3\substack{+0.5 \\ -0.3}$ & $0.4\substack{+0.2 \\ -0.2}$ & $0.2\substack{+0.1 \\ -0.1}$ \\
S${_2}$ & $(2.4\pm0.2)\times10^4$ & $0.57\pm0.09$ & $31.4\pm1.3$ & $0.06\pm0.01$ & $0.8\substack{+0.1 \\ -0.2}$ & $1.0\substack{+0.3 \\ -0.3}$ & $0.4\substack{+0.1 \\ -0.1}$ \\
N & $(3.4\pm0.2)\times10^4$ & $1.01\pm0.23$ & $22.6\pm0.4$ & $0.16\pm0.04$ & $0.7\substack{+0.2 \\ -0.2}$ & $0.7\substack{+0.3 \\ -0.3}$ & $0.6\substack{+0.2 \\ -0.2}$ \\
\enddata
\tablecomments{Identical to \autoref{tab:CF_13CO}.
}
{\addtocounter{table}{-1}}
\end{deluxetable*}

\begin{deluxetable*}{cccccccc}
\tablecaption{Magnetic field strengths and mass-to-flux ratios in NGC 2264, using N$_2$H$^{+}$.\label{tab:CF_N2H+}}
\renewcommand{\thetable}{\arabic{table}}
\tablenum{4}
\tablehead{\colhead{Leaves} & \colhead{$n_{H_2}$} & \colhead{$\sigma_{v,NT}$} & \colhead{$\delta \phi$} & \colhead{$B_{pos}$} & \colhead{$\lambda$}  & \colhead{$T/W$} & \colhead{$M/W$}\\
\colhead{} & \colhead{$\textrm{cm}^{-3}$} &\colhead{(\kms)} &\colhead{(deg)} &\colhead{(mG)} & \colhead{} & \colhead{} & \colhead{} }
\startdata
\hline
N1 & $(1.5\pm0.2)\times10^5$ & $0.76\pm0.04$ & $8.5\pm0.4$ & $0.68\pm0.05$ & $0.2\substack{+0.1 \\ -0.1}$ & $0.9\substack{+0.1 \\ -0.1}$ & $4.9\substack{+0.7 \\ -0.7}$ \\
Na1 & $(2.2\pm0.2)\times10^5$ & $0.61\pm0.12$ & $16.4\pm2.0$ & $0.34\pm0.08$ & $0.6\substack{+0.1 \\ -0.1}$ & $0.7\substack{+0.2 \\ -0.2}$ & $1.0\substack{+0.4 \\ -0.4}$  \\
Na2 & $(1.4\pm0.1)\times10^5$ & $1.06\pm0.18$ & $11.4\pm0.7$ & $0.70\pm0.13$ & $0.2\substack{+0.1 \\ -0.1}$ & $1.5\substack{+0.5 \\ -0.5}$ & $4.9\substack{+1.7 \\ -1.7}$ \\
Na3 & $(1.6\pm0.1)\times10^5$ & $0.59\pm0.04$ & $9.3\pm0.9$ & $0.48\pm0.06$ & $0.3\substack{+0.1 \\ -0.1}$ & $1.0\substack{+0.1 \\ -0.1}$ & $4.2\substack{+0.9 \\ -0.9}$ \\
Nb1 & $(8.7\pm0.6)\times10^4$ & $0.41\pm0.07$ & $28.0\pm2.7$ & $0.09\pm0.02$ & $1.0\substack{+0.2 \\ -0.2}$ & $0.7\substack{+0.2 \\ -0.2}$ & $0.3\substack{+0.1 \\ -0.1}$ \\
\hline
Na & $(5.1\pm0.4)\times10^4$ & $0.86\pm0.13$ & $21.5\pm0.4$ & $0.18\pm0.03$ & $0.7\substack{+0.1 \\ -0.1}$ & $0.6\substack{+0.2 \\ -0.2}$ & $0.6\substack{+0.2 \\ -0.2}$ \\
Nb & $(2.5\pm0.2)\times10^4$ & $0.53\pm0.15$ & $23.0\pm1.8$ & $0.07\pm0.02$ & $0.7\substack{+0.2 \\ -0.2}$ & $1.1\substack{+0.5 \\ -0.5}$ & $0.8\substack{+0.4 \\ -0.4}$ \\
\hline
N & $(3.4\pm0.2)\times10^4$ & $0.87\pm0.13$ & $22.6\pm0.4$ & $0.14\pm0.02$ & $0.8\substack{+0.1 \\ -0.1}$ & $0.5\substack{+0.2 \\ -0.2}$ & $0.4\substack{+0.1 \\ -0.1}$ \\ 
\enddata
\tablecomments{Identical to \autoref{tab:CF_13CO}.}
{\addtocounter{table}{-1}}
\end{deluxetable*}


\facilities{JCMT}
\software{Aplpy \citep{aplpy2012,aplpy2019}, Astrodendro \citep{go09}, Astropy \citep{astropy2013,astropy2018}, BTS \citep{cl18}, DisPerSE \citep{so11}, Filchap \citep{su19}, NumPy \citep{numpy}, SciPy \citep{scipy}, scousePy \citep{henshaw16,henshaw19}, Smurf \citep{be05,ch13}, Starlink \citep{cu14}}

\bibliography{main}{}
\bibliographystyle{aasjournal}
\end{document}